%% file: LLlevSem.tex
\documentclass[authoryear]{elsarticle}
\usepackage[hyperindex,breaklinks]{hyperref}

\usepackage{amssymb}
\usepackage{stmaryrd}

\usepackage{amsmath} 

\usepackage{cmll}
\usepackage{proof}
\usepackage[all]{xy}

\usepackage{tikz}
\usetikzlibrary{arrows,positioning,shapes}
\usetikzlibrary{decorations.markings}
\usepackage{tikz-inet}

\usepackage{graphicx}
\usepackage{color}

\usepackage{xspace}

\usepackage{natbib}

\usepackage[active]{srcltx}

\journal{Information and Computation}

\input{macros.tex}

\begin{document}

% For reference on how to use the "support" macro:
% 	\author[DM]{Damiano Mazza\fnref{support}}
% 	\address[DM]{CNRS-Universit\'e Paris 13, LIPN (UMR 7030)}
% 	\ead{damiano.mazza@lipn.univ-paris13.fr} 
% 
% 	\fntext[support]{Partially supported by project NOCoST (ANR, JC05\_43380).}

\begin{frontmatter}
	\title{An Abstract Approach to Stratification in Linear Logic}%\tnoteref{support}}
%	\tnotetext[support]{Work partially supported by ANR projet \textsc{Complice} and CNRS PICS ``Logique lin\'eaire et applications''.}
	
	\author[PB]{Pierre Boudes}
	\address[PB]{Universit\'e Paris 13, Sorbonne Paris Cit\'e, LIPN, CNRS}
	\ead{Pierre.Boudes@lipn.univ-paris13.fr}
 
	\author[DM]{Damiano Mazza}
	\address[DM]{CNRS, UMR 7030, LIPN, Universit\'e Paris 13, Sorbonne Paris Cit\'e}
	\ead{Damiano.Mazza@lipn.univ-paris13.fr}

	\author[LT]{Lorenzo Tortora de Falco}
	\address[LT]{Dipartimento di Filosofia, Universit\`a Roma Tre}
	\ead{tortora@uniroma3.it}

	\begin{abstract}
		\noindent We study the notion of stratification, as used in subsystems of linear logic with low complexity bounds on the cut-elimination procedure (the so-called ``light'' subsystems), from an abstract point of view, introducing a logical system in which stratification is handled by a separate modality. This modality, which is a generalization of the paragraph modality of Girard's light linear logic, arises from a general categorical construction applicable to all models of linear logic. We thus learn that stratification may be formulated independently of exponential modalities; when it is forced to be connected to exponential modalities, it yields interesting complexity properties. In particular, from our analysis stem three alternative reformulations of Baillot and Mazza's linear logic by levels: one geometric, one interactive, and one semantic.
	\end{abstract}

\end{frontmatter}

%%%%%%%%%%%%%%%%%%%%%%%%%%%%%%%%%%%%%%%%%%%%%%%%%%%%%%%%%%%%%%%%%
% Intro
%%%%%%%%%%%%%%%%%%%%%%%%%%%%%%%%%%%%%%%%%%%%%%%%%%%%%%%%%%%%%%%%%
\section*{Introduction}
\input{Intro.tex}

%%%%%%%%%%%%%%%%%%%%%%%%%%%%%%%%%%%%%%%%%%%%%%%%%%%%%%%%%%%%%%%%
% Syntax
%%%%%%%%%%%%%%%%%%%%%%%%%%%%%%%%%%%%%%%%%%%%%%%%%%%%%%%%%%%%%%%%
\section{Stratified Linear Logic}
\label{sect:StratLL}

\subsection{The logical system}
\input{StratLL.tex}

\subsection{Bounded time subsystems}
\label{sect:Bounded}
\input{Bounded.tex}

\subsection{Proof nets}
\label{sect:ProofNets}
\input{ProofNets.tex}

\subsection{Cut-elimination}
\label{sect:CutElim}
\input{CutElim.tex}

%%%%%%%%%%%%%%%%%%%%%%%%%%%%%%%%%%%%%%%%%%%%%%%%%%%%%%%%%%%%%%%%%
% Semantics
%%%%%%%%%%%%%%%%%%%%%%%%%%%%%%%%%%%%%%%%%%%%%%%%%%%%%%%%%%%%%%%%%
\section{A Categorical Construction for Stratification}
\subsection{Categorical models of linear logic}
\label{sect:Models}
\input{Models.tex}

\subsection{Categorical models of stratified linear logic}
\label{sect:StratModels}
\input{StratModels.tex}

\subsection{\Objs}\label{sect:Objs}
\input{Obj.tex}

\subsection{Building stratified models}
\input{BuildStratMod.tex}

%%%%%%%%%%%%%%%%%%%%%%%%%%%%%%%%%%%%%%%%%%%%%%%%%%%%%%%%%%%%%%%%%
% Applications
%%%%%%%%%%%%%%%%%%%%%%%%%%%%%%%%%%%%%%%%%%%%%%%%%%%%%%%%%%%%%%%%%
\section{Applications to Bounded Complexity}
\subsection{A geometric definition of \LLlev}
\label{sect:Geom}
\input{L3.tex}

\subsection{Denotational semantics of \LLlev}
\input{L3Sem.tex}

\subsection{Interactive characterization of \LLlev}
\label{sect:Interactive}
\input{Interactive.tex}

\subsection{Semantic characterization of \LLlev}
\label{sect:Completeness}
\input{RelComp.tex}

%%%%%%%%%%%%%%%%%%%%%%%%%%%%%%%%%%%%%%%%%%%%%%%%%%%%%%%%%%%%%%%%%
% Conclusions
%%%%%%%%%%%%%%%%%%%%%%%%%%%%%%%%%%%%%%%%%%%%%%%%%%%%%%%%%%%%%%%%%
% \section{Conclusions}
% Slogan
% SLL
% Bornes fortes

%%%%%%%%%%%%%%%%%%%%%%%%%%%%%%%%%%%%%%%%%%%%%%%%%%%%%%%%%%%%%%%%%
% Biblio
%%%%%%%%%%%%%%%%%%%%%%%%%%%%%%%%%%%%%%%%%%%%%%%%%%%%%%%%%%%%%%%%%
\bibliographystyle{elsarticle-harv}
\bibliography{Biblio}

%%%%%%%%%%%%%%%%%%%%%%%%%%%%%%%%%%%%%%%%%%%%%%%%%%%%%%%%%%%%%%%%%
% Appendices
%%%%%%%%%%%%%%%%%%%%%%%%%%%%%%%%%%%%%%%%%%%%%%%%%%%%%%%%%%%%%%%%%
\appendix
\section{Some Categorical Background}
\label{sect:CatBack}
\input{CatBack.tex}

% \section{Coherence Spaces are Swap-sensitive}
% \label{sect:CohSwap}
% \input{CohSwap.tex}

\end{document}

%% file: macros.tex
\newtheorem{thm}{Theorem}
\newtheorem{prop}[thm]{Proposition}
\newtheorem{lem}[thm]{Lemma}

\newdefinition{defn}{Definition~}
\newproof{pf}{Proof}

\newcommand{\refsect}[1]{Sect.~\ref{sect:#1}}
\newcommand{\refapp}[1]{Appendix~\ref{sect:#1}}
\newcommand{\refdef}[1]{Definition~\ref{def:#1}}
\newcommand{\refth}[1]{Theorem~\ref{th:#1}}
\newcommand{\refprop}[1]{Proposition~\ref{prop:#1}}
\newcommand{\reflemma}[1]{Lemma~\ref{lemma:#1}}

\newcommand{\reffig}[1]{Fig.~\ref{fig:#1}}
\newcommand{\reftab}[1]{Table~\ref{tab:#1}}

\def\cB{\mathcal B}
\def\cX{\mathcal X}

\def\catA{\mathcal A}
\def\catB{\mathcal B}
\def\catC{\mathcal C}
\def\catM{\mathcal M}
\def\catL{\mathcal L}
\def\objA{A}
\def\objB{B}

\def\ie{\textit{i.e.}\xspace}
\def\cf{\textit{cf.}\xspace}
\def\lac{\mbox{$\lambda$-calculus}\xspace}
\def\lat{\mbox{$\lambda$-term}}
\def\pps{net}
\def\ps{net}
\def\scorr{\mbox{$\parg$-correct}}
\def\dr{DR-net}

\def\sps{sequentializable \ps}
\def\ips{indexable \ps}
\def\pn{proof net}

\def\Ps{Net}
\def\Sps{Sequentializable \ps}

\def\pax{auxiliary port}
\def\pal{principal port}
\def\struct{$\flat$-formula}
\def\obj{object with involutions}
\def\objs{objects with involutions}
\def\Obj{Object with involutions}
\def\Objs{Objects with involutions}

\newcommand{\Nat}{\ensuremath{\mathbb N}}
\newcommand{\Int}{\ensuremath{\mathbb Z}}

\newcommand{\Set}{\ensuremath{\mathbf{Set}}}
\newcommand{\Cat}{\ensuremath{\mathbf{Cat}}}

\newcommand{\SymMonCat}{\ensuremath{\mathbf{SMCat}}}
\newcommand{\id}{id}
\newcommand{\Id}{Id}
\newcommand{\sem}[1]{\llbracket #1\rrbracket}
\newcommand{\altsem}[1]{\llparenthesis #1\rrparenthesis}

\newcommand{\Cox}{\mathcal I}
\newcommand{\Inv}[1]{{#1}^\Cox}
\newcommand{\SInv}{\mathop{\mathrm{Inv}}}
\newcommand{\catOne}{\mathbf{1}}
\newcommand{\term}{T}
\newcommand{\eval}{\mathrm{eval}}

\newcommand{\stratLL}{\ensuremath{\mathbf{LL}_\parg}}

\newcommand{\ELL}{\ensuremath{\mathbf{ELL}}}
\newcommand{\LLL}{\ensuremath{\mathbf{LLL}}}

\newcommand{\LLlev}{\ensuremath{\mathbf{L^3}}}
\newcommand{\LLLlev}{\ensuremath{\mathbf{L^4}}}

\newcommand{\SLL}{\ensuremath{\mathbf{SLL}}}

\newcommand{\lone}{1}
\newcommand{\ltens}{\otimes}
\newcommand{\lbot}{\bot}
\newcommand{\lpar}{\parr}
\newcommand{\llinimp}{\multimap}
\newcommand{\lzero}{0}
\newcommand{\lplus}{\oplus}
\newcommand{\ltop}{\top}
\newcommand{\lwith}{\with}
\newcommand{\parg}{\S}

\newcommand{\gpipe}{\mathrel{|}}
\newcommand{\ts}{\vdash}
\newcommand{\etq}[1]{\mbox{\scriptsize #1}}
\newcommand{\OneRule}[1]{
\begin{center}
	\makebox[\textwidth][c]{
		\makebox[\textwidth][c]{
			\begin{minipage}[c]{\textwidth}#1\end{minipage}
		}
	}
\end{center}
}
\newcommand{\TwoRules}[2]{
\begin{center}
	\makebox[\textwidth][c]{
		\makebox[0.5\textwidth][c]{
			\begin{minipage}[c]{0.5\textwidth}#1\end{minipage}
		}
		\makebox[0.5\textwidth][c]{
			\begin{minipage}[c]{0.5\textwidth}#2\end{minipage}
		}
	}
\end{center}
}
\newcommand{\ThreeRules}[3]{
\begin{center}
	\makebox[\textwidth][c]{
		\makebox[0.333\textwidth][c]{
			\begin{minipage}[c]{0.333\textwidth}#1\end{minipage}
		}
		\makebox[0.333\textwidth][c]{
			\begin{minipage}[c]{0.333\textwidth}#2\end{minipage}
		}
		\makebox[0.333\textwidth][c]{
			\begin{minipage}[c]{0.333\textwidth}#3\end{minipage}
		}
	}
\end{center}
}

\newcommand{\betared}{\rightarrow_\beta}
\newcommand{\betareds}{\betared^\ast}
\newcommand{\notaxred}{\mathrel{{\stackrel{\lnot\mathrm{ax}}{\longrightarrow}}{}_\beta}}
\newcommand{\notaxreds}{\mathrel{{\stackrel{\lnot\mathrm{ax}}{\longrightarrow}}{}_\beta^\ast}}
\newcommand{\lift}[1]{\overleftarrow{#1}}
\newcommand{\swapping}{\prec}

\newcommand{\catPN}{\ensuremath{\mathbf{PN}}}

\def\axiom{$\mathsf{axiom}$}
\def\cut{$\mathsf{cut}$}
\def\onel{$\mathsf{one}$}
\def\botl{$\mathsf{bottom}$}
\def\tensor{$\mathsf{tensor}$}
\def\parl{$\mathsf{par}$}

\def\ofcourse{$\mathsf{of\ course}$}
\def\whynot{$\mathsf{why\ not}$}
\def\weak{$\mathsf{weakening}$}
\def\flatl{$\mathsf{flat}$}
\def\paxl{$\mathsf{pax}$}
\def\pargl{$\mathsf{paragraph}$}

\newcommand{\ednd}[1]{\ensuremath{\scriptstyle{#1}}}
\tikzset{netcircle/.style={shape=circle,thick,align=center,minimum size=6mm,font=\sffamily,fill=white,text height=1.5ex, text depth=0.25ex}}
\tikzset{point/.style={circle,inner sep=0pt,minimum size=2pt,fill=red}}
\tikzset{nopoint/.style={circle,inner sep=0pt,minimum size=0pt,fill=none,draw=none}}
\tikzset{netpi/.style={shape=rectangle,draw=black,thick,align=center,minimum width=12mm,fill=white,text height=2ex, text depth=0.75ex,rounded corners}}
\tikzset{lab/.style={rectangle,inner sep=2pt,text height=2ex, text depth=0.3ex,fill=white}}
\tikzset{directed/.style={postaction={decorate,decoration={markings,
    mark=at position .5 with {\arrow[scale=1]{stealth}}}}}}
\tikzset{reverse directed/.style={postaction={decorate,
    decoration={markings,
    mark=at position .5 with {\arrowreversed[scale=1]{stealth}}}}}}
\tikzset{smallnetcircle/.style={shape=circle,thin,align=center,minimum size=4mm,font=\sffamily\scriptsize,fill=white,text height=1.5ex, text depth=0.25ex,inner sep=0mm}}
\tikzset{smallnetpi/.style={shape=rectangle,draw=black,thick,align=center,minimum width=4mm,font=\scriptsize,fill=white,text height=1.5ex, text depth=0.75ex,rounded corners, inner sep=1mm}}
\tikzset{smalllab/.style={font=\scriptsize, rectangle, inner sep=1pt, text height=2ex, text depth=0.3ex,fill=white}}

\newcommand{\netax}[2][shape=circle]{\inetcell[netcircle,#1] (#2) {ax}}
\newcommand{\netcut}[2][shape=circle]{\inetcell[netcircle,#1] (#2) {cut}}
\newcommand{\netpar}[2][shape=circle]{\inetcell[netcircle,#1] (#2) {$\parr$}}
\newcommand{\nettens}[2][shape=circle]{\inetcell[netcircle,#1] (#2) {$\otimes$}}
\newcommand{\netoc}[2][shape=circle]{\inetcell[netcircle,#1] (#2) {$\oc$}}
\newcommand{\netwn}[2][shape=circle]{\inetcell[netcircle,#1] (#2) {$\wn$}}
\newcommand{\netft}[2][shape=circle]{\inetcell[netcircle,#1] (#2) {$\flat$}}
\newcommand{\netparg}[2][shape=circle]{\inetcell[netcircle,#1] (#2) {$\parg$}}
\newcommand{\netone}[2][shape=circle]{\inetcell[netcircle,#1] (#2) {$1$}}
\newcommand{\netbot}[2][shape=circle]{\inetcell[netcircle,#1] (#2) {$\perp$}}

\newcommand{\netpax}[2][shape=circle]{\inetcell[netcircle,#1] (#2) {pax}}
\newcommand{\netid}[2][shape=circle]{\inetcell[netcircle,#1] (#2) {id}}

%% file: Intro.tex
The notion of \emph{stratification} in linear logic may be informally presented as a limitation of the dynamics of cut-elimination: in a stratified subsystem of linear logic, proofs may be seen as partitioned into \emph{strata} which never ``communicate'' with each other, in the sense that no cut between two dual formulas belonging to different strata will ever appear during cut-elimination. All extant time-bounded subsystems of linear logic (with the exception of Lafont's (\citeyear{Lafont:SLL}) soft linear logic) use some form of stratification to control the complexity of the cut-elimination procedure, which would otherwise be non-elementary (as a consequence of the well known result of \cite{Statman}, modulo the translation of intuitionistic logic in linear logic given by \cite{Girard:LL}).

In the original systems introduced by \cite{Girard:LLL}, namely elementary and light linear logic, stratification coincided with the \emph{exponential depth}, \ie, the nesting level of the logical rules introducing the exponential modality ``of course''. More recently, \cite{BaillotMazza:LLlev} introduced a more general form of stratification, still connected with the exponential modalities but no longer coinciding with depth, which keeps ensuring the desired complexity properties.

The present paper originated from a semantic investigation of this more liberal stratification. Our (successful!) attempt to define a denotational semantics for Baillot and Mazza's system naturally revealed that \emph{stratification may actually be formulated independently of exponential modalities; when it is somehow forced to be connected with them, it yields interesting complexity properties}. This is essentially because exponential modalities in linear logic are in control of duplication, the only true source of complexity in cut-elimination.

The above is the main message brought forth by this paper. We shall now proceed to describe its contents more thoroughly.

\subsection*{Background}
\paragraph*{Linear logic, stratification and computational complexity}
At the heart of our work there is the so-called Curry-Howard correspondence, which sees logical proofs as programs, and cut-elimination as their execution. From this perspective, it is not so much the expressiveness of a logical system \emph{as a language} which matters, but the complexity of its cut-elimination procedure: if a logical system has a low-complexity cut-elimination, its proofs will necessarily correspond to low-complexity programs. This approach, which has a marked proof-theoretic nature and, as such, is orthogonal to the model-theoretic methods of descriptive complexity, falls within the larger area of \emph{implicit computational complexity}, whose concrete aim is to define programming languages enjoying intrinsic complexity bounds, \ie, automatically ensured at compile time. Apart from those already mentioned above, other notable  examples of work in this field, not necessarily related to logic, are given by \cite{BellantoniCook92,Jones,Hofmann,Schwichtemberg}.

The use of linear logic as a tool for developing a Curry-Howard-based approach to implicit computational complexity was initiated by \cite{GirardScedrovScott92} and perfected by \cite{Girard:LLL}. The central idea of this latter work is that the complexity of the cut-elimination procedure is mostly owed to the presence of structural rules, in particular the contraction rule. Indeed, the cut-elimination procedure, which is in general non-elementary in the size of proofs \citep{Statman}, becomes quite manageable (\textit{e.g.}\ quadratic) in substructural logical systems lacking the contraction rule  \citep{Girard:LLL}. In linear logic, structural rules are managed by the so-called \emph{exponential} modalities. Girard showed that altering the behavior of these modalities offers a way to define logical systems in which cut-elimination is still feasible (or at most elementary) in spite of the presence of the contraction rule: \emph{light linear logic} (\LLL) exactly captures deterministic polynomial time, and \emph{elementary linear logic} (\ELL) exactly captures elementary time.\footnote{We refer here to the Curry-Howard sense of ``capturing'': in these systems, there is a formula $F$ representing functions from binary strings to binary strings such that a proof of $F$ corresponds to a function in the given complexity class and, conversely, every function in that class may be represented by a proof of $F$.} %The system \LLL\ actually uses an additional, self-dual exponential modality, called \emph{paragraph}, which is necessary for reasons of expressiveness (\ie, to represent all polytime functions), but whose logical status has otherwise remained somewhat unclear.

% In the ensuing years, a wealth of work has been done to study, extend, and find applications for these logical systems, which are collectively known as \emph{light} subsystems of linear logic. For what concerns the refinement and extension of their syntax, and the characterization of further complexity classes (like deterministic logarithmic space and polynomial space), let us mention \cite{AspertiRoversi:ILAL,DanosJoinet:ELL,Mazza:LLAndPolytime,Schoepp07,BaillotMazza:LLlev,GaboardiMarionRonchi:PSpace}. As far as applications are concerned, several type systems guaranteeing bounded-time execution for functional languages ({\it e.g.}, the pure \lac), together with efficient typing algorithms, have been devised, for instance, in \cite{BaillotTerui:DLAL,ABT06,coppola05tlca,coppola06tocl}.
% 
% 
% 
% Last but not least, there have also been efforts for studying \LLL, \ELL\ and related systems from the semantic point view, in the sense of truth semantics \citep{KanovichOkadaScedrov}, and denotational semantics \citep{BaillotPedicini01,Baillot04a,LaurentTortora:OCliques,DalLagoLaurent,Laurent:ELLCat}.

% \paragraph*{Stratification}
The restriction that Girard imposed on the exponential modalities of linear logic is a form of \emph{stratification}. Basically, the rules of linear logic are modified so that the nesting level of exponential modalities, called \emph{depth}, may not be changed during cut-elimination. Therefore, a proof may be seen as partitioned into ``strata'', one for each depth, which never interact through cut-elimination.
We observe that this is not the the only use of stratification in implicit computational complexity. For example, \cite{LeivantMarion} introduced \emph{tiers}, which are integers assigned to subterms of \lat s, to induce a stratification on the \lac, yielding characterizations of interesting complexity classes. %In the former field, stratification was used long ago by \cite{Quine} in his axiomatization of set theory known today as ``New Foundations'' (NF). In NF, naive comprehension, which postulates the existence of the set $\{x\mathrel{|}A(x)\}$, is restricted to the case where the formula $A$ is \emph{stratified}, which means that there exists a function $f$ from the terms appearing in $A$ to the integers such that, for every atomic subformula of $A$ of the form $t\in u$ (resp.\ $t=u$), one has $f(u)=f(t)+1$ (resp.\ $f(t)=f(u)$). The theory NF escapes Russel's antinomy because the formula $\lnot(x\in x)$ is \emph{not} stratified; note the similarity with the refusal of self-application in \ELL.

\paragraph*{Separating stratification from exponential depth}
Recently, \cite{BaillotMazza:LLlev} proposed a new subsystem of linear logic corresponding to elementary time, \emph{linear logic by levels} (\LLlev). This system is also based on a form of stratification, but in this case it is achieved by retaining only those linear logic proofs $\pi$ for which there exists a function from the occurrences of formulas in $\pi$ to the integers, called \emph{indexing}, which satisfies certain conditions. In a nutshell, these conditions state that axioms introduce dual occurrences of identical level, and that the level of an occurrence of formula is decreased only when it is the principal occurrence of a rule introducing an exponential modality.

Interestingly, this form of stratification turns out to be a generalization of Girard's stratification: \ELL\ is exactly the subsystem of \LLlev\ in which the function assigning to each occurrence its own depth is a valid indexing. This generalization is strict, both in the sense of proofs and provability: there exist \ELL-provable formulas which admit more proofs in \LLlev, and there exist \LLlev-provable formulas which are not provable in \ELL. Although no concrete use has currently been found for these additional formulas and proofs, \LLlev\ gives us at least one clear, and potentially interesting message: \emph{stratification does not need to coincide with exponential depth}. However, even if separated from the depth, stratification in \LLlev\ is still explicitly connected to the exponential modalities.%; for instance, the formulas corresponding to dereliction and digging are still not provable in \LLlev.

\paragraph*{Abstracting stratification through denotational semantics}
Denotational semantics originated in the work of \cite{ScottStr} and \cite{Scott} as an attempt to interpret in a non-trivial way the quotient induced on \lat s by $\beta$-equivalence. This amounts to finding an invariant of reduction, a question which may be extended to logical systems enjoying cut-elimination. Since its introduction, denotational semantics has proved to be an absolutely essential tool in computer science and proof theory, providing a wealth of information and insights into the nature of computation and formal proofs. A striking example is given by linear logic itself, which arose precisely from a denotational analysis of intuitionistic logic \citep{Girard:LL}.

After the successful introduction of denotational semantics for \LLL, \ELL\ and related systems \citep{BaillotPedicini01,Baillot04a,LaurentTortora:OCliques,DalLagoLaurent,Laurent:ELLCat}, it seemed natural to attempt to analyze the stratification underlying \LLlev\ from the denotational point of view. The result of such an analysis forms the contents of the present paper, whose message broadens that of \LLlev.

\subsection*{Stratified linear logic}
Soon after developing our semantic construction for \LLlev\ (which we present in \refsect{Objs}), we realized that it suggested a more general syntax than that of \LLlev\ itself, in which exponential modalities and strata are completely independent. This more general syntax has at least two alternative presentations in terms of sequent calculus, and one in terms of proof nets, all of them shown in \refsect{StratLL}. The induced logical system, which we call \emph{stratified linear logic} (\stratLL), has an additional modality with respect to linear logic, the self-dual \emph{paragraph} $\S$, which is in charge of controlling stratification. In \stratLL, dereliction and digging are provable, $\S A\llinimp\S B$ is provable from $A\llinimp B$, but $\S A\llinimp A$ and $A\llinimp\S A$ (or $\S A\llinimp \S\S A$) are not provable in general, which is the essence of stratification.

In \stratLL, \LLlev\ appears as a fragment, in which the exponential modalities are forced to be ``tied'' to paragraph modalities: $\oc A$ is replaced by $\oc\S A$, and $\wn A$ is replaced by $\wn\S A$. Additionally, the paragraph modality itself is a generalization of the paragraph modality of \LLL, which justifies our terminology and notation. This allows to define polytime subsystems within \LLlev\ generalizing \LLL, as already shown in \cite{BaillotMazza:LLlev}.

\subsection*{A categorical construction for stratification}
Our denotational analysis brings a new understanding of the exponential modalities of light logics: together with the control of duplication, which is their usual task in linear logic, they are also charged with the additional task of controlling stratification, which is represented by the paragraph modality.

From the semantic point of view, we interpret the paragraph modality in ``augmented'' models of linear logic. More precisely, we define a categorical construction, represented by a 2-endofunctor $\SInv(-)$ of the 2-category of symmetric monoidal categories, which takes a model of linear logic $\catL$ and yields another model of linear logic $\SInv(\catL)$; this latter model is practically equivalent to the original one, but it has ``more space'', and this extra space is what allows the definition of a non-trivial paragraph functor.

Concretely, the $\SInv$ construction may be understood by looking at its action on categories. Given a category $\catA$, we define an \emph{\obj} of $\catA$ as a pair $(A,s)$, where $A$ is an object of $\catA$ and $s$ is a $\Int$-indexed sequence of involutions of $A$, \ie, automorphisms of $A$ such that $s_k\circ s_k=\id_A$ for all $k\in\Int$. \Objs\ have a natural notion of morphism: a morphism from $(A,s)$ to $(B,t)$ is a morphism $f:A\rightarrow B$ such that $t_k\circ f\circ s_k=f$. If the category $\catA$ is symmetric monoidal, the \objs\ of $\catA$ and their morphisms may themselves be arranged in a symmetric monoidal category, which is $\SInv(\catA)$ (so $\SInv$ stands for ``involutions'').

It turns out that $\SInv(\catA)$ has all the structure needed to provide a model of linear logic as soon as $\catA$ does (\refth{ModelPres}). However, in $\SInv(\catA)$ we may now define a functor $\S(-)$ which acts on objects by ``shifting'' the sequence of involutions, \ie, $\S(A,s)=(A,(s_{k-1})_{k\in\Int})$, and which acts as the identity on morphisms. It is possible to show that such a functor is never trivial, \ie, it is never isomorphic to the identity functor, unless the original model of linear logic (in the category $\catA$) is itself trivial (\refth{NonDegenerate}).

\subsection*{Applications to bounded complexity}
An important contribution of our work, resulting from an application of the results described above, is the discovery of alternative formulations of \LLlev.

The first reformulation (\refsect{Geom}) is of ``geometric'' nature. It is a presentation of \LLlev\ in terms of proof nets by means of a \emph{correctness criterion}, extending the usual one by \cite{DanosRegnier:Mult}. Its advantage is to avoid mentioning the notion of indexing, shifting from an existential condition (there exists a function satisfying\ldots) to a universal one (every cycle satisfies\ldots), which is arguably of interest.

The second reformulation (\refsect{Interactive}) is ``interactive'', and arises when applying the $\SInv$ construction to the syntactic category of formulas and proof nets. Approximatively speaking, we prove that a cut-free linear logic proof net $\pi$ is in \LLlev\ iff it ``interacts well'' with all ``tests'' (\refth{Interactive}), where a test is a certain kind of proof net which may interact with $\pi$ by means of a cut rule (the interaction being cut-elimination).

Finally, we provide a semantic characterization of \LLlev\ (\refsect{Completeness}), which may be roughly formulated as follows: given a model of linear logic $\catL$ which satisfies a certain condition we call \emph{swap-sensitivity}, we have that a cut-free linear logic proof net $\pi$ of conclusion $A$ is in \LLlev\ iff its denotational interpretation $\sem\pi$ in $\catL$ is a morphism of $\SInv(\catL)$ from the tensor unit to a certain \obj\ $\altsem{A}$, which depends solely on the formula $A$ (\refth{Sem}). In other words, provided swap-sensitivity holds, the semantics is able to ``detect'' whether a proof net fails to admit a valid indexing, or fails to satisfy any of the two equivalent conditions mentioned above. The swap-sensitivity condition is a bit technical, but it is very mild: it is satisfied by all models of linear logic we are aware of.

\subsection*{Acknowledgments}
We would like to thank Paul-Andr\'e Melli\`es for several interesting discussions on the categorical constructions used in this paper.

This work was partially supported by ANR projects \textsc{Complice} (08-BLAN-0211-01) and \textsc{Logoi} (10-BLAN-0213-02), and by the CNRS PICS ``Logique Lin\'eaire et Applications''.

%% file: StratLL.tex
The formulas of stratifed linear logic (\stratLL) are generated by the following grammar, where $X,X^\perp$ range over a denumerable set of propositional variables:
$$\begin{array}{rcll}
	A,B & ::= & X \gpipe X^\bot & \textrm{propositional atoms} \\
	&\gpipe& \lone \gpipe A\ltens B \gpipe \lbot \gpipe A\lpar B & \textrm{multiplicative connectives and units} \\
	&\gpipe& \ltop \gpipe A\lwith B \gpipe \lzero \gpipe A\lplus B & \textrm{additive connectives and units} \\
	&\gpipe& \forall X.A \gpipe \exists X.A & \textrm{second order quantifiers} \\
	&\gpipe& \oc A \gpipe \wn A \gpipe \parg A & \textrm{exponential modalities and paragraph.}
\end{array}$$
Linear negation is defined through De Morgan laws:
\begin{displaymath}
	\begin{array}{rclcrcl}
		(X)^\bot & = & X^\bot && (X^\bot)^\bot & = & X \\
		(\lone)^\bot & = & \lbot && (\lbot)^\bot & = & \lone \\
		(A\ltens B)^\bot & = & A^\bot\lpar B^\bot && (A\lpar B)^\bot & = & A^\bot\ltens B^\bot \\
		(\ltop)^\bot & = & \lzero && (\lzero)^\bot & = & \ltop \\
		(A\with B)^\bot & = & A^\bot\lplus B^\bot && (A\lplus B)^\bot & = & A^\bot\lwith B^\bot \\
		(\forall X.A)^\bot & = & \exists X.A^\bot && (\exists X.A)^\bot & = & \forall X.A^\bot \\
		(\oc A)^\bot & = & \wn A^\bot && (\wn A)^\bot & = & \oc A^\bot\\
		&&& (\parg A)^\bot\ =\ \parg A^\bot &&&
	\end{array}
\end{displaymath}
Two connectives or modalities exchanged by negation are said to be \emph{dual}; the paragraph modality $\parg$ is self-dual. %Second order quantifiers may be added as usual, but we do not consider them in this paper.

Linear implication is defined as $A\llinimp B=A^\perp\lpar B$. Multisets of formulas will be ranged over by $\Gamma,\Delta$. The notations $\wn\Gamma$ and $\S\Gamma$ will stand for multisets containing formulas all starting with a $\wn$ or $\S$ modality, respectively.

\begin{table}[tp]
	\TwoRules
	{\infer[\etq{Axiom}]{\ts {A^\bot}^i,A^i}{}}
	{\infer[\etq{Cut}]{\ts\Gamma,\Delta}{\ts\Gamma,A^i & \ts\Delta,{A^\bot}^i}}
	\smallskip
	\TwoRules
	{\infer[\etq{One}]{\ts\lone^i}{}}
	{\infer[\etq{Bottom}]{\ts\Gamma,\lbot^i}{\ts\Gamma}}
	\smallskip
	\TwoRules
	{\infer[\etq{Tensor}]{\ts\Gamma,\Delta,{A\ltens B}^i}{\ts\Gamma,A^i & \ts\Delta,B^i}}
	{\infer[\etq{Par}]{\ts\Gamma,{A\lpar B}^i}{\ts\Gamma,A^i,B^i}}
	\smallskip
	\OneRule
	{\infer[\etq{Top}]{\ts\Gamma,\ltop^i}{}}
	\smallskip
	\TwoRules
	{\infer[\etq{With}]{\ts\Gamma,{A\lwith B}^i}{\ts\Gamma,A^i & \ts\Gamma,B^i}}
	{\infer[\etq{Plus ($k\in\{1,2\}$)}]{\ts\Gamma,{A_1\lplus A_2}^i}{\ts\Gamma,A_k^i}}
	\smallskip
	\TwoRules
	{\infer[\etq{For all ($\ast$)}]{\ts\Gamma,{\forall X.A}^i}{\ts\Gamma,A^i}}
	{\infer[\etq{Exists}]{\ts\Gamma,{\exists X.A}^i}{\ts\Gamma,A^i}}
	\smallskip
	\TwoRules
	{\infer[\etq{Promotion}]{\ts\wn\Gamma,\oc A^i}{\ts\wn\Gamma,A^i}}
	{\infer[\etq{Dereliction}]{\ts\Gamma,\wn A^i}{\ts\Gamma,A^i}}
	\smallskip
	\TwoRules
	{\infer[\etq{Weakening}]{\ts\Gamma,\wn A^i}{\ts\Gamma}}
	{\infer[\etq{Contraction}]{\ts\Gamma,\wn A^i}{\ts\Gamma,\wn A^i,\wn A^i}}
	\smallskip
	\OneRule
	{\infer[\etq{Paragraph}]{\ts\Gamma,{\parg A}^i}{\ts\Gamma,A^{i+1}}}
	\caption{Stratified sequent calculus. In every rule, $i\in\Int$. The condition ($\ast$) is that $X$ is not free in $\Gamma$}
	\label{tab:StratSeqCalc}
\end{table}
The most convenient way of formulating a sequent calculus for \stratLL\ is using \mbox{\emph{2-sequents}} \citep{Masini:TwoSeq,GuerriniMartiniMasini:TwoSeq}. A 2-sequent is a sort of bidimensional sequent; for our purposes, we define it simply as a sequent in which every occurrence of formula is associated with a relative integer, called its \emph{index}. It is important to note that indexes are not part of the logical language; occurrences of subformulas are not indexed.

The 2-sequent rules defining the admissible derivations of \stratLL\ are given in \reftab{StratSeqCalc}. Indexes are annotated as superscripts; we keep using $\Gamma,\Delta$ to range over multisets of indexed formulas.

The 2-sequent calculus of \stratLL\ admits cut-elimination; the cut-elimination rules are those of linear logic, plus a rule transforming the cut
$$\infer{\ts\Gamma,\Delta}
	{\infer{\ts\Gamma,\parg A^i}{\ts\Gamma,A^{i+1}} &
	\infer{\ts\Delta,\parg {A^\bot}^i}{\ts\Delta,{A^\bot}^{i+1}}}
$$
into the cut
$$\infer{\ts\Gamma,\Delta}{\ts\Gamma,A^{i+1} & \ts\Delta,{A^\bot}^{i+1}}
$$
For the propositional fragment, the cut-elimination property may be proved by standard arguments, using the usual notions of degree and rank of cuts. We shall give detailed cut-elimination rules for proof nets (\refsect{ProofNets}), which is why we do not linger any further on this topic here.

The additive group $\Int$ acts on multisets of indexed formulas in the obvious way: given $k\in\Int$ and $\Gamma=A_1^{i_1},\ldots,A_n^{i_n}$, we set $k\cdot\Gamma=A_1^{i_1+k},\ldots,A_n^{i_n+k}$. The action can easily be lifted to derivations: if $\delta$ is a derivation of $\ts\Gamma$, we get a derivation $k\cdot\delta$ of $\ts k\cdot\Gamma$. Then, we obtain
\begin{lem}
	\label{lemma:Action}
	The 2-sequent $\ts\Gamma$ is derivable iff the 2-sequent $\ts k\cdot\Gamma$ is derivable, for all $k\in\Int$.\qed
\end{lem}

Therefore, the ``absolute'' value of indexes in a 2-sequent does not matter at all; what matters is their relative value, with respect to each other.

A natural property of the sequent calculus of linear logic, which is the linear incarnation of the so-called Deduction Theorem, is that ``commas are $\lpar$ connectives'', \ie, $\ts A,B$ is provable iff $\ts A\lpar B$ is provable. This, which from a semantic point of view is equivalent to asking the closure of the category underlying \stratLL, is obviously false for derivations in the calculus of \reftab{StratSeqCalc}. These considerations lead us to define the actual \emph{proofs} of \stratLL\ as follows:
\begin{defn}[\stratLL\ proofs]
	\label{def:Proofs}
	A \emph{proof} of \stratLL\ is a derivation in the 2-sequent calculus of \reftab{StratSeqCalc} such that all occurrences of formulas in the conclusion have the same index (in other words, the 2-sequent derived is actually a sequent). %A formula $A$ is \emph{provable} if there is a proof (or derivation) of $\ts A^i$, for some $i\in\Int$.
\end{defn}

Observe that second order linear logic is obviously the fragment of \stratLL\ in which the modality $\parg$ is removed. In that case, indexes are useless, and provability coincides with provability in linear logic. Another, less trivial fragment of \stratLL\ will be introduced in the next section. For the moment, let us observe that \stratLL\ itself adds virtually nothing to linear logic, because the paragraph modality is close to being trivial. In fact, the reader may check that it commutes to \emph{all} logical connectives, so that, for instance, $\parg\lone$, $\parg(A\ltens B)$, $\parg 0$, $\parg(A\lplus B)$, $\parg\exists X.A$, and $\parg\oc A$ are all provably isomorphic to $\lone$, $\parg A\ltens\parg B$, $\lzero$, $\parg A\lplus \parg B$, $\exists X.\parg A$, and $\oc\parg A$, respectively.

The only purpose of the paragraph modality is to enforce stratification, as will be discussed at the end of \refsect{CutElim}, when we will have defined cut-elimination more precisely. In terms of provability, the essential meaning of stratification is in the impossibility (which is easy to check in presence of cut-elimination) of deriving any of the implications $X\llinimp\parg X$ and $\parg X\llinimp X$, with $X$ a propositional atom. When suitably linked with the exponential modalities, stratification has remarkable complexity effects, as explained in the next section.

\begin{table}[t]
	\TwoRules
	{\infer[\etq{Bottom}]{\ts\Gamma,\parg^k\lbot}{\ts\Gamma}}
	{\infer[\etq{Par}]{\ts\Gamma,\parg^k(A\lpar B)}{\ts\Gamma,\parg^kA,\parg^kB}}
	\ThreeRules
	{\infer[\etq{Top}]{\ts\Gamma,\parg^k\ltop}{}}
	{\infer[\etq{With}]{\ts\Gamma,\parg^k(A\lwith B)}{\ts\Gamma,\parg^kA & \ts\Gamma,\parg^kB}}
	{\infer[\etq{For all ($\ast$)}]{\ts\Gamma,\parg^k\forall X.A}{\ts\Gamma,\parg^kA}}
	\smallskip
	\ThreeRules
	{\infer[\etq{Promotion}]{\ts\wn\Gamma,\parg^k\oc A}{\ts\wn\Gamma,\parg^kA}}
	{\infer[\etq{Dereliction}]{\ts\Gamma,\parg^k\wn A}{\ts\Gamma,\parg^kA}}
	{\infer[\etq{Paragraph}]{\ts\parg\Gamma}{\ts\Gamma}}
	\caption{Alternative formulation of \stratLL\ in sequent calculus. The rules handling the connectives not mentioned, as well as the axiom, cut, weakening and contraction rules, are the same as those of standard linear logic, \ie, those of \reftab{StratSeqCalc} with no index annotation. In all rules, $\parg^kA$ stands for the formula $A$ with $k\geq 0$ paragraph modalities prepended to it (note that $A$ may itself start with a paragraph modality). As usual, the condition ($\ast$) is that $X$ is not free in $\Gamma$.}
	\label{tab:AltSeqCalc}
\end{table}
The fact that the paragraph modality commutes to all logical operations suggests that \stratLL\ may be reformulated by explicitly writing these commutations into its rules. We thus obtain a ``usual'' sequent calculus, \ie, without indexes on formulas, defined in \reftab{AltSeqCalc}. It is not hard to see that this is an equivalent formulation of \stratLL:
\begin{prop}
	\label{prop:Alt}
	The 2-sequent $\ts\Gamma^i$ (\ie, with all occurrences of formulas having the same index) is derivable in the calculus of \reftab{StratSeqCalc} iff $\ts\Gamma$ is derivable in the calculus of \reftab{AltSeqCalc}.
\end{prop}
\begin{pf}
	Note that the rules of \reftab{AltSeqCalc} are all derivable in the calculus of \reftab{StratSeqCalc}, as soon as we suppose that we translate a sequent $\ts\Gamma$ with a \mbox{2-sequent} of the form $\ts\Gamma^i$. Vice versa, a straightforward induction shows that, if the \mbox{2-sequent} $\ts A_1^{i_1},\ldots,A_n^{i_n}$ is derivable in the calculus of \reftab{StratSeqCalc}, then the sequent $\ts \parg^{i_1-m}A_1,\ldots,\parg^{i_n-m}A_n$ is derivable in the calculus of \reftab{AltSeqCalc}, with $m=\min_{1\leq k\leq n} i_k$.\qed
\end{pf}

The reader may have noticed that, in the alternative formulation of \stratLL, the commutations of the paragraph modality are added only to one ``half'' of the connectives of linear logic (namely what are known as the \emph{negative} connectives). This is because the implications $\lone\llinimp\parg\lone$, $\parg A\ltens\parg B\llinimp\parg(A\ltens B)$, $\lzero\llinimp\parg\lzero$, $\parg A\lplus\parg B\llinimp\parg(A\lplus B)$, and $\exists X.\parg A\llinimp\parg\exists X.A$ are all derivable from the usual rules plus the paragraph rule of \reftab{AltSeqCalc}. For the exponential modalities, none of the two directions of the needed equivalences holds naturally, so both promotion and dereliction must be modified.

This alternative formulation of \stratLL\ has the drawback of having a more complex cut-elimination procedure, involving even more commutations than usual (we prefer not to give any detail here). However, it is semantically interesting, because it allows the denotational interpretation of a proof to be defined, as customary, by induction on the proof itself. This is impossible if we have only \refdef{Proofs} as our definition of proof, because a subderivation of a proof need not be itself a proof. We shall find this alternative formulation of \stratLL\ particularly useful in \refsect{StratModels}.

%% file: Bounded.tex
Linear logic by levels (\LLlev) was introduced by \cite{BaillotMazza:LLlev} as a generalization of the ideas which led Girard to the definition of elementary linear logic (\ELL). Indeed, \LLlev\ captures elementary time: there is a representation of integers (as proofs) in the system such that the representable functions from integers to integers are exactly the elementary functions, \ie, those that can be computed by a Turing machine whose runtime is bounded by a tower of exponentials of fixed height. However, \LLlev\ is a strict supersystem of \ELL, both in terms of proofs and provability. It contains a subsystem, called \LLLlev, which captures deterministic polynomial time computation in the same sense, and which extends Girard's light linear logic (\LLL).

The original definition presents \LLlev\ as a subsystem of linear logic, using a notion of indexability (\cf\ \refdef{L3}). Here, we may reformulate it as a \emph{fragment} of \stratLL. In fact, \cite{BaillotMazza:LLlev} gives a 2-sequent formulation of \LLlev\ which is nearly identical to that of \reftab{StratSeqCalc}; the only difference is in the promotion and dereliction rules, which modify the indexes in the same way as the paragraph modality:
$$
	\infer[\etq{Promotion}]{\ts\wn\Gamma,\oc A^i}{\ts\wn\Gamma,A^{i+1}}\qquad\qquad
	\infer[\etq{Dereliction}]{\ts\Gamma,\wn A^i}{\ts\Gamma,A^{i+1}}
$$
Then, it is obvious that \LLlev\ coincides with the fragment of \stratLL\ whose formulas are obtained by forcing every exponential modality to be preceded by a paragraph, \ie, if $\oc_{\LLlev}$ and $\wn_{\LLlev}$ denote the exponential modalities of \LLlev, we have
$$\oc_{\LLlev}=\oc\parg\qquad\qquad\textrm{and}\qquad\qquad\wn_{\LLlev}=\wn\parg.$$

Anticipating on \pn s, since the exponential cut-elimination step of \LLlev\ is defined exactly as in \reffig{ExpStep}, and because the paragraph step (\reffig{PargStep}) is trivial, we obviously have that such equalities are sound also with respect to cut-elimination.

This may all be restated as follows: stratification has an effect on the complexity of cut-elimination as soon as the exponential modalities are linked to it. As a matter of fact, all extant stratification-based, complexity-bounded subsystems of linear logic may be seen as subsystems of the fragment of \stratLL\ just introduced.

For instance, one can easily check that \ELL\ corresponds to the subsystem of the above fragment obtained by restricting to \mbox{2-sequents} of the form
$$\vdash\parg\Delta^{i},\Gamma^{i+1},$$
where $\Gamma$ does not contain paragraph formulas. In other words, only two indexes are allowed at the same time, and all paragraph formulas must have a lower index. For what concerns polynomially-bounded subsystems, we refer the reader to \cite{Mazza:LLAndPolytime} for a formulation of \LLL\ as a subsystem of \ELL, and to \cite{BaillotMazza:LLlev} for a definition of \LLLlev\ as a subsystem of \LLlev. 

The only known complexity-related subsystem of linear logic which is not captured by our notion of stratification is Lafont's (\citeyear{Lafont:SLL}) soft linear logic (\SLL), a polynomially-bounded system which is based on ideas quite different from those originally put forward by Girard. Of course, as a subsystem of linear logic, \SLL\ is also a subsystem of \stratLL, but in a completely uninteresting way, \ie, our notion of stratification does not intervene in the definition (the paragraph modality is not used). Furthermore, we immediately see that \SLL\ rejects the idea that one must link exponentials to stratification in order to achieve complexity effects: in fact, this latter system allows the dereliction principle, \ie, the formula $\oc A\llinimp A$ is provable in \SLL, for all $A$; this is impossible in \stratLL\ if the $\oc$ modality is in linked in any way to the $\parg$ modality. In the end, we may say that linking exponentials to stratification is a sufficient condition for obtaining bounded cut-elimination, but it is by no means necessary.

%If we ignore additive connectives ($\lwith$ and $\lplus$), \LLL\ is obtained by further restricting the promotion rule to have at most one formula in its context, and to force the paragraph rule to be applied to a whole sequent at once:
% $$\infer[\etq{Paragraph}]{\ts\parg\Gamma^{i}}{\ts\Gamma^{i+1}}$$. Reformulating \LLL\ in presence of additive connectives is still possible, but a bit more involved; we refer the interested reader to \cite{Mazza:LLAndPolytime}.

%% file: ProofNets.tex
The proof theory of \stratLL\ may also be developed in terms of \pn s, originally introduced by \cite{Girard:LL,Girard:ProofNets}, and subsequently reformulated by other authors by means of alternative, but equivalent syntactical definitions. In this paper, we use a combination of the presentations given by \cite{DanosRegnier:PNHilb} and \cite{Tortora:AdditivesAndNormalizationI}, with a slight change in the terminology: the term ``proof structure'', introduced by~\cite{Girard:LL} and traditionally used in the literature, is here dismissed in favor of the term \emph{\ps}. On the contrary, the term \pn, i.e., a \ps\ satisfying certain structural conditions (the correctness criterion), retains its usual meaning.

We shall henceforth ignore additive connectives ($\lwith$, $\lplus$, and their respective units $\ltop$ and $\lzero$) and second order quantifiers, because their representation in proof nets is unnecessarily complex for the purposes of this work.

\begin{figure}[t]
	\input{links}
	\caption{Links.}
	\label{fig:Links}
\end{figure}

\begin{figure}[t]
        \input{boxes}	
	\caption{A box.}
	\label{fig:Boxes}
\end{figure}

In the context of proof nets, it is useful to consider \emph{\struct s}, which will be denoted by $\flat A$, where $A$ is a formula. Note that \struct s are not formulas: they cannot be composed, \ie, they cannot be subformulas of a formula. For instance, $(\flat A)\ltens B$ and $\flat\flat A$ are meaningless expressions.

In the following definition, and throughout the rest of the paper, unless explicitly stated we shall make no distinction between the concepts of \emph{formula} (or \struct) and \emph{occurrence of formula}. The same will be done for what we call \emph{links} and their occurrences.
\begin{defn}[\Ps]
	\label{def:Net}
	A \emph{\pps} is a pair $(\mathcal G,\mathsf B)$, where $\mathcal G$ is a finite graph-like object\footnote{In \cite{Tortora:Obsessional,deCarvalhoTortora}, $\mathcal G$ is called \emph{linear proof structure}.} whose nodes are occurrences of what we call \emph{links}, and whose edges are directed and labelled by formulas or \struct s; and $\mathsf B$ is a set of subgraphs of $\mathcal G$ called \emph{boxes}.
	\begin{itemize}
		\item Links (\reffig{Links}) are labelled by connectives and constants of \stratLL\ (excluding $\lwith$, $\lplus$, $\ltop$, and $\lzero$), or by one of the labels $\mathsf{ax}$, $\mathsf{cut}$, $\flat$, $\mathsf{pax}$. Two links labelled by dual connectives are said to be \emph{dual}. Each link has an arity and co-arity, which are resp.\ the number of its incoming and outgoing edges. The arity and co-arity is fixed for all links except \whynot\ links, which have co-arity $1$ and arbitrary arity. A nullary \whynot\ link is also referred to as a \weak\ link.
		
		\item The incoming edges of a link (and the formulas that label them) are referred to as its \emph{premises}; these are ordered (\ie, there is a ``left'' and a ``right'' premise) in the case of multiplicative links (\parl\ and \tensor), and unordered in the case of \cut\ and \whynot\ links. The outgoing edges of a link (and the formulas that label them) are referred to as its \emph{conclusions}.
		
		\item Premises and conclusions of links must respect a precise labeling (which depends on the link itself), given in \reffig{Links}. In particular, the edges labelled by \struct s can only be premises of \paxl\ and \whynot\ links;
		
		\item Each edge must be the conclusion of exactly one link, and the premise of at most one link. The edges that are not premises of any link, and the formulas that label them, are deemed \emph{conclusions} of the \pps. (Note that the presence of these ``pending'' edges, together with the fact that some premises are ordered, is why \pps s are not exactly graphs).
		
		\item A box is depicted as in \reffig{Boxes}, in which $\pi$ is a \pps, said to be \emph{contained} in the box. The links that are explicitly represented in \reffig{Boxes} (i.e., the \paxl\ links and the \ofcourse\ link) form the \emph{border} of the box (but are not contained in the box). The unique \ofcourse\ link in the border is called the \emph{\pal} of the box, while the \paxl\ links are called \emph{\pax s}. We have the following conditions concerning boxes:
		\begin{enumerate}
			\item[a.] each \ofcourse\ link is the \pal\ of exactly one box;
			\item[b.] each \paxl\ link is in the border of exactly one box; 
			\item[c.] any two distinct boxes are either disjoint or included in one another.
		\end{enumerate}
	\end{itemize}

	If $\sigma$ is a \pps, a link (or edge) of $\sigma$ is said to have \emph{depth} $d$ if it is contained in $d$ (necessarily nested) boxes. The depth of a box of $\sigma$ is the depth of the links forming its border.
\end{defn}
% \begin{defn}[Depth]
% 	\label{def:Depth}
% 	Let $\sigma$ be a \pps. A link (or edge) of $\sigma$ is said to have \emph{depth} $d$ if it is contained in $d$ (necessarily nested) boxes. The depth of a box of $\sigma$ is the depth of the links forming its border. %The depth of a link $l$, edge $e$, or box $\cB$ are denoted resp.\ by $\Depth l$, $\Depth e$ and $\Depth\cB$. The depth of $\sigma$, denoted by $\Depth\sigma$, is the maximum depth of its links.
% \end{defn}

Let $\sigma=(\mathcal G,\mathsf B)$ be a \pps. If we take $\mathcal G$ and ignore the conclusions, the orientation of edges, and the ordering of multiplicative links, we obtain an undirected graph in the usual sense; we call this graph the \emph{underlying graph} of $\sigma$, and denote it by $\widetilde\sigma$. In what follows, when we speak of a \emph{path} (or \emph{cycle}, or \emph{connected component}) of $\sigma$, we mean a path (or cycle, or connected component) of $\widetilde\sigma$, in the usual sense of graph theory.
\begin{defn}[Switching]
	\label{def:Switching}
	Let $\sigma$ be a \pps. A \emph{switching} of $\sigma$ is an undirected graph built from $\widetilde\sigma$ as follows:
	\begin{itemize}
		\item for each \parl\ and \whynot\ node $l$, exactly one premise is chosen (provided the link has any premises at all), and all the other premises are erased;
		\item for each box $\cB$ at depth zero of $\sigma$, all the nodes and edges of $\widetilde\sigma$ coming from nodes and edges of $\sigma$ belonging to $\cB$ are erased, together with all the edges connecting such nodes to the rest of the graph; these are replaced by a new node $l$ and, for any link $m$ of depth zero which was connected to the border of $\cB$, a new edge between $m$ and $l$ is added. In other words, the outermost boxes of $\sigma$ are ``collapsed'' into single nodes.
	\end{itemize}
\end{defn}
\begin{defn}[Balanced paths and cycles]
	\label{def:Balanced}
	Given a path $\phi$ of $\widetilde\sigma$, we may traverse it from one of its extremities to the other in any of the two possible directions (if $\phi$ is a cycle, it still has two possible orientations). Following $\phi$ in any of the two directions, we may record the number of times it traverses a \pargl\ link ``upwards'', \ie, from conclusion to premise, which we denote by $n^-$, and the number of times it traverses a \pargl\ link ``downwards'', \ie, from premise to conclusion, which we denote by $n^+$. It is clear that the numbers $n^+,n^-$ are exchanged when we reverse the direction of traversal of $\phi$. Therefore, the number $|n^+-n^-|$ depends only on the path $\phi$ itself, and we call it its \emph{balance}. In case $\phi$ is a cycle, the balance is computed by starting from any of the nodes traversed by $\phi$ and returning to the same node, in any direction.

	A \emph{balanced path} (or cycle) of $\sigma$ is a path (or cycle) whose balance is $0$.
\end{defn}

In the following, if $\pi$ is a \ps\ of conclusions $A_1,\ldots,A_n$ such that no $A_i$ is a \struct, by \emph{$\lpar$-closure} of $\pi$ we mean any \ps\ $\pi'$ obtained by adding a tree of \parl\ links to the conclusions of $\pi$ so that $\pi'$ has exactly one conclusion $A_1\lpar\cdots\lpar A_n$, where the placement of parenthesis depends on the shape of the tree. Of course, a \ps\ with more than 2 conclusions has more that one $\lpar$-closure; however, in all definitions and results below it does not matter which one we chose, whence we abusively speak of ``the'' $\lpar$-closure.
\begin{defn}[Correctness, \dr, \pn]
	\label{def:ProofNet}
	A \pps\ $(\mathcal G,\mathsf B)$ is said to be \emph{DR-correct} if:
	\begin{itemize}
		\item all of its switchings are acyclic;
		\item for all $\cB\in\mathsf B$, the \pps\ contained in $\cB$ is DR-correct.
	\end{itemize}

	A \pps\ is \emph{\scorr} if all of its cycles are balanced.

	A \emph{\dr} is a DR-correct \ps\ having no \struct s in its conclusions.

	A \emph{\pn} is a \dr\ whose $\lpar$-closure is \scorr.
\end{defn}

Note that DR-nets are nothing but ``proof nets'' in the sense of \cite{Girard:ProofNets,DanosRegnier:PNHilb,Tortora:AdditivesAndNormalizationI} (``DR'' stands indeed for ``Danos-Regnier''), which may however happen to contain \pargl\ links. We shall prove in the sequel that, instead, our \pn s correspond to \stratLL\ proofs, as per \refdef{Proofs}.

We start by introducing the notion of \emph{sequentializable} \ps, whose definition mimics the rules of sequent calculus (without indexes):
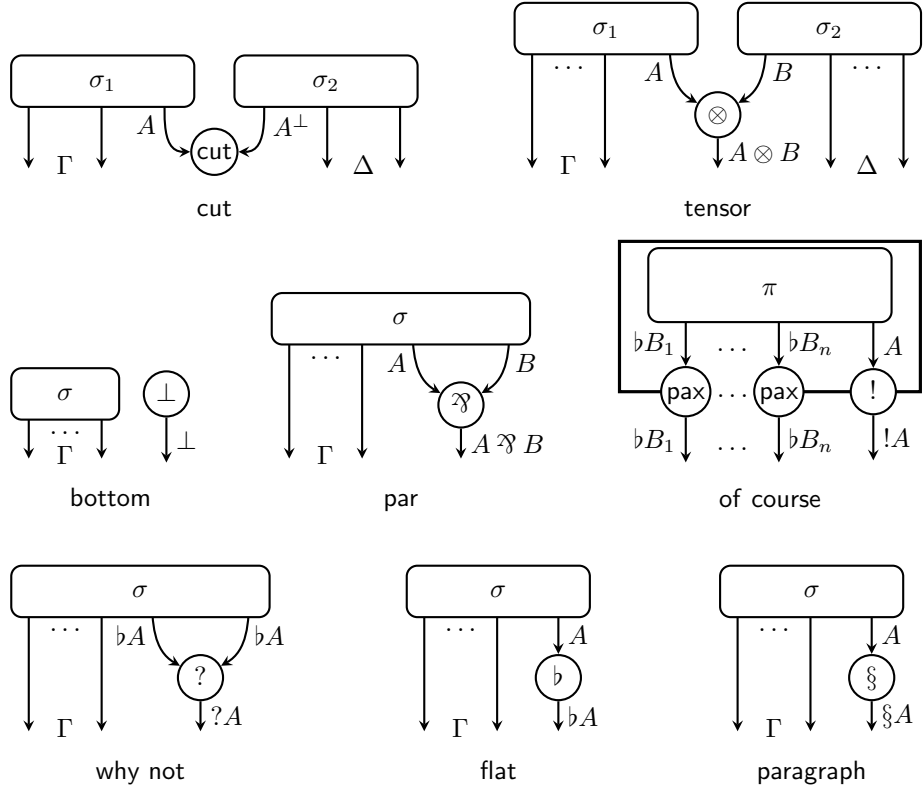
\begin{figure}[t]
        \input{SPS}
	%\pict{SPS}
	\caption{Rules for building sequentializable \ps s.}
	\label{fig:SPS}
\end{figure}

\begin{defn}[\Sps]
	\label{def:SeqNet}
	We define the set of \emph{\sps s} inductively: the empty \ps, the \ps\ consisting of a single \axiom\ link, and the \ps\ consisting of a single \onel\ link are sequentializable (daimon, axiom, and one); the juxtaposition of two sequentializable \ps s is sequentializable (mix); finally, if $\sigma$, $\sigma_1$, $\sigma_2$ are \sps s of suitable conclusions, the \ps s of \reffig{SPS} are sequentializable.
\end{defn}

The reader may object that the rules for building \sps s do not correspond exactly to index-free versions of the rules of \reftab{StratSeqCalc}. Indeed, they correspond to an index-free version of that calculus in which, apart from the additive and quantifier rules, also the promotion, dereliction, weakening and contraction rules have been removed and replaced by the rules
$$
\infer[\etq{Promotion}]{\ts\flat\Gamma,\oc A}{\ts\flat\Gamma,A}\qquad\qquad
\infer[\etq{Why not}]{\ts\Gamma,\wn A}{\ts\Gamma,\flat A,\ldots,\flat A}\qquad\qquad
\infer[\etq{Flat}]{\ts\Gamma,\flat A}{\ts\Gamma,A}
$$
so that \struct s may appear in sequents. Moreover, the \emph{mix rules}
$$
\infer[\etq{Mix}_0]{\ts\phantom{\Gamma}}{}\qquad\qquad
\infer[\etq{Mix}]{\ts\Gamma,\Delta}{\ts\Gamma & \ts\Delta}
$$
are further added to the calculus. However, ignoring the mix rules, which are introduced for the sole purpose of simplifying the definition of proof nets (otherwise, more conditions concerning \emph{connectedness} of switchings should be added to the formulation of \dr, requiring the introduction and subsequent handling of \emph{jumps}, a technical complication which we do not need in this paper), this calculus is strictly equivalent to the index-free, additive- and quantifier-free version of that of \reftab{StratSeqCalc}. In fact, it only takes a straightforward induction to prove that $\ts\flat\Gamma,\Delta$ is derivable in the former calculus iff $\ts\wn\Gamma,\Delta$ is derivable in the latter. Hence, derivability coincides for sequents not containing \struct s, which are the ones that matter (remember that \pn s have no \struct\ in their conclusions, \cf \refdef{ProofNet}).

If \sps s capture the structure of the rules of \reftab{StratSeqCalc}, we still need to account for 2-sequents, \ie, the fact that the calculus of \stratLL\ manipulates formulas carrying indexes, which influence provability in an essential way. For instance, if we took \sps s as our proofs, both $\parg X\llinimp X$ and $X\llinimp\parg X$ would become provable, in striking contrast with what observed about \stratLL. This may be done by resorting to the notion of \emph{indexing}, originally introduced by \cite{BaillotMazza:LLlev}.
\begin{figure}[t]
	\input{indexes}
	\caption{Constraints for indexing \ps s. Next to each edge we represent the integer assigned by the indexing; formulas are omitted, because irrelevant to the indexing.}
	\label{fig:Indexes}
\end{figure}
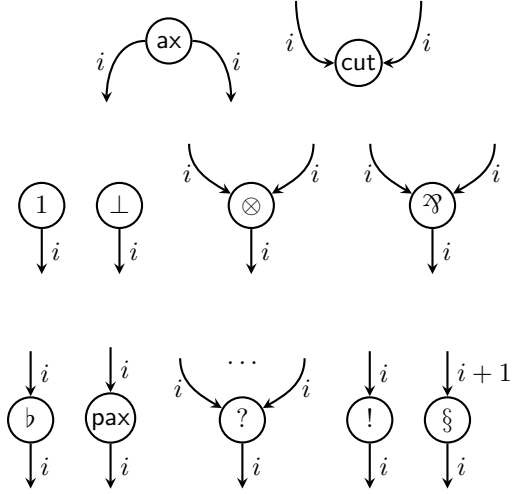
\begin{defn}[Indexing, \ips]
	\label{def:Indexing}
	Let $\sigma$ be a \ps. An \emph{indexing} for $\sigma$ is a function $I$ from the edges of $\sigma$ to $\Int$ satisfying the constraints given in \reffig{Indexes}. If, furthermore, the indexing satisfies that, for all conclusions $e,e'$ of $\sigma$, $I(e)=I(e')$, it is said to be a \emph{strong indexing}.

	A \ps\ is said to be (strongly) indexable if it admits a (strong) indexing.
\end{defn}

Observe that a net with no \struct s in its conclusions is strongly indexable iff its $\lpar$-closure is indexable (the fact that no conclusion is a \struct\ is only needed to ensure the existence of a $\lpar$-closure). This is in accordance with the fact that $\ts A_1,\ldots,A_n$ is provable in \stratLL\ iff $\ts A_1\lpar\cdots\lpar A_n$ is derivable.

%The set of indexings of an \ips\ $\sigma$ is easily seen to form an affine space over the $\Int$-module $\Int^n$, where $n$ is the number of connected components of $\sigma$. We refer the reader to \cite{BaillotMazza:LLlev} for a proof of this fact; here, we shall only use the following, simpler statement, which 
The following is a reformulation of \reflemma{Action} for \ps s:
\begin{lem}[Shift]
	\label{lemma:Translation}
	Let $I$ be an indexing for a \ps\ $\sigma$, let $\sigma_1,\ldots,\sigma_n$ be distinct connected components of $\sigma$, and let $k_1,\ldots,k_n\in\Int$. Then, the function $I'$ defined by
	$$I'(e)=\left\{\begin{array}{ll}
		I(e)+k_i & \textrm{if $e$ is an edge of $\sigma_i$, with }1\leq i\leq n \\
		I(e) & \textrm{otherwise},
	\end{array}\right.$$
	is also an indexing for $\sigma$.
\end{lem}
\begin{pf}
	Simply observe that the constraints of \reffig{Indexes} are preserved by translations, and that the indexes of edges in two distinct connected components are completely independent.\qed
\end{pf}

In light of what we said about \sps s, and by looking at how indexes are assigned in \reffig{Indexes}, it is obvious that an indexable sequentializable \ps\ corresponds to a derivation of the calculus of \reftab{StratSeqCalc}. Indeed, for example, the \sps s proving $\parg X\llinimp X$ and $X\llinimp\parg X$ are easily seen to admit no indexing. On the other hand, strongly indexable sequentializable \ps s correspond to proofs (\refdef{Proofs}), because strong indexability precisely reflects the requirement that all formulas in the conclusion have the same index.

We shall now prove that the \pn s of \refdef{ProofNet} provide a geometric characterization of strongly indexable sequentializable \ps s (modulo the absence of \struct s in the conclusions).

We start by recalling a classical result of the proof theory of linear logic:
\begin{prop}[\cite{Girard:ProofNets}]
	\label{prop:GirardSeq}
	A \ps\ is sequentializable iff it is DR-correct.\qed
\end{prop}

\begin{lem}
	\label{lemma:Indexable}
	A \ps\ is indexable iff it is \scorr.
\end{lem}
\begin{pf}
	Let $\sigma$ be indexable. The fact that all of its cycles are balanced is an immediate consequence of the following remark: if $I$ is an indexing of $\sigma$, and if $\phi$ is a path of $\sigma$, the balance of $\phi$ is obviously equal to $|I(e_1)-I(e_2)|$, where $e_1,e_2$ are the edges which mark the extremities of $\phi$; this is because indexings are defined precisely so as to ``count'' the traversals of paragraph links.

	Let now $\sigma$ be a \ps\ in which all cycles are balanced. We reason by induction on the number of links of $\sigma$. If $\widetilde\sigma$ is empty, the statement is trivially true. If $\widetilde\sigma$ has more than one connected component, the induction hypothesis immediately allows us to conclude. Hence, we may suppose that $\widetilde\sigma$ consists of exactly one, non-empty connected component. In that case, it is easy to see that there is a \emph{terminal} link $l$ of $\sigma$ which is not a \paxl\ (by ``terminal'' we mean a link at depth 0 and whose conclusions are all conclusions of $\sigma$). We may immediately discard the cases in which $l$ is an \axiom\ or a \weak\ link, because the statement is trivially seen to hold. If we ignore for the moment the possibility that $l$ is an \ofcourse\ link, we are in the following situation:
	\input{IndLemma}
	where $n\geq 1$, and the dotted arrow means that $l$ may or may not have a conclusion. Observe that every cycle of $\sigma_0$ is a cycle of $\sigma$, hence balanced; therefore, the induction hypothesis gives us an indexing $I_0$ of $\sigma_0$. In case $n=1$, $I_0$ may obviously be extended into an indexing for $\sigma$, so we may actually suppose that $n\geq 2$. Now, if $e_{i_1},e_{i_2}$ are two distinct premises of $l$ belonging to the same connected component of $\sigma_0$, by definition there is a path $\phi$ connecting them. If we had $I_0(e_{i_1})\neq I_0(e_{i_2})$, by the remark made at the beginning of this proof, the balance of $\phi$ would be non-null; this would induce a unbalanced cycle in $\sigma$, a contradiction. Hence, all premises of $l$ belonging to the same connected component of $\sigma_0$ are assigned the same index by $I_0$. It suffices now to apply the Shift \reflemma{Translation} to find an indexing $I_0'$ for $\sigma_0$ which assigns the same integer to $e_1,\ldots,e_n$; such an $I_0'$ obviously extends into an indexing for $\sigma$.

	We are left with the case in which $l$ is an \ofcourse\ link. Since all other links have been treated, we may assume that $\sigma$ actually consists of a single box $\cB$ whose \pal\ is $l$ itself. In this case, the induction hypothesis gives us an indexing for the contents of $\cB$, from which an indexing for $\sigma$ may be trivially defined.\qed
\end{pf}

Composing \reflemma{Indexable} with the fact that strong indexability coincides with indexability of the $\lpar$-closure immediately implies the following:
\begin{lem}
	\label{lemma:StronglyInd}
	Let $\sigma$ be a \ps\ with no \struct s in its conclusions. Then, $\sigma$ is strongly indexable iff its $\lpar$-closure is \scorr.\qed
\end{lem}

A combination of \refprop{GirardSeq} and \reflemma{StronglyInd} gives us the announced result:
\begin{thm}[Sequentialization]
	\label{th:Seq}
	A \ps\ with no \struct s in its conclusions is strongly indexable and sequentializable iff it is a \pn.\qed
\end{thm}

Similarly, one obtains that indexable and sequentializable \ps s with no \struct s in their conclusions (which correspond to \emph{derivations}, instead of proofs) coincide with \scorr\ \dr s. Hence, as it is usual in linear logic, the proof theory of \stratLL\ may be entirely formulated in terms of \pn s. The advantage is that no notion of index is ever explicitly mentioned.

%% file: links.tex
\begin{center}  
\begin{tikzpicture}[node distance=6mm and 6mm, baseline = (lab.base)]
\netax                                    {ax};
\node[nopoint, below left=of ax] (concaxl) {};
\node[nopoint, below right=of ax] (concaxr) {};
\draw[->,>=stealth,thick]
     (ax) to [out=180,in=90]  node [left] {$A^\perp$}
     (concaxl);
\draw[->,>=stealth,thick]
     (ax) to [out=0,in=90]  node [right] {$A$}
     (concaxr);
\node[below=of ax,yshift=3mm,lab] (lab) {\axiom};
\end{tikzpicture}\quad
\begin{tikzpicture}[node distance=6mm and 6mm, baseline = (lab.base)]
\netcut                                    {cut};
\node[nopoint, above left=of cut] (concaxl) {};
\node[nopoint, above right=of cut] (concaxr) {};
\draw[->,>=stealth,thick]
     (concaxl) to [in=180,out=270]  node [left] {$A^\perp$}
     (cut);
\draw[->,>=stealth,thick]
     (concaxr) to [in=0,out=270]  node [right] {$A$}
     (cut);
\node[below=of cut,yshift=6mm,lab] (lab) {\cut};
\end{tikzpicture}\\
\medskip
\begin{tikzpicture}[node distance=6mm and 6mm, baseline = (lab.base)]
\netone                      {one};
\node[nopoint, below=of one] (conc) {};
\draw[->,>=stealth,thick]
     (one) edge node [right] {$1$}
     (conc);
\node[below=of conc,yshift=4mm,lab] (lab) {\onel};
\end{tikzpicture}\quad
\begin{tikzpicture}[node distance=6mm and 6mm, baseline = (lab.base)]
\netbot                      {bot};
\node[nopoint, below=of bot] (conc) {};
\draw[->,>=stealth,thick]
     (bot) edge node [right] {$\perp$}
     (conc);
\node[below=of conc,yshift=4mm,lab] (lab) {\botl};
\end{tikzpicture}\quad
\begin{tikzpicture}[node distance=6mm and 6mm, baseline = (lab.base)]
\nettens                                 {tens};
\node[nopoint, above left=of tens] (hl) {};
\node[nopoint, above right=of tens] (hr) {};
\node[nopoint, below=of tens] (conc) {};
\draw[->,>=stealth,thick]
     (hl) to [in=150,out=270]  node [left] {$A$}
     (tens);
\draw[->,>=stealth,thick]
     (hr) to [in=30,out=270]  node [right] {$B$}
     (tens);
\draw[->,>=stealth,thick]
     (tens) edge node [right] {$A\otimes B$}
     (conc);
\node[below=of conc,yshift=4mm,lab] (lab) {\tensor};
\end{tikzpicture}\quad
\begin{tikzpicture}[node distance=6mm and 6mm, baseline = (lab.base)]
\netpar                                 {par};
\node[nopoint, above left=of par] (hl) {};
\node[nopoint, above right=of par] (hr) {};
\node[nopoint, below=of par] (conc) {};
\draw[->,>=stealth,thick]
     (hl) to [in=150,out=270]  node [left] {$A$}
     (par);
\draw[->,>=stealth,thick]
     (hr) to [in=30,out=270]  node [right] {$B$}
     (par);
\draw[->,>=stealth,thick]
     (par) edge node [right] {$A\parr B$}
     (conc);
\node[below=of conc,yshift=4mm,lab] (lab) {\parl};
\end{tikzpicture}\\
\medskip
\begin{tikzpicture}[node distance=6mm and 6mm, baseline = (lab.base)]
\netft                               {ft};
\node[nopoint, above=of ft] (h) {};
\node[nopoint, below=of ft] (conc) {};
\draw[->,>=stealth,thick]
     (h) edge node [right] {$A$}
     (ft);
\draw[->,>=stealth,thick]
     (ft) edge node [right] {$\flat A$}
     (conc);
\node[below=of conc,yshift=4mm,lab] (lab) {\flatl};
\end{tikzpicture}\quad
\begin{tikzpicture}[node distance=6mm and 6mm, baseline = (lab.base)]
\netpax                               {pax};
\node[nopoint, above=of pax] (h) {};
\node[nopoint, below=of pax] (conc) {};
\draw[->,>=stealth,thick]
     (h) edge node [right] {$\flat A$}
     (pax);
\draw[->,>=stealth,thick]
     (pax) edge node [right] {$\flat A$}
     (conc);
\node[below=of conc,yshift=4mm,lab] (lab) {\paxl};
\end{tikzpicture}\quad
\begin{tikzpicture}[node distance=6mm and 6mm, baseline = (lab.base)]
\netwn                                 {wn};
\node[nopoint, above left=of wn] (hl) {};
\node[nopoint, above right=of wn] (hr) {};
\node[nopoint, below=of wn] (conc) {};
\draw[->,>=stealth,thick]
     (hl) to [in=150,out=270]  node [left] {$\flat A$}
     (wn);
\draw[->,>=stealth,thick]
     (hr) to [in=30,out=270]  node [right] {$\flat A$}
     (wn);
\draw[->,>=stealth,thick]
     (wn) edge node [right] {$\wn A$}
     (conc);
%\draw[thick,dotted,>=stealth]
%    (wn) ++(3mm,4mm) arc (60:120:6mm);
\node[above=of wn,yshift=-3mm] (dots) {$\ldots$};
\node[below=of conc,yshift=4mm,lab] (lab) {\whynot};
\end{tikzpicture}\quad
\begin{tikzpicture}[node distance=6mm and 6mm, baseline = (lab.base)]
\netoc                               {oc};
\node[nopoint, above=of oc] (h) {};
\node[nopoint, below=of oc] (conc) {};
\draw[->,>=stealth,thick]
     (h) edge node [right] {$A$}
     (oc);
\draw[->,>=stealth,thick]
     (oc) edge node [right] {$\oc A$}
     (conc);
\node[below=of conc,yshift=4mm,lab] (lab) {\ofcourse};
\end{tikzpicture}\quad
\begin{tikzpicture}[node distance=6mm and 6mm, baseline = (lab.base)]
\netparg                              {parg};
\node[nopoint, above=of parg] (h) {};
\node[nopoint, below=of parg] (conc) {};
\draw[->,>=stealth,thick]
     (h) edge node [right] {$A$}
     (parg);
\draw[->,>=stealth,thick]
     (parg) edge node [right] {$\parg A$}
     (conc);
\node[below=of conc,yshift=4mm,lab] (lab) {\pargl};
\end{tikzpicture}
\end{center}

%%% Local Variables: 
%%% mode: latex
%%% TeX-master: "LLlevSem"
%%% End: 

%% file: boxes.tex
\begin{center}
\begin{tikzpicture}[node distance=6mm and 6mm, baseline = (current bounding box.east)]
%\mygridto{(4,7)}
\draw[very thick] (-0.5,1) rectangle (3.5,3.5);
\node[lab,fill=white,rectangle, text width=12mm, align=center, text depth=1.1ex]     (dots) at (1,1)  {$\ldots$};
\netpax[left=of dots,xshift=10mm]                            {pax1};
\netpax[right=of dots,xshift=-10mm]                           {paxn};
\netoc[right=of paxn]                            {oc};
\node[above=of pax1, text height=10mm] (pax11) {};
\node[above=of paxn, text height=10mm] (paxn1) {};
\node[above=of oc, text height=10mm] (oc1) {};
\node[nopoint,below=of oc]   (occonc) {};
\node[nopoint,below=of pax1]                      (cp1){};
\node[nopoint,below=of paxn]                      (cpn){};
\node[above=of dots, yshift=-5mm] (dots1) {$\ldots$};
\node[below=of dots, yshift=3mm] (dots2) {$\ldots$};
\draw[thick,>=stealth]
    (pax11) edge[->] node [left] {$\flat B_1$} (pax1)
    (paxn1) edge[->] node [right] {$\flat B_n$} (paxn)
    (oc1) edge[->]   node [right] {$A$} (oc)
    (pax1) edge[->] node [left] {$\flat B_1$} (cp1)
    (paxn) edge[->] node [right] {$\flat B_n$} (cpn)
    (oc) edge[->] node [right] {$\oc A$} (occonc);
\node[netpi,above=of dots,xshift=5mm, text width=30mm,text height=5mm,text depth=2.5mm] (pi) {$\pi$};
\end{tikzpicture}
\end{center}

%%% Local Variables: 
%%% mode: latex
%%% TeX-master: "LLlevSem"
%%% End: 

%% file: SPS.tex
\begin{center}
\begin{tikzpicture}[node distance=6mm and 3mm, baseline = (lab.base)]
%\mygridto{(6,3)}
%origine au centre (fausse conclusion) 
\node[nopoint] (parc) at (3,1) {};
%gamma
\node[lab,fill=white,rectangle, text width=12mm, align=center, text depth=1.1ex, left=of parc, xshift=-10mm] (Gamma)  {$\Gamma$};
\node[nopoint,left=of Gamma,xshift= 5mm]   (gamma1) {};
\node[nopoint,right=of Gamma,xshift=-5mm]  (gamman) {};
\node[above=of gamma1, yshift=3mm, text height=10mm] (gamma11) {};
\node[above=of gamman, yshift=3mm, text height=10mm] (gamma1n) {};
\node[above=of Gamma, yshift=3mm] (dots1) {$\ldots$};
\draw[thick,>=stealth]
    (gamma11) edge[->] node [left] {} (gamma1)
    (gamma1n) edge[->] node [right] {} (gamman);
%delta
\node[lab,fill=white,rectangle, text width=12mm, align=center, text depth=1.1ex, right=of parc, xshift=10mm] (Delta)  {$\Delta$};
\node[nopoint,left=of Delta,xshift= 5mm]   (delta1) {};
\node[nopoint,right=of Delta,xshift=-5mm]  (deltan) {};
\node[above=of delta1, yshift=3mm, text height=10mm] (delta11) {};
\node[above=of deltan, yshift=3mm, text height=10mm] (delta1n) {};
\node[above=of Delta, yshift=3mm] (dots1) {$\ldots$};
\draw[thick,>=stealth]
    (delta11) edge[->] node [left] {} (delta1)
    (delta1n) edge[->] node [right] {} (deltan);
%cut
\netcut[above=of parc, yshift=-7mm]{par};
\node[above left=of par, yshift=2mm] (A) {};
\node[above right=of par, yshift=2mm] (B) {};
\draw[->,>=stealth,thick]
     (A) to [in=180,out=270]  node [left] {$A$}
     (par);
\draw[->,>=stealth,thick]
     (B) to [in=0,out=270]  node [right] {$A^\perp$}
     (par);
%reseau sigma1
\node[netpi,above=of Gamma, yshift=-1mm, xshift=5mm, text width=22mm,text height=3mm,text depth=1.5mm] (sigma1) {$\sigma_1$};
%reseau sigma2
\node[netpi,above=of Delta, yshift=-1mm, xshift=-5mm, text width=22mm,text height=3mm,text depth=1.5mm] (sigma2) {$\sigma_2$};
%label
\node[lab] (lab) at (3,0.5) {\cut};
\end{tikzpicture}\hfill
\begin{tikzpicture}[node distance=6mm and 3mm, baseline = (lab.base)]
%\mygridto{(6,3)}
%conclusion tenseur
\node[nopoint] (parc) at (3,1) {};
%gamma
\node[lab,fill=white,rectangle, text width=12mm, align=center, text depth=1.1ex, left=of parc, xshift=-10mm] (Gamma)  {$\Gamma$};
\node[nopoint,left=of Gamma,xshift= 5mm]   (gamma1) {};
\node[nopoint,right=of Gamma,xshift=-5mm]  (gamman) {};
\node[above=of gamma1, yshift=10mm, text height=10mm] (gamma11) {};
\node[above=of gamman, yshift=10mm, text height=10mm] (gamma1n) {};
\node[above=of Gamma, yshift=3mm] (dots1) {$\ldots$};
\draw[thick,>=stealth]
    (gamma11) edge[->] node [left] {} (gamma1)
    (gamma1n) edge[->] node [right] {} (gamman);
%delta
\node[lab,fill=white,rectangle, text width=12mm, align=center, text depth=1.1ex, right=of parc, xshift=10mm] (Delta)  {$\Delta$};
\node[nopoint,left=of Delta,xshift= 5mm]   (delta1) {};
\node[nopoint,right=of Delta,xshift=-5mm]  (deltan) {};
\node[above=of delta1, yshift=10mm, text height=10mm] (delta11) {};
\node[above=of deltan, yshift=10mm, text height=10mm] (delta1n) {};
\node[above=of Delta, yshift=3mm] (dots1) {$\ldots$};
\draw[thick,>=stealth]
    (delta11) edge[->] node [left] {} (delta1)
    (delta1n) edge[->] node [right] {} (deltan);
%tenseur
\nettens[above=of parc, yshift=-2mm]{par};
\node[above left=of par, yshift=3mm] (A) {};
\node[above right=of par, yshift=3mm] (B) {};
\draw[->,>=stealth,thick]
     (A) to [in=150,out=270]  node [left] {$A$}
     (par);
\draw[->,>=stealth,thick]
     (B) to [in=30,out=270]  node [right] {$B$}
     (par);
\draw[thick,>=stealth]
    (par) edge[->] node [right] {$A\otimes B$} (parc);
%reseau sigma1
\node[netpi,above=of Gamma, yshift=6mm, xshift=5mm, text width=22mm,text height=3mm,text depth=1.5mm] (sigma1) {$\sigma_1$};
%reseau sigma2
\node[netpi,above=of Delta, yshift=6mm, xshift=-5mm, text width=22mm,text height=3mm,text depth=1.5mm] (sigma2) {$\sigma_2$};
%label
\node[lab] (lab) at (3,0.5) {\tensor};
\end{tikzpicture}\\
\begin{tikzpicture}[node distance=6mm and 3mm, baseline = (lab.base)]
%\mygridto{(4,3)}
\node[lab,fill=white,rectangle, text width=12mm, align=center, text depth=1.1ex]     (Gamma) at (1,1)  {$\Gamma$};
\node[nopoint,left=of Gamma,xshift= 5mm]    (gamma1) {};
\node[nopoint,right=of Gamma,xshift=-5mm]  (gamman) {};
\node[above=of gamma1, text height=10mm] (gamma11) {};
\node[above=of gamman, text height=10mm] (gamma1n) {};
\node[above=of Gamma, yshift=-7mm] (dots1) {$\ldots$};
\draw[thick,>=stealth]
    (gamma11) edge[->] node [left] {} (gamma1)
    (gamma1n) edge[->] node [right] {} (gamman);
\node[netpi,above=of Gamma, yshift=-4mm, text width=12mm,text height=3mm,text depth=1.5mm] (sigma) {$\sigma$};
\netbot[right =of sigma, xshift=0mm]{bot};
\node[nopoint,below=of bot] (botc) {};
\draw[thick,>=stealth]
    (bot) edge[->] node [right] {$\perp$} (botc);
%label
\node[lab] (lab) at (1.6,0.5) {\botl};
\end{tikzpicture}
\hfill
\begin{tikzpicture}[node distance=6mm and 3mm, baseline = (lab.base)]
%\mygridto{(4,3)}
\node[lab,fill=white,rectangle, text width=12mm, align=center, text depth=1.1ex]     (Gamma) at (1,1)  {$\Gamma$};
\node[nopoint,left=of Gamma,xshift= 5mm]    (gamma1) {};
\node[nopoint,right=of Gamma,xshift=-5mm]  (gamman) {};
\node[above=of gamma1, yshift=10mm, text height=10mm] (gamma11) {};
\node[above=of gamman, yshift=10mm, text height=10mm] (gamma1n) {};
\node[above=of Gamma, yshift=3mm] (dots1) {$\ldots$};
\draw[thick,>=stealth]
    (gamma11) edge[->] node [left] {} (gamma1)
    (gamma1n) edge[->] node [right] {} (gamman);
%par
\node[nopoint,right=of gamman, xshift=10mm] (parc) {};
\netpar[above=of parc, yshift=-2mm]{par};
\node[above left=of par, yshift=3mm] (A) {};
\node[above right=of par, yshift=3mm] (B) {};
\draw[->,>=stealth,thick]
     (A) to [in=150,out=270]  node [left] {$A$}
     (par);
\draw[->,>=stealth,thick]
     (B) to [in=30,out=270]  node [right] {$B$}
     (par);
\draw[thick,>=stealth]
    (par) edge[->] node [right] {$A\parr B$} (parc);
%reseau sigma
\node[netpi,above=of Gamma, yshift=6mm, xshift=10mm, text width=32mm,text height=3mm,text depth=1.5mm] (sigma) {$\sigma$};
%label
\node[lab] (lab) at (2,0.5) {\parl};
\end{tikzpicture}
\hfill
\begin{tikzpicture}[node distance=6mm and 6mm, baseline = (lab.base)]
%\mygridto{(4,4)}
\draw[very thick] (-0.5,1) rectangle (3.5,3);
\node[lab,fill=white,rectangle, text width=12mm, align=center, text depth=1.1ex]     (dots) at (1,1)  {$\ldots$};
\netpax[left=of dots,xshift=10mm]                            {pax1};
\netpax[right=of dots,xshift=-10mm]                           {paxn};
\netoc[right=of paxn]                            {oc};
\node[above=of pax1, text height=10mm] (pax11) {};
\node[above=of paxn, text height=10mm] (paxn1) {};
\node[above=of oc, text height=10mm] (oc1) {};
\node[nopoint,below=of oc]   (occonc) {};
\node[nopoint,below=of pax1]                      (cp1){};
\node[nopoint,below=of paxn]                      (cpn){};
\node[above=of dots, yshift=-5mm] (dots1) {$\ldots$};
\node[below=of dots, yshift=3mm] (dots2) {$\ldots$};
\draw[thick,>=stealth]
    (pax11) edge[->] node [left] {$\flat B_1$} (pax1)
    (paxn1) edge[->] node [right] {$\flat B_n$} (paxn)
    (oc1) edge[->]   node [right] {$A$} (oc)
    (pax1) edge[->] node [left] {$\flat B_1$} (cp1)
    (paxn) edge[->] node [right] {$\flat B_n$} (cpn)
    (oc) edge[->] node [right] {$\oc A$} (occonc);
\node[netpi,above=of dots,xshift=5mm, text width=30mm,text height=5mm,text depth=2.5mm] (pi) {$\pi$};
%label
\node[lab] (lab) at (1.5,-0.4) {\ofcourse};
\end{tikzpicture}\\

\begin{tikzpicture}[node distance=6mm and 3mm, baseline = (lab.base)]
%\mygridto{(4,3)}
\node[lab,fill=white,rectangle, text width=12mm, align=center, text depth=1.1ex]     (Gamma) at (1,1)  {$\Gamma$};
\node[nopoint,left=of Gamma,xshift= 5mm]    (gamma1) {};
\node[nopoint,right=of Gamma,xshift=-5mm]  (gamman) {};
\node[above=of gamma1, yshift=10mm, text height=10mm] (gamma11) {};
\node[above=of gamman, yshift=10mm, text height=10mm] (gamma1n) {};
\node[above=of Gamma, yshift=3mm] (dots1) {$\ldots$};
\draw[thick,>=stealth]
    (gamma11) edge[->] node [left] {} (gamma1)
    (gamma1n) edge[->] node [right] {} (gamman);
%wn
\node[nopoint,right=of gamman, xshift=10mm] (parc) {};
\netwn[above=of parc, yshift=-2mm]{par};
\node[above left=of par, yshift=3mm] (A) {};
\node[above right=of par, yshift=3mm] (B) {};
\draw[->,>=stealth,thick]
     (A) to [in=150,out=270]  node [left] {$\flat A$}
     (par);
\draw[->,>=stealth,thick]
     (B) to [in=30,out=270]  node [right] {$\flat A$}
     (par);
\draw[thick,>=stealth]
    (par) edge[->] node [right] {$\wn A$} (parc);
%reseau sigma
\node[netpi,above=of Gamma, yshift=6mm, xshift=10mm, text width=32mm,text height=3mm,text depth=1.5mm] (sigma) {$\sigma$};
%label
\node[lab] (lab) at (2,0.5) {\whynot};
\end{tikzpicture}
\hfill
\begin{tikzpicture}[node distance=6mm and 3mm, baseline = (lab.base)]
%\mygridto{(4,3)}
\node[lab,fill=white,rectangle, text width=12mm, align=center, text depth=1.1ex]     (Gamma) at (1,1)  {$\Gamma$};
\node[nopoint,left=of Gamma,xshift= 5mm]    (gamma1) {};
\node[nopoint,right=of Gamma,xshift=-5mm]  (gamman) {};
\node[above=of gamma1, yshift=10mm, text height=10mm] (gamma11) {};
\node[above=of gamman, yshift=10mm, text height=10mm] (gamma1n) {};
\node[above=of Gamma, yshift=3mm] (dots1) {$\ldots$};
\draw[thick,>=stealth]
    (gamma11) edge[->] node [left] {} (gamma1)
    (gamma1n) edge[->] node [right] {} (gamman);
%flat
\node[nopoint,right=of gamman, xshift=5mm] (parc) {};
\netft[above=of parc, yshift=-2mm]{par};
\node[above=of par, yshift=-1mm] (A) {};
\draw[->,>=stealth,thick]
     (A) edge[->]  node [right] {$A$}
     (par);
\draw[thick,>=stealth]
    (par) edge[->] node [right] {$\flat A$} (parc);
%reseau sigma
\node[netpi,above=of Gamma, yshift=6mm, xshift=5mm, text width=22mm,text height=3mm,text depth=1.5mm] (sigma) {$\sigma$};
%label
\node[lab] (lab) at (1.5,0.5) {\flatl};
\end{tikzpicture}
\hfill
\begin{tikzpicture}[node distance=6mm and 3mm, baseline = (lab.base)]
%\mygridto{(4,3)}
\node[lab,fill=white,rectangle, text width=12mm, align=center, text depth=1.1ex]     (Gamma) at (1,1)  {$\Gamma$};
\node[nopoint,left=of Gamma,xshift= 5mm]    (gamma1) {};
\node[nopoint,right=of Gamma,xshift=-5mm]  (gamman) {};
\node[above=of gamma1, yshift=10mm, text height=10mm] (gamma11) {};
\node[above=of gamman, yshift=10mm, text height=10mm] (gamma1n) {};
\node[above=of Gamma, yshift=3mm] (dots1) {$\ldots$};
\draw[thick,>=stealth]
    (gamma11) edge[->] node [left] {} (gamma1)
    (gamma1n) edge[->] node [right] {} (gamman);
%flat
\node[nopoint,right=of gamman, xshift=5mm] (parc) {};
\netparg[above=of parc, yshift=-2mm]{par};
\node[above=of par, yshift=-1mm] (A) {};
\draw[->,>=stealth,thick]
     (A) edge[->]  node [right] {$A$}
     (par);
\draw[thick,>=stealth]
    (par) edge[->] node [right] {$\parg A$} (parc);
%reseau sigma
\node[netpi,above=of Gamma, yshift=6mm, xshift=5mm, text width=22mm,text height=3mm,text depth=1.5mm] (sigma) {$\sigma$};
%label
\node[lab] (lab) at (1.5,0.5) {\pargl};
\end{tikzpicture}
\end{center}

%%% Local Variables: 
%%% mode: latex
%%% TeX-master: "LLlevSem"
%%% End: 

%% file: indexes.tex
\begin{center}  
\begin{tikzpicture}[node distance=6mm and 6mm, baseline = (lab.base)]
\netax                                    {ax};
\node[nopoint, below left=of ax] (concaxl) {};
\node[nopoint, below right=of ax] (concaxr) {};
\draw[->,>=stealth,thick]
     (ax) to [out=180,in=90]  node [left] {$i$}
     (concaxl);
\draw[->,>=stealth,thick]
     (ax) to [out=0,in=90]  node [right] {$i$}
     (concaxr);
\node[below=of ax,yshift=3mm,lab] (lab) {};
\end{tikzpicture}\quad
\begin{tikzpicture}[node distance=6mm and 6mm, baseline = (lab.base)]
\netcut                                    {cut};
\node[nopoint, above left=of cut] (concaxl) {};
\node[nopoint, above right=of cut] (concaxr) {};
\draw[->,>=stealth,thick]
     (concaxl) to [in=180,out=270]  node [left] {$i$}
     (cut);
\draw[->,>=stealth,thick]
     (concaxr) to [in=0,out=270]  node [right] {$i$}
     (cut);
\node[below=of cut,yshift=6mm,lab] (lab) {};
\end{tikzpicture}\\
\medskip
\begin{tikzpicture}[node distance=6mm and 6mm, baseline = (lab.base)]
\netone                      {one};
\node[nopoint, below=of one] (conc) {};
\draw[->,>=stealth,thick]
     (one) edge node [right] {$i$}
     (conc);
\node[below=of conc,yshift=4mm,lab] (lab) {};
\end{tikzpicture}\quad
\begin{tikzpicture}[node distance=6mm and 6mm, baseline = (lab.base)]
\netbot                      {bot};
\node[nopoint, below=of bot] (conc) {};
\draw[->,>=stealth,thick]
     (bot) edge node [right] {$i$}
     (conc);
\node[below=of conc,yshift=4mm,lab] (lab) {};
\end{tikzpicture}\quad
\begin{tikzpicture}[node distance=6mm and 6mm, baseline = (lab.base)]
\nettens                                 {tens};
\node[nopoint, above left=of tens] (hl) {};
\node[nopoint, above right=of tens] (hr) {};
\node[nopoint, below=of tens] (conc) {};
\draw[->,>=stealth,thick]
     (hl) to [in=150,out=270]  node [left] {$i$}
     (tens);
\draw[->,>=stealth,thick]
     (hr) to [in=30,out=270]  node [right] {$i$}
     (tens);
\draw[->,>=stealth,thick]
     (tens) edge node [right] {$i$}
     (conc);
\node[below=of conc,yshift=4mm,lab] (lab) {};
\end{tikzpicture}\quad
\begin{tikzpicture}[node distance=6mm and 6mm, baseline = (lab.base)]
\netpar                                 {par};
\node[nopoint, above left=of par] (hl) {};
\node[nopoint, above right=of par] (hr) {};
\node[nopoint, below=of par] (conc) {};
\draw[->,>=stealth,thick]
     (hl) to [in=150,out=270]  node [left] {$i$}
     (par);
\draw[->,>=stealth,thick]
     (hr) to [in=30,out=270]  node [right] {$i$}
     (par);
\draw[->,>=stealth,thick]
     (par) edge node [right] {$i$}
     (conc);
\node[below=of conc,yshift=4mm,lab] (lab) {};
\end{tikzpicture}\\
\medskip
\begin{tikzpicture}[node distance=6mm and 6mm, baseline = (lab.base)]
\netft                               {ft};
\node[nopoint, above=of ft] (h) {};
\node[nopoint, below=of ft] (conc) {};
\draw[->,>=stealth,thick]
     (h) edge node [right] {$i$}
     (ft);
\draw[->,>=stealth,thick]
     (ft) edge node [right] {$i$}
     (conc);
\node[below=of conc,yshift=4mm,lab] (lab) {};
\end{tikzpicture}\quad
\begin{tikzpicture}[node distance=6mm and 6mm, baseline = (lab.base)]
\netpax                               {pax};
\node[nopoint, above=of pax] (h) {};
\node[nopoint, below=of pax] (conc) {};
\draw[->,>=stealth,thick]
     (h) edge node [right] {$i$}
     (pax);
\draw[->,>=stealth,thick]
     (pax) edge node [right] {$i$}
     (conc);
\node[below=of conc,yshift=4mm,lab] (lab) {};
\end{tikzpicture}\quad
\begin{tikzpicture}[node distance=6mm and 6mm, baseline = (lab.base)]
\netwn                                 {wn};
\node[nopoint, above left=of wn] (hl) {};
\node[nopoint, above right=of wn] (hr) {};
\node[nopoint, below=of wn] (conc) {};
\draw[->,>=stealth,thick]
     (hl) to [in=150,out=270]  node [left] {$i$}
     (wn);
\draw[->,>=stealth,thick]
     (hr) to [in=30,out=270]  node [right] {$i$}
     (wn);
\draw[->,>=stealth,thick]
     (wn) edge node [right] {$i$}
     (conc);
%\draw[thick,dotted,>=stealth]
%    (wn) ++(3mm,4mm) arc (60:120:6mm);
\node[above=of wn,yshift=-3mm] (dots) {$\ldots$};
\node[below=of conc,yshift=4mm,lab] (lab) {};
\end{tikzpicture}\quad
\begin{tikzpicture}[node distance=6mm and 6mm, baseline = (lab.base)]
\netoc                               {oc};
\node[nopoint, above=of oc] (h) {};
\node[nopoint, below=of oc] (conc) {};
\draw[->,>=stealth,thick]
     (h) edge node [right] {$i$}
     (oc);
\draw[->,>=stealth,thick]
     (oc) edge node [right] {$i$}
     (conc);
\node[below=of conc,yshift=4mm,lab] (lab) {};
\end{tikzpicture}\quad
\begin{tikzpicture}[node distance=6mm and 6mm, baseline = (lab.base)]
\netparg                              {parg};
\node[nopoint, above=of parg] (h) {};
\node[nopoint, below=of parg] (conc) {};
\draw[->,>=stealth,thick]
     (h) edge node [right] {$i + 1$}
     (parg);
\draw[->,>=stealth,thick]
     (parg) edge node [right] {$i$}
     (conc);
\node[below=of conc,yshift=4mm,lab] (lab) {};
\end{tikzpicture}
\end{center}

%%% Local Variables: 
%%% mode: latex
%%% TeX-master: "LLlevSem"
%%% End: 

%% file: IndLemma.tex
\begin{center}
\begin{tikzpicture}[node distance=6mm and 3mm]
%\mygridto{(4,3)}
\node[lab,fill=white,rectangle, text width=12mm, align=center, text depth=1.1ex]     (Gamma) at (1,1)  {};
\node[nopoint,left=of Gamma,xshift= 5mm]    (gamma1) {};
\node[nopoint,right=of Gamma,xshift=-5mm]  (gamman) {};
\node[above=of gamma1, yshift=10mm, text height=10mm] (gamma11) {};
\node[above=of gamman, yshift=10mm, text height=10mm] (gamma1n) {};
\node[above=of Gamma] (dots1) {$\ldots$};
\draw[thick,>=stealth]
    (gamma11) edge[->] node [left] {} (gamma1)
    (gamma1n) edge[->] node [right] {} (gamman);
%par
\node[nopoint,right=of gamman, xshift=10mm] (parc) {};
\inetcell[netcircle, above=of parc, yshift=-2mm] (par) {$l$};
\node[above left=of par, yshift=3mm] (A) {};
\node[above right=of par, yshift=3mm] (B) {};
\node[above=of par, yshift=-1mm] (aritydots) {$\ldots$};
\draw[->,>=stealth,thick]
     (A) to [in=150,out=270]  node [left] {$e_1$}
     (par);
\draw[->,>=stealth,thick]
     (B) to [in=30,out=270]  node [right] {$e_n$}
     (par);
\draw[thick,>=stealth]
    (par) edge[->,dotted] node [right] {} (parc);
%reseau sigma0
\node[netpi,above=of Gamma, yshift=6mm, xshift=10mm, text width=32mm,text height=3mm,text depth=1.5mm] (sigma0) {$\sigma_0$};
\node[left=of dots1,xshift=-5mm] (sigma) {$\sigma =$};
\end{tikzpicture}
\end{center}

%%% Local Variables: 
%%% mode: latex
%%% TeX-master: "LLlevSem"
%%% End: 

%% file: CutElim.tex
\begin{figure}[t]
        \input{axstep.tex}
	\caption{Axiom step.}
	\label{fig:AxStep}
\end{figure}
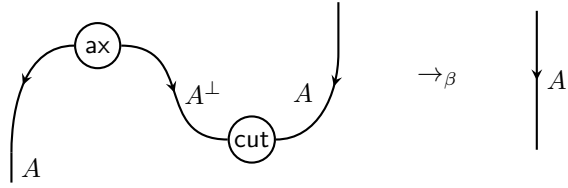
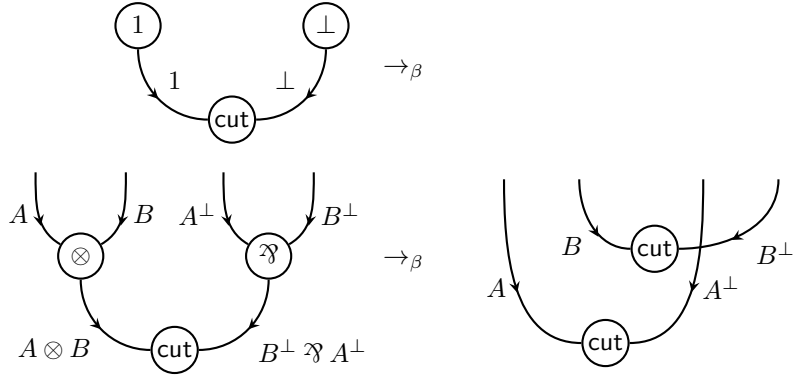
\begin{figure}[t]
	\input{multstep.tex}
	\caption{Multiplicative steps.}
	\label{fig:MultStep}
\end{figure}
\begin{figure}[t]
        \input{expstep.tex}
	\caption{Exponential step; $\flat\Gamma$ is a multiset of \struct s, so one \paxl\ link, \whynot\ link, or wire in the picture may in some case stand for several (including zero) \paxl\ links, \whynot\ links, or wires.}
	\label{fig:ExpStep}
\end{figure}
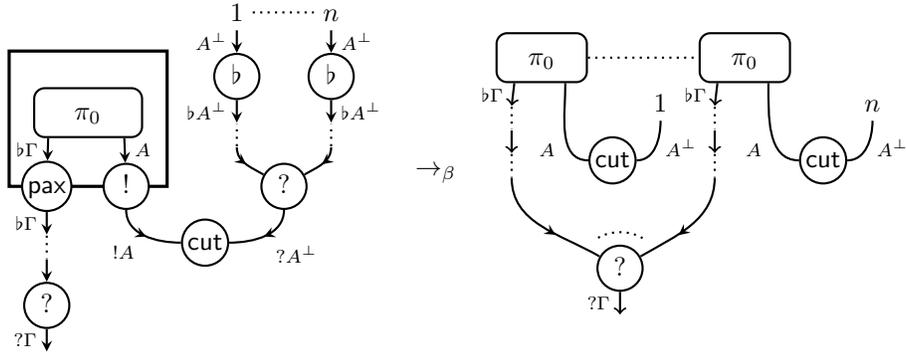
\begin{figure}[t]
	\input{pargnoboxstep}
	\caption{Paragraph step.}
	\label{fig:PargStep}
\end{figure}
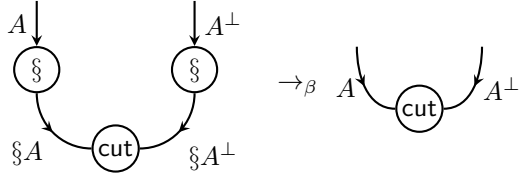

Formulating the cut-elimination procedure in \pn s is quite simple: there are only five rules (or \emph{steps}, as they are often called), taking the form of the graph-rewriting rules given in Figures~\ref{fig:AxStep} through~\ref{fig:PargStep}. Observe that, to be properly formulated, the exponential step requires that no \struct\ is a conclusion, for otherwise the \ps\ on the right-hand side of \reffig{ExpStep} may have more conclusions than the \ps\ on the left-hand side. This is why such a condition is imposed on \dr s and \pn s.

When a \pn\ (or \dr) $\pi$ is transformed into $\pi'$ by the application of one cut-elimination step, we write $\pi\betared\pi'$, and we say that $\pi$ \emph{$\beta$-reduces} to $\pi'$. The preservation of \dr s under cut-elimination is a classical result of linear logic:
\begin{prop}
	\label{prop:preservation}
	Let $\pi$ be a \dr, and let $\pi\betared\pi'$. Then, the \ps\ $\pi'$ is a \dr.
\end{prop}

Confluence and strong normalization of reduction of \pn s (and of \dr s) may be proved by standard arguments, which may be found for instance in \cite{Girard:LL,Danos:PhDThesis}. The key observation is that the paragraph step is trivial, and adds virtually nothing with respect to the dynamics of cut-elimination in propositional multiplicative exponential linear logic.

By looking at cut-elimination in \pn s, we may find a deeper meaning to the concept of stratification. Indeed, as observed in \cite{BaillotMazza:LLlev}, it is immediate to see that indexings (which exist for \pn s by \refth{Seq}) are preserved under cut-elimination, in the sense that the residue of a link may be given the same index as its ancestor (``residue'' and ``ancestor'' are here meant to take the standard meaning of rewriting theory, \cf\ \refdef{Lift} for a formal definition). This remark allows (using Theorem~\ref{th:Seq}) to immediately extend Proposition~\ref{prop:preservation} to \pn s:
\begin{prop}
	Let $\pi$ be a \pn, and let $\pi\betared\pi'$. Then, the \ps\ $\pi'$ is a \pn.
\end{prop}

One can thus meaningfully speak of the \emph{level} of a link of a \pn, and observe that interaction (by means of a cut) may only ever happen between dual links of identical level. A \pn\ of \stratLL\ may therefore be seen as partitioned into \emph{strata}, which do not communicate with each other.

Thanks to the stability of levels, it is possible to consider a cut-elimination protocol which attempts to eliminate cuts at level $i$ before touching any cut at level $i+1$, starting from the minimum level. In full \stratLL, this protocol has no special property; however, in the fragment corresponding to \LLlev, where the exponential modalities (which control duplication) are forced to modify levels, this protocol enjoys an elementary bound, as proved in \cite{BaillotMazza:LLlev} (or a polynomial bound in the case of the subsystem \LLLlev). Furthermore, since in \ELL\ and \LLL\ levels coincide with exponential depths (\ie, the nesting of boxes, see \refdef{Net}), this very same protocol is also the one for which the elementary and polynomial bounds were originally proved by Girard.

%% file: axstep.tex
\begin{center}  
\begin{tikzpicture}[node distance=8mm and 8mm]
\netax                                    {ax};
\netcut [below right=of ax, xshift=8mm]   {cut};
\node   [above right=of cut, yshift=8mm]  (cutr)     {};
\node   [below left=of ax, yshift=-8mm]   (axl)      {};
\node   [below left=of ax, yshift=-4mm]   (axl2)      {};
\node   [below right=of cutr]             (beta)     {$\betared$};
\node   [right=of beta]                   (fil)      {};
\node   [above=of fil]                    (fil1)     {};
\node   [below=of fil]                    (fil2)     {};
\draw[directed,thick]
     (ax.east) .. controls +(right:10mm) 
                       and +(left:10mm) ..
                       node [right] {$A^\perp$} 
     (cut.west);
\draw[directed,thick]
     (cutr) -- +(down:8mm) to [in=0,out=270]
                     node [above left] {$A$}
     (cut.east);
\draw[directed,thick]
     (ax) to [out=180,in=90] 
     (axl2) edge node [right] {$A$}
     (axl);
\draw[thick]
     (fil1) edge [directed] node [right] {$A$} 
     (fil2);
\end{tikzpicture}
\end{center}

%%% Local Variables: 
%%% mode: latex
%%% TeX-master: "LLlevSem"
%%% End: 

%% file: multstep.tex
\begin{align*}
\begin{tikzpicture}[node distance = 8mm and 8mm,baseline = (current bounding box.east)]
\netcut {cut};
\netone [above left=of cut] {one};
\netbot [above right=of cut] {bot};
\draw[thick,directed]
     (one.south) to [out=270,in=180] node [above right] {$1$}
     (cut.west);
\draw[thick,directed]
     (bot.south) to [out=270,in=0] node [above left] {$\bot$}
     (cut.east);
\end{tikzpicture}\quad
&\betared & \\
\begin{tikzpicture}[node distance = 8mm and 8mm,baseline = (current bounding box.east)]
\netcut {cut};
\nettens [above left=of cut] {tens};
\netpar [above right=of cut] {parr};
\node [above =of tens, xshift=-6mm]  (A) {};
\node [above =of tens, xshift=6mm]   (B) {};
\node [above =of parr, xshift=-6mm]  (Ao) {};
\node [above =of parr, xshift=6mm]   (Bo) {};
\draw[thick,directed]
     (tens.south) to [out=270,in=180] node [below left] {$A\otimes B$}
     (cut.west);
\draw[thick,directed]
     (parr.south) to [out=270,in=0] node [below right] {$B^\perp\parr A^\perp$}
     (cut.east);
\draw[reverse directed, thick]
     (tens) to [in=270,out=150] node [left] {$A$}
     (A);
\draw[reverse directed, thick]
     (parr) to [in=270,out=150] node [left] {$A^\perp$}
     (Ao);
\draw[reverse directed, thick]
     (tens) to [in=270,out=30] node [right] {$B$}
     (B);
\draw[thick,reverse directed]
     (parr) to [in=270,out=30] node [right] {$B^\perp$}
     (Bo);
\end{tikzpicture}
&\betared & 
\begin{tikzpicture}[node distance = 8mm and 6mm,baseline = (current bounding box.east)]
\node (tens) {};
\node [right=of tens,xshift=16mm](parr) {};
\netcut [left=of parr] {cut2};
\netcut [below left=of cut2,xshift=4mm] {cut1};
\node [above =of tens, xshift=-6mm]  (A) {};
\node [above =of tens, xshift=4mm]   (B) {};
\node [above =of parr, xshift=-4mm]  (Ao) {};
\node [above =of parr, xshift=6mm]   (Bo) {};
\draw[reverse directed, thick]
     (cut1) to [in=270,out=180] node [left] {$A$}
     (A);
\draw[reverse directed, thick]
     (cut2) to [in=270,out=180] node [below left] {$B$}
     (B);
\draw[reverse directed, thick]
     (cut1) to [in=270,out=0] node [right] {$A^\perp$}
     (Ao);
\draw[thick,reverse directed]
     (cut2) to [in=270,out=0] node [below right] {$B^\perp$}
     (Bo);
\end{tikzpicture}
\end{align*}

%%% Local Variables: 
%%% mode: latex
%%% TeX-master: "LLlevSem"
%%% End: 

%% file: expstep.tex
\begin{center}
\begin{tikzpicture}[node distance=3mm and 4mm, baseline = (current bounding box.east)]
%\draw[step=4mm,very thin,gray] (0,0) grid (5,3);
\draw[very thick] (-0.5,0) rectangle (1.6,1.8);
\netpax                                  {pax};
\node[netpi,above=of pax,xshift=5.5mm, text width=12mm] (pi) {$\pi_0$};
\netoc[right=of pax]                     {oc};
\netcut[below right=of oc, xshift=2mm]   {cut};
\netwn[above right=of cut, xshift=2mm]   {wn1};
\node[nopoint,above left=of wn1]                 (l1){};
\node[nopoint,above right=of wn1]                (r1){};
\node[nopoint,above=of l1]                       (l2){};
\node[nopoint,above=of r1]                       (r2){};
\netft[above=of l2]                      {l};
\netft[above=of r2]                      {r};
\node[above=of l,text height=1.5ex]                        (l3){$1$};
\node[above=of r,text height=1.5ex]                        (r3){$n$};
\node[nopoint,below=of pax]                      (p1){};
\node[nopoint,below=of p1]                       (p2){};
\netwn[below=of p2]                      {wn2};
\node[nopoint,below=of wn2]                      (p3){};
\draw[thick,>=stealth]
    (pi.south west) ++(2mm,0) edge[->] node [left] {\ednd{\flat\Gamma}}
    (pax)
    (pi.south east) ++(-2.5mm,0) edge[->] node [right] {\ednd{A}}
    (oc)
    (pax) edge [->] node [left] {\ednd{\flat\Gamma}}
    (p1)
    (p2) edge [->]
    (wn2)
    (wn2) edge [->] node [left] {\ednd{\wn\Gamma}}
    (p3)
    (l3) edge [->] node [left] {\ednd{A^\perp}}
    (l)
    (r3) edge [->] node [right] {\ednd{A^\perp}}
    (r)
    (l) edge [->] node [left] {\ednd{\flat A^\perp}}
    (l2)
    (r) edge [->] node [right] {\ednd{\flat A^\perp}}
    (r2);
\draw[thick,directed]
    (l1) to [out=270,in=150] 
    (wn1);
\draw[thick,directed]
    (r1) to [out=270,in=30]
    (wn1);
\draw[thick,directed]
    (wn1) to [out=270,in=0] node [below right] {\ednd{\wn A^\perp}}
    (cut.east);
\draw[thick,directed]
    (oc) to [out=270,in=180,->] node [below left] {\ednd{\oc A}}
    (cut.west);
\draw[thick,dotted]
    (l1) edge (l2)
    (r1) edge (r2)
    (p1) edge (p2)
    (l3) edge (r3);
\end{tikzpicture}
\hfill$\betared$\hfill
\begin{tikzpicture}[node distance=3mm and 2mm, baseline = (current bounding box.east)]
%\draw[step=4mm,very thin,gray] (0,0) grid (5,3);

\node[nopoint] (l1) at (0.3,1) {};
\node[nopoint, above=of l1] (l2) {};
\node[nopoint, above=of l2] (l3) {};
\node[nopoint, above=of l3] (l4) {};
\node[netpi,above=of l4,xshift=4mm, text width=8mm] (pi1) {$\pi_0$};
\netcut[below right=of pi1,xshift=-1mm,yshift=-5mm] {cut1};
\node[above right=of cut1] (1) {$1$};

\node[nopoint] (r1) at (3,1) {};
\node[nopoint, above=of r1] (r2) {};
\node[nopoint, above=of r2] (r3) {};
\node[nopoint, above=of r3] (r4) {};
\node[netpi,above=of r4,xshift=4mm, text width=8mm] (pin) {$\pi_0$};
\netcut[below right=of pin,xshift=0mm,yshift=-5mm] {cutn}; 
\node[above right=of cutn] (n) {$n$};

\netwn[below=of cut1, xshift=1mm,yshift=-5mm] {wn};
\node[nopoint, below=of wn] (p) {};

\draw[thick,->]
    (pi1.south west) ++(2.5mm,0) edge node [left] {\ednd{\flat\Gamma}}
    (l4)
    (l3) edge [->] (l2)
    (pin.south west) ++(2.5mm,0) edge node [left] {\ednd{\flat\Gamma}}
    (r4)
    (r3) -- (r2)
    (wn) edge node [left] {\ednd{\wn\Gamma}} (p);
\draw[thick]
    (pi1.south east) ++(-3mm,0) to [out=270,in=180] node [below left] {\ednd{A}}
    (cut1.west);
\draw[thick]
    (pin.south east) ++(-3mm,0) to [out=270,in=180] node [below left] {\ednd{A}}
    (cutn.west);
\draw[thick]
    (cut1.east) to [out=0,in=270] node [right] {\ednd{A^\perp}}
    (1);
\draw[thick]
    (cutn.east) to [out=0,in=270] node [right] {\ednd{A^\perp}}
    (n);
\draw[thick,reverse directed]
    (wn) to [in=270,out=150] 
    (l1);
\draw[thick,reverse directed]
    (wn) to [in=270,out=30] 
    (r1);
\draw[thick,dotted,>=stealth]
    (pi1) edge (pin)
    (l2) edge (l1)
    (l4) edge (l3)
    (r2) edge (r1)
    (r4) edge (r3)
    (wn) ++(3mm,4mm) arc (60:120:6mm);
\end{tikzpicture}
\end{center}

%%% Local Variables: 
%%% mode: latex
%%% TeX-master: "LLlevSem"
%%% End: 

%% file: pargnoboxstep.tex
\begin{center}
\begin{tikzpicture}[node distance = 6mm and 6mm, baseline = (current bounding box.east)]
\netcut {cut};
\netparg [above left=of cut] {p1};
\netparg [above right=of cut] {p2};
\node[nopoint, above=of p1] (pp1) {};
\node[nopoint, above=of p2] (pp2) {};
\draw[thick,directed]
     (p1.south) to [out=270,in=180] node [below left] {$\parg A$}
     (cut.west);
\draw[thick,directed]
     (p2.south) to [out=270,in=0] node [below right] {$\parg A^\perp$}
     (cut.east);
\draw[->,thick,>=stealth]
     (pp1) edge node [left] {$A$} (p1)
     (pp2) edge node [right] {$A^\perp$} (p2);
\end{tikzpicture}\quad
$\betared$ 
\begin{tikzpicture}[node distance = 6mm and 6mm,baseline = (current bounding box.east)]
\netcut {cut};
\node[nopoint, above left=of cut] (pp1) {};
\node[nopoint, above right=of cut] (pp2) {};
\draw[thick,directed]
     (pp1.south) to [out=270,in=180] node [left] {$A$}
     (cut.west);
\draw[thick,directed]
     (pp2.south) to [out=270,in=0] node [right] {$A^\perp$}
     (cut.east);
\end{tikzpicture}
\end{center}

%%% Local Variables: 
%%% mode: latex
%%% TeX-master: "LLlevSem"
%%% End: 

%% file: Models.tex
Intuitively, a denotational semantics of a logical system is an interpretation $\sem\cdot$ of proofs as some kind of mathematical objects which:
\begin{enumerate}
	\item[i.] is stable under cut-elimination: if $\pi\betared\pi'$, then $\sem\pi=\sem{\pi'}$;
	\item[ii.] gives rise to a congruence: if $\sem{\pi_1}=\sem{\pi_2}$ and if $\pi_1',\pi_2'$ are obtained resp.\ from $\pi_1,\pi_2$ by applying the same inference rule, then $\sem{\pi_1'}=\sem{\pi_2'}$.
\end{enumerate}
A denotational semantics is \emph{non-trivial} if there exist two distinct proofs of the same formula $\pi_1,\pi_2$ such that $\sem{\pi_1}\neq\sem{\pi_2}$; usually, this is implicitly assumed to be the case.

We may formalize the above idea using higher-order graphs and categories (see \refapp{CatBack} for a brief introduction to $2$-categories; a \emph{$2$-graph} is basically a $2$-category without any notion of composition or identity). In fact, the syntax may be seen as a $2$-graph $\mathcal S$: the $0$-cells are formulas, the $1$-cells are proofs, and the $2$-cells are cut-elimination steps. Then, the interpretation map $\sem{\cdot}$ described above is nothing but a morphism of $2$-graphs from $\mathcal S$ to a category $\catA$, seen as a degenerate $2$-category (\ie, in which the only $2$-cells are identities), seen in turn as a $2$-graph. This guarantees that requirement (i) above is met. For what concerns requirement (ii), one may suppose that inference rules are interpreted by functors or natural transformations, whereas axioms ar interpreted by identity arrows and the cut rule by composition.

% Category theory is an elegant and powerful framework in which the above intuition can be formalized. The simplest approach is to interpret formulas and proofs as objects and morphisms of a category: given an arbitrary assignment of objects to propositional atoms (such an assignment will be a parameter of the interpretation), axioms are interpreted by identity morphisms, the cut rule by composition, and the other inference rules by functors or natural transformations; the above requirements are thus met automatically (moreover, one also automatically models $\eta$-expansion).

A \emph{categorical model} of a logical system is a category having enough structure so that a denotational semantics, as defined above, immediately arises from it. One of the most well-known examples is probably that of propositional\footnote{While propositional logical systems admit in many cases satisfactory categorical axiomatizations, there is currently no general, widely accepted categorical formulation of a second order denotational model. This is why, in this paper, we drop second-order quantifiers whenever we deal with denotational semantics. This is consistent with the fact that we only defined propositional proof nets.} intuitionistic logic, whose models correspond to Cartesian closed categories. In the case of propositional linear logic, several authors have proposed their notion of categorical model; let us mention for instance~\cite{Lafont:PhDThesis,Seely,Bierman,BentonEtAl,Benton,deCarvalho:PhDThesis}, and~\cite{Mellies:Models,CurienHerbelinKrivineMellies} for surveys. Most of these axiomatizations are equivalent, and can be very elegantly synthesized by the notion of \emph{linear-non-linear adjunction} (\refdef{LNL}).

Giving the full categorical background needed to formulate linear-non-linear adjunctions is out of the scope of this paper; for the sake of self-containedness, \refapp{CatBack} recalls the basic definitions and properties, assuming as understood only the concept of symmetric monoidal category. For a more detailed exposition, we refer the reader to Paul-Andr\'e Melli\`es excellent survey to be found in \cite{CurienHerbelinKrivineMellies}. Here, we content ourselves with fixing some (standard) notations. If $\catA$ is a category and $A,A'$ objects of $\catA$, we denote as usual by $\catA[A,A']$ the class of morphisms of $\catA$ of source $A$ and target $A'$, and we write $f:A\rightarrow A'$ to mean that $f\in\catA[A,A']$. The identity arrow of $A$ is denoted as usual by $\id_A$ or, when there is no ambiguity, by $A$ itself. Applying this notation to the category of categories, we write $F:\catA\rightarrow\catB$ to say that $F$ is a functor from the category $\catA$ to the category $\catB$, whereas the identity functor on $\catA$ is denoted by $\Id_\catA$ or $\catA$ itself. The image of a functor $F(-)$ may be abbreviated by $F-$. Composition, usually denoted by $f\circ g$ with $f:B\rightarrow C$ and $g:A\rightarrow B$, may be abbreviated by $fg$. We use the notation $$\varphi:F\Rightarrow G:\catA\rightarrow\catB$$
to say that $\varphi$ is a natural transformation between the functors $F,G:\catA\rightarrow\catB$. So, for instance, a (lax) symmetric monoidal functor between two symmetric monoidal categories $(\catA,\otimes,1)$ and $(\catB,\odot,I)$ is a triple $(F,\mu,\eta)$ where $F:\catA\rightarrow\catB$ and
\begin{align*}
	\mu &: F(-)\odot F(-)\Rightarrow F(-\otimes -) : \catA\times\catA\rightarrow\catB, \\
	\eta &: K_I \Rightarrow K_{F1} : \catA\rightarrow\catB,
\end{align*}
such that $\mu$ and $\nu$ satisfy certain conditions (spelled out in \refapp{CatBack}). Here, $K_C$ denotes the constant functor mapping all objects on a fixed object $C$ and all morphisms on $\id_C$. Observe that, in this case, the natural transformation $\eta$ is degenerate, \ie, it is reduced to a single morphism of $\catB[I,F1]$.

% In what follows, we say that a monoidal structure $(\catA,\times,\term)$ is \emph{Cartesian} if, for all objects $A,B$ of $\catA$, $A\times B$ is a product of $A$ and $B$, and $\term$ is a terminal object; in other words, $\catA$ is actually a Cartesian category, and the monoidal structure gives a choice for product objects and the terminal object. Observe that such a monoidal structure is necessarily symmetric.
\begin{defn}[Linear-non-linear adjuction]
	\label{def:LNL}
	\sloppy{A \emph{linear-non-linear adjunction} is a symmetric monoidal adjunction between symmetric monoidal functors}
	\begin{displaymath}
		\xymatrix{
			(\catM,\times,\term) \ar@/^1pc/[rr]^{(L,l,t)} & \perp & (\catL,\ltens,\lone) \ar@/^1pc/[ll]^{(M,m,u)}
		}
	\end{displaymath}
	such that $(\catM,\times,\term)$ is Cartesian (with $T$ being the terminal object).
\end{defn}
% \begin{defn}[Linear category~\cite{BentonEtAl}]
% 	\label{def:LLModel}
% 	A \emph{model} of linear logic is a category $\bC$ with the following structure:
% 	\begin{itemize}
% 		\item a $\ast$-autonomous structure $(\otimes,1,(-)^\perp)$;
% 		\item a lax monoidal comonad $(\oc(-),dig,der)$;
% 		\item two natural transformations
% 		$$c:\oc(-)\Rightarrow\oc(-)\otimes\oc(-)\qquad\qquad w:\oc(-)\Rightarrow\mathcal K_1$$
% 		(where $\mathcal K_1$ is the constant functor yielding $1$ for all objects and $id_1$ for all arrows of $\bC$), such that, for every object $A$ of $\bC$, $(\oc A,c_A,w_A)$ is a commutative comonoid.
% 	\end{itemize}
% 	Moreover, the following conditions must be satified:
% 	\begin{itemize}
% 		\item for every $A$, the morphisms $c_A$ and $w_A$ are coalgebra morphisms;
% 		\item for every $A$, the morphism $dig_A$ is a comonoid morphism.
% 	\end{itemize}
% \end{defn}

% For modeling linear logic, we need one further ingredient, namely the notion of $\ast$-autonomous category \citep{Barr}, which we recall here:
\begin{defn}[$\ast$-autonomous category]
	\label{def:StarAut}
	A \emph{symmetric monoidal closed category} is a symmetric monoidal category $(\catA,\otimes,\lone)$ such that, for every object $X$ of $\catA$, the functor $-\otimes X$ has a right adjoint $R_X$. This latter induces a bifunctor $\llinimp:\catA^{\mathrm{op}}\times\catA\rightarrow\catA$, such that $R_X=X\llinimp -$, and the adjunction corresponds to the existence of a natural isomorphism
	$$\Psi:[-\otimes -,-]\Rightarrow[-,-\llinimp -]:(\catA^{\mathrm{op}}\times\catA^{\mathrm{op}})\times\catA\rightarrow\Set.$$
	Given two objects $A,B$ of $\catA$, it is customary to denote by $\eval_{A,B}$ the morphism $\Psi^{-1}_{A\llinimp B,A,B}(\id_{A\llinimp B}):(A\llinimp B)\otimes A\rightarrow B$.

	A \emph{$\ast$-autonomous category} \citep{Barr} is a symmetric monoidal closed category $(\catA,\otimes,\lone)$ together with a \emph{dualizing object}, \ie, an object $\bot$ of $\catA$ such that, for every object $A$, $$\Psi_{A,A\llinimp\bot,\bot}(\eval_{A,\bot}\circ\gamma_{A,A\llinimp\bot}):A\rightarrow(A\llinimp\bot)\llinimp\bot$$
	is an isomorphism, where $\gamma$ is the symmetry of $\otimes$. In that case, we denote by $(-)^\bot$ the contravariant endofunctor $-\llinimp\bot$.
\end{defn}

% \begin{definition}[$\ast$-autonomous category]
% 	\label{def:StarAut}
% 	\sloppy{A $\ast$-autonomous category is a symmetric monoidal category $(\catA,\ltens,\lone)$ together with a functor $(-)^\bot:\catA^{\mathrm{op}}\rightarrow\catA$ such that the following natural isomorphisms hold:}
% 	\begin{itemize}
% 		\item $(-)^{\bot\bot}\cong Id_{\catA}$;
% 		\item $\catA[-\ltens B,(C)^\bot]\cong\catA[-,(B\ltens C)^\bot]$, for all objects $B,C$ of $\catA$;
% 		\item $\catA[A\ltens-,(C)^\bot]\cong\catA[A,(-\ltens C)^\bot]$, for all objects $A,C$ of $\catA$;
% 		\item $\catA[A\ltens B,(-)^\bot]\cong\catA[A,(B\ltens -)^\bot]$, for all objects $A,B$ of $\catA$.
% 	\end{itemize}
% \end{definition}
% We remark that the above four conditions together imply that $\ltens$ has a right adjoint, namely \mbox{$(-\ltens(-)^\bot)^\bot$}, so a $\ast$-autonomous category is in particular a \emph{closed} monoidal category. In fact, $\ast$-autonomous categories may be alternatively presented as closed symmetric monoidal categories with a \emph{dualizing object}, which corresponds to $\lone^\bot$ in the above formulation.

\begin{defn}[Model of linear logic]
	\label{def:Model}
	A \emph{model} of linear logic is given by the following data:
	\begin{itemize}
		\item a $\ast$-autonomous category $(\catL,\ltens,\lone,\lbot)$ which has also binary products and a terminal object, denoted by $A\lwith B$ and $\ltop$, respectively;
		\item a Cartesian category $(\catM,\times,\term)$;
		\item a linear-non-linear adjunction $(L,l,t)\dashv(M,m,u)$ between $(\catM,\times,\term)$ and $(\catL,\ltens,\lone)$.
	\end{itemize}
	The category $\catL$ is said to be the \emph{linear} category.
\end{defn}

The linear category is where the model ``lives''; formulas and proofs are interpreted in it. The motivation behind the structures required by \refdef{Model} may be spelled out as follows:
\begin{itemize}
	\item The $\ast$-autonomous structure is used to interpret the multiplicative layer of linear logic: tensor, linear negation, and par are interpreted by the functors $\otimes$, $(-)^\bot$, and $(-)^\bot\llinimp-$, respectively.
	\item The additive connectives are interpreted by products and coproducts in $\catL$. Indeed, since $\catL$ has products, for all objects $A,B$, the object $(A^\bot\lwith B^\bot)^\bot$ may be seen to be a coproduct of $A$ and $B$, and $\ltop^\bot$ may be seen to be an initial object. We note in passing that right adjoints are \emph{continuous}, \ie, they preserve limits; this implies that $M$ transports products in $\catL$ to products in $\catM$, \ie, $M(A\lwith B)\cong MA\times MB$ and $M\ltop\cong\term$.
	\item The linear-non-linear adjunction offers the structure needed to interpret the exponential layer:
	\begin{itemize}
		\item since $L\dashv M$, we know that the endofunctor $\oc=L\circ M$ of $\catL$ is a comonad: the promotion rule (\ie, boxes in proof nets) is interpreted by $\oc(-)$ and its comultiplication; derelection is interpreted by the counit;
		\item the structural rules are interpreted thanks to the Cartesian structure of $\catM$, using \refprop{SymMonAdj}. In fact, from the diagonal maps and the terminal arrows of $\catM$, we can define two natural transformations $\kappa:M\Rightarrow M\times M$ and $\omega:M\Rightarrow K_\term$, where by $K_\term$ we denote here the constant functor yielding $\term$ and its identity arrow. Now, the fact that $(L,l,t)$ is strong means that $l$ and $t$ are isomorphisms; contraction and weakening are then interpreted by applying $L$ to $\kappa$ and $\omega$, and by composing with $l^{-1}$ and $t^{-1}$, respectively:
		\begin{displaymath}
			\begin{array}{rl}
				\textrm{contraction:} & \xymatrix{\oc(-) \ar@{=>}[r]^-{L\kappa} & L(M(-)\times M(-)) \ar@{=>}[r]^-{l^{-1}} & \oc(-)\ltens\oc(-)} \\
				\textrm{weakening:} & \xymatrix{\oc(-) \ar@{=>}[r]^-{L\omega} & LK_\term \ar[r]^-{t^{-1}} & \lone}
			\end{array}
		\end{displaymath}
	\end{itemize}
\end{itemize}

The terminology \emph{linear} is justified by the fact that $\catL$ is a sort of ``linear world'', where structural rules are forbidden; the functor $M$ allows to transport proofs into a ``Cartesian world'', where duplication and erasing are available; then, the functor $L$ ``linearizes'' the proofs back to the linear world. The so-called exponential isomorphisms of linear logic, \ie, $\oc(A\lwith B)\cong\oc A\ltens\oc B$ and $\oc\ltop\cong\lone$, already express this idea; the linear-non-linear adjunction refines it by implementing it as a ``return trip'' from the linear world to itself, passing through a non-linear stage.

%% file: StratModels.tex
It is fairly straightforward to define a notion of model for stratified linear logic. Essentially, all that is needed is to define how the paragraph modality is interpreted; the remaining connectives will be handled by the same structures as \refdef{Model}.

\begin{defn}[Model of stratified linear logic]
	\label{def:StratModel}
	A \emph{model} of \stratLL\ is given by the same data as a model of linear logic, plus an endofunctor $\parg$ of the linear category $(\catL,\otimes,\lone,\bot)$ which commutes to all logical operations, \ie:
	\begin{itemize}
		\item $\parg\bot\cong\bot$;
		\item $\parg\circ\otimes$ is naturally isomorphic to $\otimes\circ(\parg\times\parg)$;
		\item $\parg\circ\llinimp$ is naturally isomorphic to $\llinimp\circ(\parg^{\mathrm{op}}\times\parg)$;
		\item $\parg$ preserves all finite products;
		\item $\parg\circ\oc$ is naturally isomorphic to $\oc\circ\parg$.
	\end{itemize}
	A model of \stratLL\ is said to be \emph{degenerate} when $\parg$ is isomorphic to the identity functor of $\catL$ (which obviously verifies all of the above conditions).
\end{defn}

Given a category with the structure of \refdef{StratModel}, it is immediate to interpet the proofs of propositional \stratLL, provided we use the alternative formulation of \reftab{AltSeqCalc}, modulo \refprop{Alt}. The interpretation (which is always parametric in an assignment of objects to propositional atoms) is defined as usual by induction: axioms and cuts are interpreted by identities and composition; the structural rules, and the positive logical rules are identical to those of linear logic, so the interpretation is already defined; for the negative rules and the exponential rules, the standard interpretation of linear logic is composed with the suitable isomorphisms given by \refdef{StratModel}; finally, the paragraph rule is interpreted by applying the $\parg$ functor.

%% file: Obj.tex
We now introduce the categorical construction at the center of our work. This is based on the notion of \emph{\obj}, which we define as an object (of some category) equipped with denumerably many involutions, indexed by \Int. Given a category $\catA$ and an object $A$ of $\catA$, an \emph{involution} of $A$ is an automorphism $s$ of $A$ which is its own inverse, \ie, such that $s\circ s=\id_A$.
\begin{defn}[\Obj]
	\label{def:Obj}
	Let $\catA$ be a category. An \emph{\obj} of $\catA$ is a pair $(A,s)$ where $A$ is an object of $\catA$ and $s=(s_k)_{k\in\Int}$ is a $\Int$-indexed sequence of involutions of $A$.

	If $(A,s),(B,t)$ are \objs\ of $\catA$, a \emph{morphism} of \objs\ from the first to the second is an arrow $f:A\rightarrow B$ of $\catA$ such that, for all $k\in\Int$, the following diagram commutes:
\begin{center}
\begin{tikzpicture}
\matrix[row sep=10mm, column sep=9mm] {
 \node (A)   {$A$}; &
 \node (D)   {$B$}; \\
 \node (B)   {$A$}; & 
 \node (C)   {$B$}; \\
};
\path [->]
      (A) edge node [left]  {\ednd{s_k}} (B)
      (B) edge node [below] {\ednd{f}} (C)
      (A) edge node [above] {\ednd{f}} (D)
      (D) edge node [right] {\ednd{t_k}} (C);
\end{tikzpicture}
\end{center}
	or, equivalently, $t_k\circ f\circ s_k = f$ for all $k\in\Int$.
\end{defn}

We remark that we may restrict the above definition so that, in an \obj\ $(A,s)$, the sequence $s$ is required to be equal to $\id_A$ almost everywhere. This restriction, which limits the cardinality of the set of all \objs, is preserved by all constructions we shall consider, but is by no means necessary.

The composition of two morphisms of \objs\ is easily seen to be a morphism of \objs, and identity arrows are also easily seen to be morphisms of \objs. Hence, the \objs\ of a category $\catA$ and their morphisms form themselves a category, denoted by $\Inv\catA$.

The reason behind such a notation is the following. Consider the free Coxeter group on $\Int$, which we denote by $\Cox$: the elements of $\Cox$ are finite sequences of integers not containing two consecutive equal elements; its group law, denoted by $\circ$, is defined by concatenation modulo the equation $n\circ n=\id$ for all $n\in\Int$, where $\id$ is the empty sequence. Now, $\Cox$ can be seen as a groupoid with only one object and whose morphisms are exactly the elements of $\Cox$ itself; then, we invite the reader to check that the category of \objs\ of a category $\catA$ is exactly $\catA^\Cox$, that is, the category of functors from $\Cox$ to $\catA$, with natural transformations as morphisms.

Therefore, $\Inv\catA$ is nothing but the image of $\catA$ through the object map of the endofunctor $\Inv{(-)}$ of \Cat, the right adjoint of the product functor $-\times\Cox$ in the category of categories. Actually, $\Inv{(-)}$ is a 2-endofunctor, \ie, it also acts on natural transformations. We may explicitly describe it as follows.

Let $F:\catA\rightarrow\catB$ be a functor. We define a functor $\Inv F:\Inv\catA\rightarrow\Inv\catB$ as follows: if $(A,s)$ is an object of $\Inv\catA$, we set $\Inv F(A,s)=(FA,Fs)$, where by $Fs$ we mean the sequence $(F(s_k))_{k\in\Int})$; if $f:(A,s)\rightarrow(B,t)$ is an arrow of $\Inv\catA$, we set $\Inv Ff=Ff$. Functors preserve involutions, so $(FA,Fs)$ is an \obj; moreover, given $k\in\Int$, we have $F(t_k)\circ Ff\circ F(s_k)=F(t_k\circ f\circ s_k)=Ff$, which proves that $Ff$ is a morphism between the \objs\ $\Inv F(A,s)$ and $\Inv F(B,t)$; the fact that $\Inv F$ preserves composition and identities follows immediately from the fact that $F$ does.

Let now $\varphi:F\Rightarrow G:\catA\rightarrow\catB$ be a natural transformation. First of all, we check that, for each object $(A,s)$ of $\Inv\catA$, the arrow $\varphi_A:FA\rightarrow GA$ of $\cB$ is actually a morphism between the \objs\ $\Inv F(A,s)$ and $\Inv G(A,s)$, \ie, an arrow of $\Inv\catB$. As a matter of fact, given any $k\in\Int$, we have $\varphi_A\circ F(s_k)=G(s_k)\circ\varphi_A$, simply because $\varphi$ is a natural transformation from $F$ to $G$. Therefore, if we set $\Inv\varphi_{(A,s)}=\varphi_A$, we have defined a family of arrows of $\Inv\catB$, whose naturality is an immediate consequence of the naturality of $\varphi$. We have thus obtained a natural transformation $\Inv\varphi:\Inv F\Rightarrow\Inv G:\Inv\catA\rightarrow\Inv\catB$, as desired.

The following technical result will be essential in the sequel. It allows, given a natural transformation $\varphi:F\Rightarrow G:\catA\rightarrow\catB$, to ``pull back'' the natural transformation $\Inv\varphi:\Inv F\Rightarrow\Inv G:\Inv\catA\rightarrow\Inv\catB$ along any functor $H:\catC\rightarrow\Inv\catA$. 
\begin{lem}
	\label{lemma:NatTrans}
	Let $\varphi:F\Rightarrow G:\catA\rightarrow\catB$ be a natural transformation, let $\catC$ be another category and $H:\catC\rightarrow\Inv\catA$ a functor. Let $C$ be an object of $\catC$, and set $\varphi'_C=\varphi_{UH(C)}$, where $U:\Inv\catA\rightarrow\catA$ is the forgetful functor sending an \obj\ $(A,s)$ to $A$ and being the identity on morphisms. Then, $\varphi':\Inv F\circ H\Rightarrow\Inv G\circ H:\catC\rightarrow\Inv\catB$.% is a natural transformation from the functor $\Inv F\circ H$ to the functor $\Inv G\circ H$ (which are both of source $\catC$ and target $\Inv\catB$).
\end{lem}
\begin{pf}
	We first check that, given a generic object $C$ of $\catC$, $\varphi_{UH(C)}$ is a morphism of $\Inv\catB$ from $\Inv F H(C)$ to $\Inv G H(C)$. Let $H(C)=(A,s)$; then $UH(C)=A$, $\Inv F H(C)=(FA,Fs)$, and $\Inv G H(C)=(GA,Gs)$. Then, it is enough to verify, for all $k\in\Int$, the commutation in $\catB$ of
\begin{center}
\begin{tikzpicture}
\matrix[row sep=10mm, column sep=12mm] {
 \node (A)   {$FA$}; &
 \node (D)   {$GA$}; \\
 \node (B)   {$FA$}; & 
 \node (C)   {$GA$}; \\
};
\path [->]
      (A) edge node [left]  {\ednd{Fs_k}} (B)
      (B) edge node [below] {\ednd{\varphi_A}} (C)
      (A) edge node [above] {\ednd{\varphi_A}} (D)
      (D) edge node [right] {\ednd{Gs_k}} (C);
\end{tikzpicture}
\end{center}
	which holds by naturality of $\varphi$.

	Let us now check the naturality of $\varphi'$. For this, let $f:C\rightarrow C'$ be an arrow of $\catC$; we need to verify the commutation in $\Inv\catB$ of the diagram
\begin{center}
\begin{tikzpicture}
\matrix[row sep=10mm, column sep=14mm] {
\node (A)               {$\Inv F H(C)$};  &
\node (D)               {$\Inv G H(C)$};  \\
\node (B)               {$\Inv F H(C')$}; &
\node (C)               {$\Inv G H(C')$}; \\
};
\path [->]
      (A) edge node [left]  {\ednd{\Inv F H(f)}} (B)
      (B) edge node [below] {\ednd{\varphi'_C}} (C)
      (A) edge node [above] {\ednd{\varphi'_C}} (D)
      (D) edge node [right] {\ednd{\Inv G H(f)}} (C);
\end{tikzpicture}
\end{center}
	But if we set $H(C)=(A,s)$ and $H(C')=(A',s')$, the diagram becomes
\begin{center}
\begin{tikzpicture}
\matrix[row sep=10mm, column sep=14mm] {
 \node (A)   {$(FA,Fs)$};   &
 \node (D)   {$(GA,Gs)$}; \\
 \node (B)   {$(FA',Fs')$};  &
 \node (C)   {$(GA',Gs')$};   \\
};
\path [->]
      (A) edge node [left]  {\ednd{\varphi_A}} (B)
      (B) edge node [below] {\ednd{\varphi_A}} (C)
      (A) edge node [above] {\ednd{FH(f)}} (D)
      (D) edge node [right] {\ednd{GH(f)}} (C);
\end{tikzpicture}
\end{center}
	which commutes because $H(f)$, as an arrow of $\Inv\catA$, is actually an arrow of $\catA$ from $A$ to $A'$, $\varphi$ is natural from $F$ to $G$, and commutation in $\catB$ implies commutation in $\Inv\catB$, because the arrows and composition of this latter category are arrows and composition of the former.\qed
\end{pf}

With the help of \reflemma{NatTrans}, we now lift $\Inv{(-)}$ to a 2-endofunctor of \SymMonCat, the 2-category of symmetric monoidal categories, symmetric monoidal functors and monoidal natural transformations. We shall denote such a 2-endofunctor by $\SInv$.

First of all, observe that, if we fix a category $\catC$, the covariant endofunctor $(-)^\catC$ of $(\Cat,\times,\catOne)$ is strong symmetric monoidal, where $\times$ is the product of categories and $\catOne$ the terminal category (having one object and its identity morphism only). Indeed, for any categories $\catA,\catB$, we have natural isomorphisms $\Phi^\catC_{\catA,\catB}:\catA^\catC\times\catB^\catC\rightarrow (\catA\times\catB)^\catC$ and $\Phi^\catC_0:\catOne\rightarrow\catOne^\catC$, which make $((-)^\catC,\Phi^\catC,\Phi^\catC_0)$ strong symmetric monoidal. In case $\catC=\Cox$, we drop the superscript from notations, and write simply $\Phi$ and $\Phi_0$; additionally, we denote by $\Phi_\catA$ the (invertible) functor $\Phi_{\catA,\catA}$.

We start by defining $\SInv$ on objects. Let $(\catA,\otimes,\lone)$ be a symmetric monoidal category. By definition, $\otimes$ is a bifunctor of $\catA$, and $\lone$ an object of $\catA$, which may be seen as a functor (which we still denote by $\lone$) from the terminal category $\catOne$ to $\catA$. If we apply $\Inv{(-)}$ to these functors, we obtain a functor $\Inv\otimes:\Inv{(\catA\times\catA)}\rightarrow\Inv\catA$ and a functor $\Inv\lone:\Inv\catOne\rightarrow\Inv\catA$. Then, we set $\SInv\catA=(\Inv\catA,\otimes',\lone')$, where
\begin{eqnarray*}
	\otimes' &=& \Inv\otimes\circ\Phi_\catA:\Inv\catA\times\Inv\catA\rightarrow\Inv\catA \\
	\lone' &=& \Inv\lone\circ\Phi_0:\catOne\rightarrow\Inv\catA.
\end{eqnarray*}
As a functor from the terminal category to $\Inv\catA$, $\lone'$ actually represents an object of $\Inv\catA$, which is readily seen to be $(\lone,\id_\lone)$, with $\id_\lone$ representing the constant $\Int$-indexed sequence everywhere equal to the identity arrow of $1$.

We claim that $\SInv\catA$ is a symmetric monoidal category. To show this, it will be enough to apply \reflemma{NatTrans} to the natural isomorphisms coming with the monoidal structure of $\catA$.

\smallskip
\noindent\textbf{Associator:} If we denote by $A_\catA$ the invertible functor realizing the isomorphism between $(\catA\times\catA)\times\catA$ and $\catA\times(\catA\times\catA)$ (\ie, the associator of the monoidal category $(\Cat,\times,\catOne)$ instantiated on $\catA,\catA,\catA$), we see that the associator $\alpha$ of $\catA$ is a natural isomorphism of the following type:
\begin{center}
  \begin{tikzpicture}[node distance = 8mm and 6mm]
  \node (A)                           {$(\catA\times\catA)\times\catA$};
  \node (B) [below right=of A]        {$\catA\times(\catA\times\catA)$};
  \node (notC) [right=of B]           {};
  \node (C) [right=of notC]           {$\catA\times\catA$};
  \node (E) [above right=of C]        {$\catA$};
  \node (notD) [above=of notC]        {};
  \node (D) [above =of notD]          {$\catA\times\catA$};
\path [thick,->, bend left=10]
          (A) edge node [above left]    {\ednd{\otimes\times\catA}} (D)
          (D) edge node [above right]   {\ednd{\otimes}} (E);
\path [thick,->, bend right=10]
          (A) edge node [below left]    {\ednd{A_\catA}} (B)
          (C) edge node [below right]   {\ednd{\otimes}} (E);
\path [thick,->]
          (B) edge node [below]    {\ednd{\catA\times\otimes}} (C);
\draw [double,->, >=latex',double distance=2pt] 
          (D) -- node [right]        {$\alpha$} (notC);
\end{tikzpicture}
\end{center}
where by $\catA$ we also denote the identity functor of $\catA$. Then, if we set
\begin{eqnarray*}
	F &=& \otimes\circ(\otimes\times\catA), \\
	G &=& \otimes\circ(\catA\times\otimes)\circ A_\catA, \\
	H &=& \Phi_\catA\circ(\Phi_\catA\times\Inv\catA),
\end{eqnarray*}
we may apply \reflemma{NatTrans} and obtain a natural transformation $$\alpha':\Inv F H\Rightarrow\Inv G H:(\Inv\catA\times\Inv\catA)\times\Inv\catA\rightarrow\Inv\catA,$$
which may be seen to have the right type for being the associator of $\SInv\catA$ by looking at the following diagram:

\begin{center}
\begin{tikzpicture}[node distance = 4mm and 1.7mm]
  \node (A)                           {$\Inv\catA\times\Inv\catA$};
  \node (C) [below=of A]              {$\Inv{((\catA\times\catA)\times\catA)}$};
  \node (B) [below left=of C]         {$\Inv{(\catA\times\catA)}\times\Inv\catA$};
  \node (D) [below right=of C]        {$\Inv{(\catA\times\catA)}$};
  \node (notE1) [below=of B] {};
  \node (notE2) [below=of notE1] {};
  \node (notE3) [below=of notE2] {};
  \node (E) [below left=of notE1]     {$(\Inv\catA\times\Inv\catA)\times\Inv\catA$};
  \node (G) [below =of notE3]         {$\Inv{(\catA\times\catA)}\times\Inv\catA$};
  \node (H) [below right=of G]        {$\Inv{((\catA\times\catA)\times\catA)}$};
  \node (I) [above right=of H]        {$\Inv{(\catA\times(\catA\times\catA))}$};
  \node (J) [above right=of I]        {$\Inv{(\catA\times\catA)}$};
  \node (F) [above=of J]              {$\Inv\catA$};
  \node (notK) [below=of H] {};
  \node (K) [below=of notK]           {$\Inv\catA\times(\Inv\catA\times\Inv\catA)$};
  \node (notL) [below=of I] {};
  \node (L) [below=of notL]           {$\Inv\catA\times\Inv{(\catA\times\catA)}$};
  \node (M) [below=of J]              {$\Inv\catA\times\Inv\catA$};

\path [->,bend left=20]
         (B) edge node [left]           {\ednd{\Inv\otimes\times\Inv\catA}} (A)
         (A) edge node [above right]    {\ednd{\Phi_\catA}} (D);
\path [thick,->, bend left=10] 
          (E) edge node [above left]    {\ednd{\Phi_\catA\times\Inv\catA}} (B)
          (B) edge node [below right]   {\ednd{\Phi_\catA}} (C)
          (C) edge node [below left]    {\ednd{\Inv{(\otimes\catA)}}} (D)
          (D) edge node [above right]   {\ednd{\Inv\otimes}} (F);
\path [thick,->, bend right=10] 
          (E) edge node [above right]   {\ednd{\Phi_\catA\times\Inv\catA}} (G)
          (G) edge node [above right]   {\ednd{\Phi_\catA}} (H)
          (H) edge node [above left]    {\ednd{A_\catA^{\mathcal I}}} (I)
          (I) edge node [above left]    {\ednd{\Inv{(\catA\times\otimes)}}} (J)
          (J) edge node [right]         {\ednd{\Inv\otimes}} (F);
\path [->, bend right=20]
          (E) edge [bend right=40] node [below left] {\ednd{A_{\Inv\catA}}} (K)
          (K) edge node [below right]   {\ednd{\Inv\catA\times\Phi_\catA}} (L)
          (L) edge node [below right]   {\ednd{\Inv\catA\times\Inv\otimes}} (M);
\path [->]
          (L) edge node [left]          {\ednd{\Phi_\catA}} (I)
          (M) edge node [right]         {\ednd{\Phi_\catA}} (J);
\draw [double,->, >=latex',double distance=2pt] 
          (C) -- node [right]        {$\alpha'$} (H);
\path
          (A) edge [draw=white] node [inner sep=1pt, draw=black,shape=circle] {$1$} (C)
          (H) edge [draw=white] node [inner sep=1pt, draw=black,shape=circle,left, xshift=-1cm] {$2$} (K)
          (I) edge [draw=white] node [inner sep=1pt, draw=black,shape=circle, below, yshift=-0.2cm] {$3$} (M);
\end{tikzpicture}
\end{center}

Indeed, diagrams (1) and (3) commute because of the naturality of $\Phi$, and diagram (2) commutes because $\Inv{(-)}$ is monoidal (remember that $A$ is the associator of $(\Cat,\times,\catOne)$). The fact that $\alpha'$ is an isomorphism follows immediately from its definition (\cf\ \reflemma{NatTrans}), because $\alpha$ is an isomorphism and the arrows of $\Inv\catA$ are arrows of $\catA$.

\smallskip
\noindent\textbf{Unitors:} We proceed as above, checking only the case of the left unitor, the right unitor being completely analogous. If we denote by $\Lambda_\catA$ the invertible functor realizing the isomorphism from $\catOne\times\catA$ to $\catA$ (\ie, the left unitor of $(\Cat,\times,\catOne)$ instantiated on $\catA$), we see that the left unitor $\lambda$ of $\catA$ is of the following type:
\begin{center}
  \begin{tikzpicture}[node distance = 8mm and 6mm]
  \node (A)                           {$\catOne\times\catA$};
  \node (notB) [below right=of A]     {};
  \node (B) [above right=of A]          {$\catA\times\catA$};
  \node (C) [below right=of B,xshift=2.8mm]     {$\catA$};
\path [thick,->, bend left=10]
          (A) edge node [above left]    {\ednd{\lone\times\catA}} (B)
          (B) edge node [above right]   {\ednd{\otimes}} (C);
\path [thick,->, bend right=60]
          (A) edge [draw=none] node [name=bb]    {} (C)
          (A) edge node [below]    {\ednd{\Lambda_\catA}} (C);
\draw [double,->, >=latex',double distance=2pt] 
          (B) -- node [right]        {$\lambda$} (bb);
\end{tikzpicture}
\end{center}
Then, if we apply \reflemma{NatTrans} to the data
\begin{eqnarray*}
	F &=& \otimes\circ(\lone\times\catA), \\
	G &=& \Lambda_\catA, \\
	H &=& \Phi_\catA\circ(\Phi_0\times\Inv\catA),
\end{eqnarray*}
we obtain the left unitor $\lambda'$ of $\SInv\catA$, modulo the commutations shown in the following diagram
\begin{center}
  \begin{tikzpicture}[node distance = 6mm and 6mm]
  \node (A)                           {$\Inv\catA\times\Inv\catA$};
  \node (C) [below =of A]             {$\Inv{(\catOne\times\catA)}$};
  \node (B) [below left=of C,yshift=3mm]         {$\Inv\catOne\times\Inv\catA$};
  \node (D) [below right=of C,yshift=3mm]        {$\Inv{(\catA\times\catA)}$};
  \node (E) [below left=of B]         {$\catOne\times\Inv\catA$};
  \node (F) [below right=of D]        {$\Inv\catA$};
  \node (G) [below right=of E]        {$\Inv\catOne\times\Inv\catA$};
  \node (H) [right=of G,yshift=-6mm]   {$\Inv{(\catOne\times\catA)}$};
\path [thick,->, bend left=5]
          (E) edge node [above left]    {\ednd{\Phi_0\times\Inv\catA}} (B)
          (B) edge node [below right]   {\ednd{\Phi_\catA}} (C)
          (C) edge node [below left]    {\ednd{\Inv{(\lone\times\catA)}}} (D)
          (D) edge node [above right]   {\ednd{\Inv\otimes}} (F);
\path [thick,->, bend right=10]
          (E) edge node [above right]   {\ednd{\Phi_0\times\Inv\catA}} (G)
          (G) edge node [above]   {\ednd{\Phi_\catA}} (H)
          (H) edge node [above left]   {\ednd{\Lambda_\catA^{\mathcal I}}} (F);
\path [->, bend right=70]
          (E) edge node [below,name=lambda]         {\ednd{\Lambda_{\Inv\catA}}} (F);
\path [->,bend left=30]
          (B) edge node [above left]    {\ednd{\Inv\lone\times\Inv\catA}} (A)
          (A) edge node [above right]   {\ednd{\Phi_\catA}} (D);
\draw [double,->, >=latex',double distance=2pt] 
          (C) -- node [right]        {$\lambda'$} (H);
\path[draw=white] 
          (A) edge[draw=none] node [inner sep=1pt, draw=black,shape=circle] {$1$} (C)
          (H) edge[draw=none] node [inner sep=1pt, draw=black,shape=circle,yshift=1mm] {$2$} (lambda);
\end{tikzpicture}
\end{center}
Indeed, (1) holds because of the naturality of $\Phi$, and (2) because $\Inv{(-)}$ is monoidal.

\smallskip
\noindent\textbf{Symmetry:} Again, if we denote by $\Gamma_\catA$ the endofunctor of $\catA\times\catA$ which ``swaps'' the two occurrences of $\catA$ (\ie, the involutive braiding of $(\Cat,\times,\lone)$ seen as a symmetric monoidal category, instantiated on $\catA,\catA$), we see that the involutive braiding $\gamma$ of $\catA$ is of the following type:
\begin{center}
  \begin{tikzpicture}[node distance = 6mm and 6mm]
  \node (A)                           {$\catA\times\catA$};
  \node (B) [below right=of A]        {$\catA\times\catA$};
  \node (C) [above right=of B,xshift=3.3mm]        {$\catA$};
\path [thick,->, bend left=60]
          (A) edge node [above]    {\ednd{\otimes}} (C)
          (A) edge[draw=none] node [name=haut]    {} (C);
\path [thick,->, bend right=10]
          (A) edge node [below left]    {\ednd{\Gamma_\catA}} (B)
          (B) edge node [below right]   {\ednd{\otimes}} (C);
\draw [double,->, >=latex',double distance=2pt] 
          (haut) -- node [right]        {$\gamma'$} (B);
\end{tikzpicture}
\end{center}

Then, once more, we obtain the involutive braiding $\gamma'$ of $\SInv\catA$ by applying \reflemma{NatTrans} to the following data:
\begin{eqnarray*}
	F &=& \otimes, \\
	G &=& \otimes\circ\Gamma_\catA, \\
	H &=& \Phi_\catA,
\end{eqnarray*}
modulo the commutation shown in the following diagram, which holds because $\Inv{(-)}$ is symmetric monoidal:
\begin{center}
\begin{tikzpicture}[node distance = 6mm and 6mm]
  \node (A)                           {$\Inv{(\catA\times\catA)}$};
  \node (B) [below left=of A,xshift=-9mm]         {$\Inv\catA\times\Inv\catA$};
  \node (C) [below right=of B]        {$\Inv{(\catA\times\catA)}$};
  \node (almostC) [right=of C,xshift=-6mm]        {};
  \node (D) [right=of C,yshift=3mm]              {$\Inv{(\catA\times\catA)}$};
  \node (E) [above right=of D,,yshift=-3mm]        {$\Inv\catA$};
  \node (F) [below=of C]              {$\Inv\catA\times\Inv\catA$};
  \path [thick,->, bend left=10]
          (B) edge node [above left]    {\ednd{\Phi_\catA}} (A)
          (A) edge node [above right]   {\ednd{\Inv\otimes}} (E);
\path [thick,->, bend right=10]
          (B) edge node [above right]   {\ednd{\Phi_\catA}} (C)
          (C) edge node [below]         {\ednd{\Gamma_\catA^{\mathcal I}}} (D)
          (D) edge node [below right]    {\ednd{\Inv\otimes}} (E);
\path [->,bend right=30]
          (B) edge node [below left]    {\ednd{\Gamma_{\Inv\catA}}} (F)
          (F) edge node [below right]   {\ednd{\Phi_\catA}} (D);
\draw [double,->, >=latex',double distance=2pt] 
          (A) -- node [right]        {$\lambda'$} (almostC);
\path[draw=none] 
          (C) edge[draw=none] node [inner sep=1pt] {$=$} (F);
\end{tikzpicture}
\end{center}

We have just defined the object map of $\SInv$; let us define it on morphisms. Let $(F,m,u)$ be a symmetric monoidal functor from $(\catA,\otimes,\lone)$ to $(\catB,\odot,U)$. We set $\SInv F=(\Inv F,m',u')$, where
\begin{eqnarray*}
	m' &=& \Inv m\circ\iota_{\Phi_\catA} \\
	u' &=& \Inv u\circ\iota_{\Phi_0}
\end{eqnarray*}
in which $\Phi_\catA$ and $\Phi_0$ denote the same functors as above (coming from the monoidality of $\Inv{(-)}$), $\iota_{\Phi_\catA}$ and $\iota_{\Phi_0}$ their respective identity natural transformations, and $\circ$ is horizontal composition of natural transformations. We can see that $m'$ has the right type as follows: we have $m:\odot\circ(F\times F)\Rightarrow F\circ\otimes:\catA\times\catA\rightarrow\catB$, so $\Inv m:\Inv\odot\circ\Inv{(F\times F)}\Rightarrow \Inv F\circ\Inv\otimes:\Inv{(\catA\times\catA)}\rightarrow\Inv\catB$; composing horizontally with $\iota_{\Phi_\catA}:\Phi_\catA\Rightarrow\Phi_\catA:\Inv\catA\times\Inv\catA\rightarrow\Inv{(\catA\times\catA)}$ yields
$$m':\Inv\odot\circ\Inv{(F\times F)}\circ\Phi_\catA\Rightarrow\Inv F\circ\Inv\otimes\circ\Phi_\catA:\Inv\catA\times\Inv\catA\rightarrow\Inv\catB.$$
Horizontal source and target are as expected; the vertical target is equal to $\Inv F\circ\otimes'$, as desired; for what concerns the vertical source, if we remember that $\Phi_\catA$ is the component of the natural transformation of the monoidality of $\Inv{(-)}$, we have $\Inv\odot\circ\Inv{(F\times F)}\circ\Phi_\catA=\Inv\odot\circ\Phi_A\circ(\Inv F\times\Inv F)=\odot'\circ(\Inv F\times\Inv F)$, as expected.

If we spell them out, the components of $m'$ and $u'$ are actually extremely simple: if $(A,s),(A',s')$ are \objs\ of $\catA$, we have $m'_{(A,s),(A',s')}=m_{A,A'}$, and $u'$ is actually $u$ itself. This shows immediately the monoidality of $(\Inv F,m',u')$, because if a diagram commutes in $\catB$, it commutes \emph{a fortiori} in $\Inv\catB$ (composition of morphisms is the same).

The definition of $\SInv$ on cells is trivial: if $\catA,\catB$ are symmetric monoidal categories, $F,G:\catA\rightarrow\catB$ symmetric monoidal functors, and $\varphi:F\Rightarrow G$ a monoidal natural transformation, we simply put $\SInv\varphi=\Inv\varphi$; the monoidality of $\SInv\varphi$ follows immediately from that of $\varphi$ and from the definitions of $\SInv F$ and $\SInv G$.

%%% Local Variables: 
%%% mode: latex
%%% TeX-master: "LLlevSem"
%%% End: 

%% file: BuildStratMod.tex
We may now state the fundamental properties of the $\SInv$ construction:
\begin{lem}
	\label{lemma:LinNonLin}
	The $\SInv$ functor preserves linear-non-linear adjunctions.
\end{lem}
\begin{pf}
	The fact that $\SInv$ is a 2-endofunctor of \SymMonCat\ immediately implies that it preserves monoidal adjunctions (\cf\ \refapp{CatBack}). The only thing that is left to check is that, whenever a monoidal category $\catM$ is Cartesian,  $\SInv\catM$ is still Cartesian. This is an easy verification, which we leave to the reader.\qed
\end{pf}
\begin{lem}
	\label{lemma:StarAut}
	The $\SInv$ functor preserves $\ast$-autonomous categories.
\end{lem}
\begin{pf}
	We start by proving that $\SInv$ preserves monoidal closure. Let $(\catA,\otimes,\lone)$ be a symmetric monoidal closed category, with the right adjoint of $\otimes$ denoted by $\llinimp$, as usual. Let $L,R:\catA^{\mathrm{op}}\times\catA^{\mathrm{op}}\times\catA\rightarrow\Set$ denote the hom-functors $\catA[-\otimes -,-]$ and $\catA[-,-\llinimp -]$, respectively. On morphisms, $L$ and $R$ act as follows: if $s:A'\rightarrow A$, $t:B'\rightarrow B$, and $u:C\rightarrow C'$ are morphisms of $\catA$, we have
	\begin{eqnarray*}
		L(s,t,u) &=& \lambda f.uf(s\otimes t):\catA[A\otimes B,C]\rightarrow\catA[A'\otimes B',C'], \\
		R(s,t,u) &=& \lambda f.(t\llinimp u)fs:\catA[A,B\llinimp C]\rightarrow\catA[A',B'\llinimp C'],
	\end{eqnarray*}
	where we denoted by $\lambda x.e(x)$ the (set-theoretic) function mapping $x$ to $e(x)$, and we wrote composition in $\catA$ by simple juxtaposition.
	By \refdef{StarAut}, we have a natural isomorphism $\Psi:L\Rightarrow R$. Naturality means that $\Psi\circ L=R\circ\Psi$; more explicitly, for all $s:A'\rightarrow A$, $t:B'\rightarrow B$, $u:C\rightarrow C'$, and $f\in[A\otimes B,C]$,
	$$\Psi_{A,B,C}(uf(s\otimes t))=(t\llinimp u)\,\Psi_{A,B,C}(f)\,s.$$
	Now, if we define the bifunctor $\llinimp'$ of $\SInv\catA$ by $(A,s)\llinimp'(B,t)=(A\llinimp B,(s_k\llinimp t_k)_{k\in\Int})$ on objects and $f\llinimp' g=f\llinimp g$ on morphisms (checking that this is indeed a functor from $(\SInv\catA)^{\mathrm{op}}\times\SInv\catA$ to $\SInv\catA$ is straightforward), we claim that the functors $L'=(\SInv\catA)[-\otimes'-,-]$ and $R'=(\SInv\catA)[-,-\llinimp'-]$ are still naturally isomorphic. In fact, we shall prove that a natural isomorphism is given by $\Psi$ itself, restricted to morphisms of \objs.

	For all objects $(A,s),(B,t),(C,u)$ of $\SInv\catA$, by definition, $(\SInv\catA)[(A,s)\otimes'(B,t),(C,u)]$ and $(\SInv\catA)[(A,s),(B,t)\llinimp'(C,u)]$ are subsets of $\catA[A\otimes B,C]$ and $\catA[A,B\llinimp C]$, respectively. Then, given $f\in(\SInv\catA)[(A,s)\otimes'(B,t),(C,u)]$, we define $\Psi'_{(A,s),(B,t),(C,u)}(f)=\Psi_{A,B,C}(f)$. It is now enough to prove that $\Psi_{A,B,C}(f)\in(\SInv\catA)[(A,s),(B,t)\llinimp'(C,u)]$. In other words, since $(B,t)\llinimp'(C,u)$ is by definition equal to $(B\llinimp C,(t_k\llinimp u_k)_{k\in\Int})$, we need to check that, for all $k\in\Int$, $\Psi_{A,B,C}(f)=(t_k\llinimp u_k)\circ\Psi_{A,B,C}(f)\circ s_k$. But this is an immediate consequence of the fact that $f$ is a morphism of \objs, which means that $f=u_kf(s_k\otimes t_k)$, and of the naturality of $\Psi$ as described above.

	Let now $(\catA,\otimes,\lone,\bot)$ be a $\ast$-autonomous category. We have just proved that $\SInv(\catA,\otimes,\lone)$ is symmetric monoidal closed; we claim that $\bot'=(\bot,\id_\bot)$, where $\id_\bot$ is the $\Int$-indexed sequence identically equal to $\id_\bot$, is a dualizing object. By the definitions given above, it is actually easy to check that, for all objects $(A,s),(B,t)$ of $\SInv\catA$, we have $\Psi'_{(A,s),(A,s)\llinimp'(B,t),(B,t)}(\eval_{(A,s),(B,t)}\circ\gamma'_{(A,s),(A,s)\llinimp'(B,t)})=\Psi_{A,A\llinimp B,B}(\eval_{A,B}\circ\gamma_{A,A\llinimp B})$. Therefore, if we set, for a given object $A$, $h_A=\eval_{A,\bot}\circ\gamma_{A,A\llinimp\bot}$, it is enough to check that $\Psi_{A,A\llinimp\bot,\bot}(h_A)$ is a morphism of \objs, because the fact that it is an isomorphism in $\SInv\catA$ follows immediately from the fact that it is an isomorphism in $\catA$. For this, we use the fact that $h_A$ is a morphism of \objs, which means that $h_A=h_A(s_k\otimes(s_k\llinimp\id_\bot))$ for all $k\in\Int$, and the naturality of $\Psi$, as above.\qed
% 	{\setlength{\arraycolsep}{2pt}\begin{eqnarray*}
% 		\Psi_{A,A\llinimp\bot,\bot}(h_A) &=& \Psi_{A,A\llinimp\bot,\bot}(h_A(s_k\otimes(s_k\llinimp\id_\bot)))= \\
% 		&=& ((s_k\llinimp\id_\bot)\llinimp\id_\bot)\,\Psi_{A,A\llinimp\bot,\bot}(h_A)\,s_k,
% 	\end{eqnarray*}}
% 	as desired.\qed
\end{pf}

As a consequence, we have
\begin{thm}[Preservation of models]
	\label{th:ModelPres}
	Models of linear logic are preserved by $\SInv$, \ie, if $\catL$ is the linear category of a model of linear logic, then $\SInv\catL$ is also the linear category of a model of linear logic.\qed
\end{thm}

\def\cF{\mathcal F}
Observe that, up to and including \reflemma{LinNonLin}, we never used the fact that the endomorphisms $s_k$ in the objects $(A,s)$ are involutions. Indeed, we may define more generally a notion of ``object with endomorphisms'' as a pair $(A,s)$ where $s$ is a $\Int$-indexed sequence of endomorphisms of $A$, not necessarily involutive, or even invertible. The notion of morphism for these objects is the same as that of \refdef{Obj}, \ie, an arrow $f$ of the original category such that $f s_k=t_k f$ for all $k\in\Int$, where $s$ and $t$ are the sequences of endomorphisms of the source and target object, respectively. This gives again a category, of objects with endomorphisms and their morphisms (the free Coxeter groupoid $\Cox$ is replaced by the free monoid on $\Int$). From this, a 2-endofunctor of \SymMonCat, call it $\cF$, may be defined, which shares with $\SInv$ every property mentioned in this section, except \reflemma{StarAut}. Indeed, this last result uses in a somewhat hidden but crucial way the fact that, given \objs\ $(A,s),(B,t)$, a morphism between them is an arrow $f:A\rightarrow B$ such that $f=t_kfs_k$, for all $k\in\Int$, which is equivalent to $fs_k=t_kf$ because $s_k$ (or $t_k$) is an involution.

It is easy to see that requiring the endomorphisms to be involutions is in general necessary for $\cF$ to preserve $\ast$-autonomous categories (\ie, to have \reflemma{StarAut}), or even monoidal closure, which is an essential requirement for a model of linear logic. In fact, consider the category $(\Set,\times,1)$ of sets and functions, with the usual Cartesian product, and where $1$ is a singleton. This is the prototypical Cartesian closed category, which is therefore symmetric monoidal closed. We claim that $\cF(\Set)$, defined mimicking the definition of $\SInv(\Set)$, is not closed. If it were, we would have that the set $\cF(\Set)[(A,s),(B,t)]$, where $A,B$ are sets and $s,t$ $\Int$-indexed sequences of functions on them, would be in bijection with the set  $\cF(\Set)[(1,\id_1),(B^A,t^s)]$, where $B^A$ is the set of all functions from $A$ to $B$, and, given $f\in B^A$ and $k\in\Int$, $(t^s)_k(f)=t_kfs_k$. Now, $\cF(\Set)[(1,\id_1),(B^A,t^s)]$ is in bijection with the set of all functions $f:A\rightarrow B$ such that $f=t_kfs_k$ for all $k\in\Int$. But this set is too ``small''; indeed, given some $g\in\cF(\Set)[(A,s),(B,t)]$, we do know that $gs_k=t_kg$ for all $k\in\Int$, but there is no reason for the equality $g=t_kgs_k$ to hold as well, if $s_k$ and $t_k$ are arbitrary functions from $A$ and $B$ to themselves.

The model resulting from the $\SInv$ construction is in some sense ``isomorphic'' to the original one. Nevertheless, the construction is of interest because it yields a model of \emph{stratified} linear logic which is never degenerate, unless the original model is trivial. In order to prove this, we first need to recall a basic categorical result concerning models of linear logic.
\begin{figure}
   \input{lemmedegen.tex}
   \caption{The linear logic proof nets $\rho$ (bottom, left hand side) and $\rho'$ (top, left hand side).}
   \label{fig:lemmanets}
\end{figure}
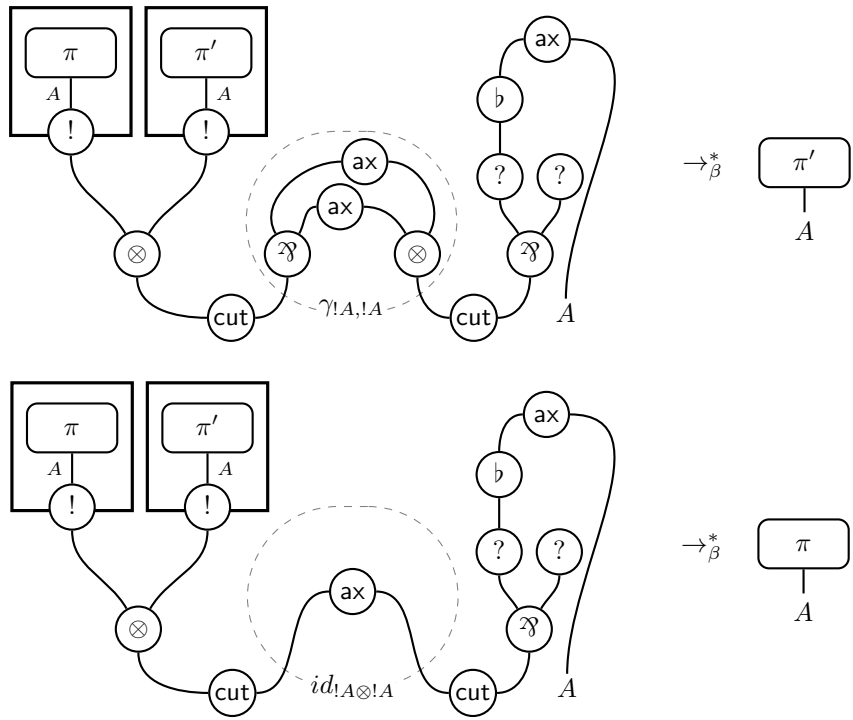
\begin{lem}
	\label{lemma:Achiral}
	Let $(\catL,\ltens,\lone)$ be the linear category of a non-trivial model of linear logic, with the symmetry of $\ltens$ denoted by $\gamma$. Then, there exists an object $\Xi$ of $\catL$ such that $\gamma_{\Xi,\Xi}\neq\id_{\Xi\ltens\Xi}$.
\end{lem}
\begin{pf}
	By the non-triviality of the model, there is a formula $A$ and two proof nets $\pi,\pi'$ of conclusion $A$ such that their respective interpretations $\sem\pi,\sem{\pi'}$ are different morphisms of $\catL[1,\sem A]$, where $\sem A$ is the interpretation of $A$. Consider now the proof nets $\rho,\rho'$ defined in \reffig{lemmanets}. Observe that the subnets of $\rho$ and $\rho'$ which are circled in \reffig{lemmanets} are interpreted by $\id_{\sem{\oc A}\ltens\sem{\oc A}}$ and $\gamma_{\sem{\oc A},\sem{\oc A}}$, respectively. Therefore, by invariance of the model under cut-elimination, and by the fact that it induces a congruence, assuming $\id_{\sem{\oc A}\ltens\sem{\oc A}}=\gamma_{\sem{\oc A},\sem{\oc A}}$ would imply $\sem\rho=\sem{\rho'}$ and hence $\sem\pi=\sem{\pi'}$, against our hypothesis. So the object $\Xi=\sem{\oc A}$ satisfies the requirement.\qed
\end{pf}

\begin{thm}
	\label{th:NonDegenerate}
	Every non-trivial model of linear logic yields, through the $\SInv$ construction, a non-degenerate model of \stratLL.
\end{thm}
\begin{pf}
	Let $\catL$ be the linear category of a non-trivial model of linear logic. By \refth{ModelPres}, $\SInv\catL$ also hosts a model of linear logic; by \refdef{StratModel}, in order to define a model of \stratLL\ we only need to find an interpretation of the paragraph modality. This may be done by fixing some $n\in\Int$ and considering the endofunctor $\parg_n$ of $\SInv\catL$ defined as follows. If $(A,s)$ is an object of $\SInv\catL$, we set $\parg_n(A,s)=(A,(s_{k-n})_{k\in\Int})$, \ie, the image of $(A,s)$ through $\parg_n$ has the same underlying object, but the sequence of involutions is ``shifted'' by $n$; on morphisms, $\parg_n$ is the identity. For convenience, we set $\parg=\parg_1$, but what we shall say holds for all $n\in\Int$.

	It is immediate to see that the functor $\parg$ satisfies all of the requirements of \refdef{StratModel}; we leave the verification to the reader. The fact that the model is non-degenerate is a consequence of \reflemma{Achiral}. Indeed, composed with the non-triviality of the original model, that result ensures us that in $\catL$ there exists an object $\Xi$ such that $\gamma_{\Xi,\Xi}\neq\id_{\Xi\ltens \Xi}$. Consider then the \obj\ $(\Xi\ltens \Xi,\xi)$, where $\xi_0=\gamma_{\Xi,\Xi}$ and $\xi_k=\id_{\Xi\ltens \Xi}$ for all $k\neq 0$. Let $f:(\Xi\ltens \Xi,\xi)\rightarrow\parg(\Xi\ltens \Xi,\xi)$ in the category $\SInv\catL$. By definition, $f$ is an endomorphism of $\Xi\ltens \Xi$ in $\catL$, which must further satisfy $f\circ\gamma_{\Xi,\Xi}=f=f\circ\id_{\Xi\ltens \Xi}$ (and also $\gamma_{\Xi,\Xi}\circ f=f$). But then $f$ is not monic (nor epic), and thus cannot be an isomorphism.\qed
\end{pf}

%% file: lemmedegen.tex
\begin{center}
\begin{tikzpicture}[node distance=4mm and 3mm, baseline = (current bounding box.east)]
%\mygridto{(11,5)}
\node[nopoint] (origine1) at (0,5.2) {};
\node[netpi,below right=of origine1, xshift=-5mm, text width=6mm] (pi) {$\pi$};
\draw[very thick]  (pi) ++(-8mm,6mm) rectangle +(16mm,-17mm);
\netoc[below=of pi] {pioc};

\node[nopoint] (origine2) at (1.3,5.2) {};
\node[netpi,below right=of origine2, text width=6mm] (pi2) {$\pi'$};
\draw[very thick]  (pi2) ++(-8mm,6mm) rectangle +(16mm,-17mm);
\netoc[below=of pi2] {pi2oc};

\node[nopoint] (origine3) at (1.3,2.5) {};
\nettens[below=of origine3] {tens1};
\netcut[below right=of tens1,xshift=5mm] {cut1};

\node[rectangle,rounded corners=12mm,dashed,draw=gray,      
      text width=4mm, inner sep=12mm] (ici) at (4.15, 2.21) {};
\node[nopoint, below=of ici, yshift=9mm, fill=white] (leg) {$\gamma_{\oc A,\oc A}$};

\netpar[above right=of cut1] {par2};
\netax[above=of par2,xshift=7mm,yshift=-4mm] {ax1};
\netax[right=of ax1,yshift=6mm,xshift=-6mm] {ax2};
\nettens[right=of par2, xshift=8mm] {tens2};
\netcut[below right=of tens2] {cut2};
\netpar[above right=of cut2] {par1};
\netwn[above=of par1,xshift=-4mm] {wn1};
\netwn[above=of par1,xshift=4mm] {wn2};
\netft[above=of wn1] {flat1};
\netax[right=of flat1,yshift=8mm,xshift=-3mm] {ax};
\node[nopoint] (conc) at (7,1) {$A$};
\draw[thick]
    (pi) edge node [left] {\ednd{A}} (pioc)
    (pioc) to [out=270,in=120]  (tens1)
    (pi2) edge node [right] {\ednd{A}} (pi2oc)
    (pi2oc) to [out=270,in=60]  (tens1)
    (tens1) to [out=270,in=180] (cut1)
    (par2) to [out=270,in=0]    (cut1)
    (ax1) to [out=180, in=60]   (par2)
    (ax2) to [out=180, in=120]  (par2)
    (ax1) to [out=0,in=120]     (tens2)
    (ax2) to [out=0,in=60]      (tens2)
    (tens2) to [out=270,in=180] (cut2)
    (par1) to [out=270,in=0]    (cut2)
    (wn1) to [out=270,in=120]   (par1)
    (wn2) to [out=270,in=60]    (par1)
    (flat1) --                  (wn1)
    (ax) to [out=180,in=90]     (flat1)
    (ax) to [out=0, in=90]      (conc);
\end{tikzpicture} 
$\betareds$
\quad
\begin{tikzpicture}[node distance=6mm and 3mm, baseline = (current bounding box.east)]
\node[nopoint] (origine1) at (0,5) {};
\node[netpi,below right=of origine1, xshift=-5mm, text width=6mm] (pi) {$\pi'$};
\node[below=of pi,yshift=2.8mm] (conc) {$A$};
\draw[thick] (pi) -- (conc);
\end{tikzpicture}
\end{center}

\begin{center}
\begin{tikzpicture}[node distance=4mm and 3mm, baseline = (current bounding box.east)]
%\mygridto{(11,5)}
\node[nopoint] (origine1) at (0,5.2) {};
\node[netpi,below right=of origine1, xshift=-5mm, text width=6mm] (pi) {$\pi$};
\draw[very thick]  (pi) ++(-8mm,6mm) rectangle +(16mm,-17mm);
\netoc[below=of pi] {pioc};

\node[nopoint] (origine2) at (1.3,5.2) {};
\node[netpi,below right=of origine2, text width=6mm] (pi2) {$\pi'$};
\draw[very thick]  (pi2) ++(-8mm,6mm) rectangle +(16mm,-17mm);
\netoc[below=of pi2] {pi2oc};

\node[nopoint] (origine3) at (1.3,2.5) {};
\nettens[below=of origine3] {tens1};
\netcut[below right=of tens1,xshift=5mm] {cut1};

\node[rectangle,rounded corners=12mm,dashed,draw=gray,      
      text width=4mm, inner sep=12mm] (ici) at (4.15, 2.21) {};
\node[nopoint, below=of ici, yshift=10mm,fill=white] (leg) {$\id_{\oc A\otimes\oc A}$};

\netax[above right=of cut1, xshift=8.5mm, yshift=5mm] {ax1};
\netcut[below right=of ax1, xshift=8.5mm, yshift=-5mm] {cut2};
\netpar[above right=of cut2] {par1};
\netwn[above=of par1,xshift=-4mm] {wn1};
\netwn[above=of par1,xshift=4mm] {wn2};
\netft[above=of wn1] {flat1};
\netax[right=of flat1,yshift=8mm,xshift=-3mm] {ax};
\node[nopoint] (conc) at (7,1) {$A$};
\draw[thick]
    (pi) edge node [left] {\ednd{A}} (pioc)
    (pioc) to [out=270,in=120]  (tens1)
    (pi2) edge node [right] {\ednd{A}} (pi2oc)
    (pi2oc) to [out=270,in=60]  (tens1)
    (tens1) to [out=270,in=180] (cut1)
    (ax1) to [out=180, in=0]   (cut1)
    (ax1) to [out=0,in=180]     (cut2)
    (par1) to [out=270,in=0]    (cut2)
    (wn1) to [out=270,in=120]   (par1)
    (wn2) to [out=270,in=60]    (par1)
    (flat1) --                  (wn1)
    (ax) to [out=180,in=90]     (flat1)
    (ax) to [out=0, in=90]      (conc);
\end{tikzpicture} 
$\betareds$
\quad
\begin{tikzpicture}[node distance=6mm and 3mm, baseline = (current bounding box.east)]
\node[nopoint] (origine1) at (0,5) {};
\node[netpi,below right=of origine1, xshift=-5mm, text width=6mm] (pi) {$\pi$};
\node[below=of pi,yshift=2.8mm] (conc) {$A$};
\draw[thick] (pi) -- (conc);
\end{tikzpicture}
\end{center}

%%% Local Variables: 
%%% mode: latex
%%% TeX-master: "LLlevSem"
%%% End: 

%% file: L3.tex
Recall how in \refsect{Bounded} we reformulated linear logic by levels (\LLlev), which is in some sense the largest extant stratification-based, bounded-complexity subsystem of linear logic, as the fragment of \stratLL\ in which the exponential modalities are forced to be ``tied'' to the paragraph modality. In light of this, using the results of \refsect{ProofNets}, we may give a new, geometric definition of \LLlev, equivalent to the original one given in \cite{BaillotMazza:LLlev} and based on indexings.

% In the following, when we speak of a \dr\ \emph{of linear logic}, we mean a \dr\ as in \refdef{ProofNet}, but containing no \pargl\ link. Under this assumption, \dr s and \pn s actually coincide, but we use the term \dr\ to stress that $\parg$-correctness is void. Then, we shall impose on linear logic \dr s a criterion adapted from $\parg$-correctness, in order to characterize \LLlev\ within linear logic.

We start by recalling the definition of \LLlev, as given in \cite{BaillotMazza:LLlev}.
\begin{defn}[Exponential indexing]
	\label{def:L3Indexing}
	An \emph{exponential indexing} (called \emph{indexing} in \cite{BaillotMazza:LLlev}) for a \ps\ $\pi$ is a function $I$ from the edges of $\pi$ to $\Int$ which:
	\begin{itemize}
		\item satisfies the constraints of \reffig{Indexes}, except for \ofcourse\ and \whynot\ links, for which, if $e_1,\ldots,e_n$ and $e'$ are the premises and conclusion of such a link, we have $I(e_1)=\cdots=I(e_n)=I(e')+1$;
		\item satisfies $I(e)=I(e')$ for all conclusions $e,e'$ of $\pi$.
	\end{itemize}
\end{defn}
In other words, an exponential indexing is just as a strong indexing of \refdef{Indexing} in which exponential links behave like \pargl\ links.
\begin{defn}[Linear logic by levels \citep{BaillotMazza:LLlev}]
	\label{def:L3}
	\sloppy{\LLlev\ is the set of all \dr s admitting an exponential indexing.}
\end{defn}

Let us now give the geometric definition.
\begin{defn}[$\oc\wn$-balanced cycle]
	\label{def:ExpBalanced}
	We define a $\oc\wn$-balanced cycle in a \dr\ just as a balanced cycle (\refdef{Balanced}), except that instead of counting only \pargl\ links, we also count exponential links (\ie, \ofcourse\ and \whynot\ links).
\end{defn}
% \begin{defn}[Geometric definition of \LLlev]
% 	\label{def:L3Geom}
% 	\LLlev\ is the logical system consisting of all \dr s whose $\lpar$-closure is such that all cycles are $\oc\wn$-balanced.
% \end{defn}

% The equivalence of \refdef{L3Geom} and \refdef{L3} is a corollary of \refth{Seq}, modulo the following technical results, which are immediate.
\begin{defn}[Shifted formula and shifted proof]
	\label{def:Plus}
	Let $A$ be a formula. We define $A^+$ as the formula obtained by adding a paragraph modality after every exponential modality appearing in $A$ (\ie, every occurrence of $\oc$ and $\wn$ is replaced by $\oc\parg$ and $\wn\parg$, respectively).
	
	Let $\pi$ be a \ps\ of conclusions $A_1,\ldots,A_n$. We define the \ps\ $\pi^+$ by adding a \pargl\ link above every \ofcourse\ and \flatl\ link of $\pi$ (in sequent calculus derivations, we add a paragraph rule before every promotion and dereliction rule). It is immediate to see that $\pi^+$ has conclusions $A_1^+,\ldots,A_n^+$.
\end{defn}

\begin{figure}
  \centering
  \input{exampleshift.tex}
  \caption{Example of a shifted \pn.}
  \label{fig:shift}
\end{figure}
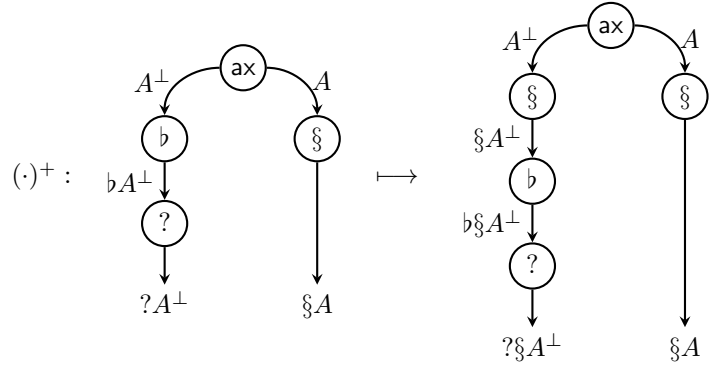

Figure~\ref{fig:shift} shows a \pn\ $\pi$ of \LLlev\  (at left hand side) and (at right hand side) the shifted \ps $\pi^+$, which indeed is a \pn\ of \stratLL.

The following technical results are immediate:
\begin{lem}
	\label{lemma:stratLLandL3}
	Let $\pi$ be a \ps\ with no \struct\ in its conclusions. Then:
	\begin{enumerate}
		\item $\pi$ is DR-correct iff $\pi^+$ is;
		\item if $\pi_0$ and $(\pi^+)_0$ denote the $\lpar$-closures of $\pi$ and $\pi^+$, respectively, we have $(\pi^+)_0=\pi_0^+$;
		\item $\pi$ admits an exponential indexing (\refdef{L3Indexing}) iff $\pi^+$ is strongly indexable (\refdef{Indexing});
		\item $\pi$ contains a cycle which is not $\oc\wn$-balanced (\refdef{ExpBalanced}) iff $\pi^+$ contains a cycle which is not balanced (\refdef{Balanced}).\qed
	\end{enumerate}
\end{lem}
\begin{thm}[Geometric characterization of \LLlev]
	\label{th:Geom}
	A \dr\ $\pi$ is in \LLlev\ (as per \refdef{L3}) iff all cycles in its $\lpar$-closure $\pi_0$ are $\oc\wn$-balanced.
\end{thm}
\begin{pf}
	Let $(\pi^+)_0$ be the $\lpar$-closure of $\pi^+$. By point 3 of \reflemma{stratLLandL3}, $\pi$ satisfies \refdef{L3} iff $\pi^+$ is strongly indexable. But, thanks to \reflemma{StronglyInd} and point 1 of \reflemma{stratLLandL3}, this is equivalent to the fact that all cycles of $(\pi^+)_0$ are balanced. Now, since $(\pi^+)_0=\pi_0^+$ (point 2 of \reflemma{stratLLandL3}), we may conclude by the equivalence of point 4 of \reflemma{stratLLandL3}.\qed
\end{pf}

%% file: exampleshift.tex
\begin{center}
$(\cdot)^+ :$\quad
\begin{tikzpicture}[node distance=5mm and 6mm, baseline = (current bounding box.east)]
\netax                                    {ax};
\netft [below left=of ax]                 {ft};
\netwn [below =of ft]                     {wn};
\node[lab, below =of wn] (concl)          {$\wn A^\perp$};
\netparg[right=of ft,xshift=8mm]          {parg};
\node[lab] (concr) at (parg |- concl)     {$\parg A$};
\draw[->,>=stealth,thick]
     (ax) to [out=180,in=90]  node [left] {$A^\perp$}
     (ft);
\draw[->,>=stealth,thick]
     (ax) to [out=0,in=90]  node [right]  {$A$}
     (parg);
\draw[->,>=stealth,thick] 
     (ft) edge node [left] {$\flat A^\perp$} (wn);
\draw[->,>=stealth,thick] 
     (wn) -- (concl);
\draw[->,>=stealth,thick]
     (parg) -- (concr);
\end{tikzpicture}
\quad
$\longmapsto$ \quad
\begin{tikzpicture}[node distance=5mm and 6mm, baseline = (current bounding box.east)]
\netax                                    {ax};
\netparg [below left=of ax]               {parg2};
\netft [below =of ft]                     {ft};
\netwn [below =of ft]                     {wn};
\node[lab, below =of wn] (concl)          {$\wn \parg A^\perp$};
\netparg[right=of parg2,xshift=8mm]       {parg};
%\node[lab, right =of concl,xshift=7.9mm] (concr)   {$\parg A$};
\node[lab] (concr) at (parg |- concl)   {$\parg A$};
\draw[->,>=stealth,thick]
     (ax) to [out=180,in=90]  node [left] {$A^\perp$}
     (parg2);
\draw[->,>=stealth,thick]
     (ax) to [out=0,in=90]  node [right] {$A$}
     (parg);
\draw[->,>=stealth,thick] 
     (ft) edge node [left] {$\flat \parg A^\perp$} (wn);
\draw[->,>=stealth,thick] 
     (wn) -- (concl);
\draw[->,>=stealth,thick]
     (parg) -- (concr);
\draw[->,>=stealth,thick]
     (parg2) edge node [left] {$\parg A^\perp$} (ft);
\end{tikzpicture}
\end{center}

%%% Local Variables: 
%%% mode: latex
%%% TeX-master: "LLlevSem"
%%% End: 

%% file: L3Sem.tex
As a subsystem of linear logic, \LLlev\ may be semantically interpreted in any model of linear logic. However, such a denotational semantics is absolutely uninformative by itself, because it is incapable of distinguishing a linear logic proof which is part of \LLlev\ from one that is not.

We shall see how the $\SInv$ construction may be used to build more informative denotational semantics of \LLlev. For this, we shall use the models defined in the proof of \refth{NonDegenerate}, living in the category $\SInv\catL$ where $\catL$ is the linear category of a model of linear logic. We remind that such models interpret the paragraph modality with the functor defined by $\parg(A,s)=(A,(s_{k-1})_{k\in\Int})$ on objects and acting as the identity on morphisms. To be able to fully use \refth{NonDegenerate}, in the sequel we shall assume that the starting model of linear logic is non-trivial.

Recall that denotational interpretations are always parametric in an assignment of objects to propositional atoms. For all of our future purposes, it will be enough to consider what we shall call the \emph{default assignment}, which was already defined in the proof of \refth{NonDegenerate}, and which we give again here.
\begin{defn}[Default assignment]
	\label{def:DefAss}
	Let $\catL$ be the linear category of a non-trivial model of linear logic. By \reflemma{Achiral}, there exists an object $\Xi$ of $\catL$ such that $\gamma_{\Xi,\Xi}\neq\id_{\Xi\ltens \Xi}$. A \emph{default assignment} maps every propositional atom to the \obj\ $(\Xi\ltens \Xi,\xi)$, where $\xi_0=\gamma_{\Xi,\Xi}$ and $\xi_k=\id_{\Xi\ltens \Xi}$ for all $k\neq 0$.

	Of course, there is one default assignment for each object $\Xi$ of $\catL$ such that $\gamma_{\Xi,\Xi}\neq\id_{\Xi\ltens \Xi}$; however, since the choice of a particular $\Xi$ is irrelevant, we shall abusively speak of ``the'' default assignment, and denote by $\sem{\cdot}$ the interpretation map (from formulas and proofs of \stratLL\ to objects and morphisms of $\SInv\catL$, respectively) under that default assignment.
\end{defn}

Observe that, even in non-degenerate models, the $\parg$ functor still acts as the identity on morphisms. This means that, if $\pi^-$ is a \dr\ obtained from the \pn\ $\pi$ by removing any number of \pargl\ links, we have $\sem{\pi}=\sem{\pi^-}$. In other words, paragraphs only matter in formulas, not in proofs.

\begin{defn}[Shifted interpretation of formulas]
	\label{def:AltSem}
	\sloppy{Given a non-trivial model of linear logic whose linear category is $\catL$, we define the \emph{shifted interpretation map} $\altsem{\cdot}$ from formulas to objects of $\SInv\catL$ by $\altsem{A}=\sem{A^+}$, where the formula $A^+$ is introduced in \refdef{Plus}.}
\end{defn}

If we are given a \pn\ $\pi$ of conclusion $A$, the interpretation $\sem{\pi}$ is always well defined, as a morphism of $(\SInv\catL)[(\lone,\id_1),\sem A]$. We shall be interested in the following question:
$$\textrm{do we have }\sem\pi\in(\SInv\catL)[(\lone,\id_\lone),\altsem A]\textrm{?}$$
This is meaningful, because $A^+$ and $A$ differ only for the addition of paragraph modalities and thus, by definition of the $\parg$ functor, if $\sem A=(S,\sigma)$, then $\altsem A=(S,\sigma')$, \ie, the underlying object $S$ is the same. Therefore, a morphism of type $(\lone,\id_\lone)\rightarrow(S,\sigma)$ in $\SInv(\catL)$, which is first of all a morphism of type $\lone\rightarrow S$ in $\catL$, may \emph{a priori} be also a morphism of type $(\lone,\id_\lone)\rightarrow(S,\sigma')$ in $\SInv(\catL)$.

To state the question more concisely, we shall use the notation $\sem\pi\in\altsem A$ as a short hand for $\sem\pi\in(\SInv\catL)[(\lone,\id_\lone),\altsem A]$. There is one interesting case in which the answer is always positive:
\begin{thm}[Soundness]
	\label{th:Soundness}
	If a \dr\ $\pi$ of conclusion $A$ belongs to \LLlev, then $\sem\pi\in\altsem A$.
\end{thm}
\begin{pf}
  By points~1 and~3 of \reflemma{stratLLandL3}, we know that $\pi^+$ (see \refdef{Plus}) is a strongly indexable \dr. Then, by \reflemma{StronglyInd}, $\pi^+$ is a \stratLL\ \pn\ of conclusion $A^+$. Since the $\SInv$ construction yields models of \stratLL\ (\refth{ModelPres}), we have that $\sem{\pi^+}\in(\SInv\catL)[(\lone,\id_\lone),\sem{A^+}]$, which means $\sem{\pi^+}\in\altsem A$. But, as remarked after \refdef{DefAss}, we have $\sem{\pi^+}=\sem\pi$.\qed
\end{pf}

Therefore, testing whether $\sem\pi\in\altsem A$ is always a way to obtain negative information about the membership of a \dr\ $\pi$ to \LLlev; if the answer is negative, then $\pi$ cannot belong to \LLlev.

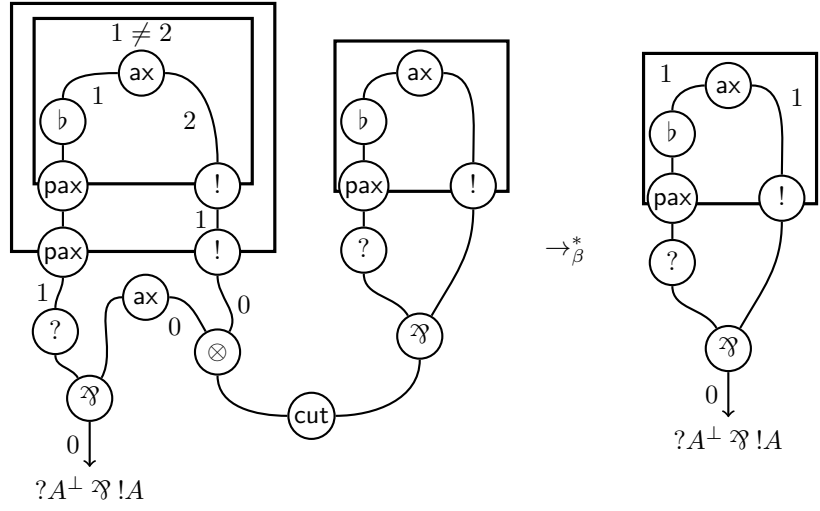
\begin{figure}
\input{examplenotll3.tex}
  \caption{The left hand side \dr\ is not in \LLlev\ (there is a mismatch of indexes on the left-most top-most axiom) while its cut-free form is in \LLlev.}
  \label{fig:examplenotl3notbetastable}
\end{figure}

There is no hope for the converse of \refth{Soundness} to hold in general, for the simple reason that, while \LLlev\ is of course stable under cut-elimination, \emph{non}-membership to \LLlev\ is \emph{not} itself preserved under cut-elimination. It is indeed easy to exhibit a non-cut-free \dr\ $\pi$ violating the conditions for belonging to \LLlev, such that $\pi\betareds\pi'$ with $\pi'$ in \LLlev, see for instance \reffig{examplenotl3notbetastable}. Since denotational semantics ``sees through'' cut-elimination, we would have $\sem{\pi}\in\altsem{A}$ whilst having $\pi\not\in\LLlev$.

In \refsect{Completeness} we shall give, under the hypothesis of absence of cuts, a sufficient condition for the converse of \refth{Soundness} to hold. For the moment, we observe that, for some cut-free \dr s, the converse is always realized. It is the case, for example, of the \dr\ proving the dereliction principle, $\oc A\llinimp A$, which does not belong to \LLlev. Indeed, if $\pi$ is such a \dr\ in the case in which $A$ is a propositional atom, we have $\sem{\pi}=\varepsilon_{\Xi\ltens \Xi}:\oc(\Xi\ltens \Xi)\rightarrow \Xi\ltens \Xi$, \ie, the counit of the comonad $\oc(-)$. Then, if we had $\sem{\pi}\in\SInv\catL[\altsem{\oc A},\altsem A]$, we would have in particular $\gamma_{\Xi,\Xi}\circ\varepsilon_{\Xi\ltens \Xi}=\varepsilon_{\Xi\ltens \Xi}$, which elementary computations similar to those of \reffig{lemmanets} show to be inconsistent with the non-triviality of the original model.

%% file: examplenotll3.tex
\begin{center}
\begin{tikzpicture}[node distance=2mm and 3mm, baseline = (current bounding box.east)]
%\mygridto{(10,5)}
\node[nopoint] (start1) at (0,5) {};
\draw[very thick]  (start1) ++(3mm,3mm) rectangle +(29mm,-22mm);
\draw[very thick]  (start1) ++(0mm,5mm) rectangle +(35mm,-33mm);
\netax[below right=of start1,xshift=12mm] {ax1};
\netft[below left=of ax1,xshift=-3mm] {ft1};
\netpax[below=of ft1] {pax1};
\netpax[below=of pax1] {pax2};
\netoc[right=of pax1,xshift=11mm] {oc1};
\netoc[right=of pax2,xshift=11mm] {oc2};
\nettens[below=of oc2,yshift=-5mm] {tens};
\draw[thick]
    (ax1) to [out=180,in=90,->] node [below right] {$1$}
    (ft1)
    (ax1) to [out=0,in=90,->]  node [below left] {$2$}
    (oc1)
    (ft1) -- (pax1)
    (pax1) -- (pax2)
    (oc1) edge node [left] {$1$} (oc2) 
    (oc2) to [out=270,in=60] node [right] {$0$}
    (tens);

\node[nopoint] (start2) at (4,5) {};
\draw[very thick]  (start2) ++(3mm,0mm) rectangle +(23mm,-20mm);
\netax[below right=of start2,xshift=9mm] {ax2};
\netft[below left=of ax2] {ft2};
\netpax[below=of ft2] {pax12};
\netoc[right=of pax12,xshift=5mm] {oc12};
\netwn[below=of pax12] {wn2};
\netpar[below right=of wn2,yshift=-5mm] {par};
\draw[thick]
    (ax2) to [out=180,in=90,->]
    (ft2)
    (ax2) to [out=0,in=90,->]
    (oc12)
    (ft2) -- (pax12)
    (pax12) -- (wn2)
    (wn2) to [out=270,in=120] 
    (par)
    (oc12) to [out=270,in=60]
    (par);

\netcut[below right=of tens,xshift=5mm,yshift=-2mm] {cut};
\draw[thick]
    (tens) to [out=270,in=180] (cut)
    (par)  to [out=270,in=0]   (cut);

\netax[below of=pax2,yshift=-4mm,xshift=11mm] {ax3};
\netwn[below=of pax2,xshift=-1mm,yshift=-2mm] {wn1};
\netpar[below left=of ax3, yshift=-7mm] {par2};
\node[below =of par2,yshift=-4mm] (conc) {$\wn A^\perp\parr\oc A$};
\draw[thick]
   (pax2) to [out=270, in=90] node [left] {$1$}(wn1)
   (ax3) to [out=0, in=120] node [below left] {$0$} (tens)
   (ax3) to [out=180,in=60] (par2)
   (wn1) to [out=270,in=120] (par2)
   (par2) edge[->] node [left] {$0$}  (conc);
\node[above=of ax1,yshift=-2.8mm] (diff) {$1\neq 2$};
\end{tikzpicture}
\quad $\betareds$ \quad
\begin{tikzpicture}[node distance=2mm and 3mm, baseline = (current bounding box.east)]
\node[nopoint] (start2) at (4,5) {};
\draw[very thick]  (start2) ++(3mm,0mm) rectangle +(23mm,-20mm);
\netax[below right=of start2,xshift=9mm] {ax2};
\netft[below left=of ax2] {ft2};
\netpax[below=of ft2] {pax12};
\netoc[right=of pax12,xshift=5mm] {oc12};
\netwn[below=of pax12] {wn2};
\netpar[below right=of wn2,yshift=-5mm] {par};
\node[below =of par,yshift=-4mm] (conc) {$\wn A^\perp\parr\oc A$};
\draw[thick]
    (ax2) to [out=180,in=90,->] node [above left] {$1$}
    (ft2)
    (ax2) to [out=0,in=90,->] node [above right] {$1$}
    (oc12)
    (ft2) -- (pax12)
    (pax12) -- (wn2)
    (wn2) to [out=270,in=120] 
    (par)
    (oc12) to [out=270,in=60]
    (par)
    (par) edge [->] node [left] {$0$} (conc);
\end{tikzpicture}
\end{center}

%%% Local Variables: 
%%% mode: latex
%%% TeX-master: "LLlevSem"
%%% End: 

%% file: Interactive.tex
A particularly simple, and indeed rather uninformative model of linear logic is the so-called \emph{syntactic model}. Although of little value \emph{per se}, it lends itself to an interesting application of the $\SInv$ construction. The linear category of this model, which we denote by \catPN, is defined as follows:
\begin{itemize}
	\item its objects are the formulas of \stratLL;
	\item a morphism from a formula $A$ to a formula $B$ is a cut-free \dr\ with atomic axioms (\ie, $\eta$-expanded), of conclusions $A^\bot,B$;
	\item composition is given by forming a cut and computing the cut-free form. This operation is well defined because cut-elimination always terminates and, in the propositional case, preserves atomic axioms; it is associative thanks to confluence. The identities of the category are $\eta$-expansions of axioms.
\end{itemize}

Note that, in this paper, we did not define \pn s for additive connectives, so technically speaking \catPN\ is not a linear category because it lacks products and coproducts (and we cannot properly define the Cartesian category of the linear-non-linear adjunction). This is not a concern here, as we shall not need all of the structure of the model in what follows.

All semantic constructions (monoidal structure, exponentials) are defined in \catPN\ using the very syntactic constructions they are supposed to model, so everything works without surprises. The presence of the $\parg$ modality (and of \pargl\ links in \dr s) should not confuse the reader: the syntactic model is a model of \emph{linear logic}; paragraphs are ignored by the constructions of the model. Of course, it is also a model of \stratLL, but it is a degenerate one: $\parg A$ is isomorphic to $A$ for every formula $A$. It would be non-degerate if we had defined morphisms to be \emph{\pn s}, instead of \dr s, but this would not be interesting for our purposes: the interactive characterization of \LLlev\ (\refth{Interactive}) would be less clear, because the correctness of paragraphs would be already accounted for by the model, and not by interaction.

Let us now describe the category $\SInv(\catPN)$ in detail:
\begin{itemize}
	\item its objects are pairs $(A,\sigma)$ where $A$ is a formula and $\sigma$ is a $\Int$-indexed sequence of cut-free, $\eta$-expanded \dr s of conclusions $A^\bot,A$, such that when $\sigma_k$ is cut with itself, it reduces to $\id_A$ (we may assume that $\sigma$ is almost everywhere equal to $\id_A$, so the set of objects of $\SInv(\catPN)$ is still denumerable);
	\item a morphism from $(A,\sigma)$ to $(B,\tau)$ is a cut-free, $\eta$-expanded \dr\ $\pi$ of conclusions $A^\bot,B$ such that, for all $k\in\Int$, the \dr : 
          \begin{center}
            \input{InvChain.tex}
          \end{center}
	reduces to $\pi$ itself. Composition and identity morphisms are the same as $\catPN$.
\end{itemize}

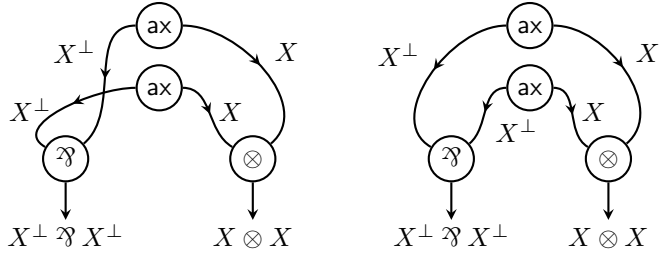
\begin{figure}[t]
	\input{IdAndFlip}
	\caption{The morphisms $\id_{X\ltens X}$ (left) and $\gamma_{X,X}$ (right) in \catPN.}
	\label{fig:IdAndFlip}
\end{figure}

The default interpretations, which, in the case of the syntactic model, we denote by $\sem{\cdot}_s$ and $\altsem{\cdot}_s$, may be described as follows.
\begin{defn}[Default atomic substitution]
	if $A$ is a formula, we denote by $A^\bullet$ the formula obtained from $A$ by substituting every atom $Z$ with $X\ltens X$ (and its dual $Z^\bot$ with $X^\bot\lpar X^\bot$), where $Z$ ranges over all atoms.

	Similarly, If $\pi$ is an $\eta$-expanded net of conclusion $A$, we denote by $\pi^\bullet$ the net of conclusion $A^\bullet$ obtained by replacing every atomic axiom of $\pi$ with the net $\id_{X\ltens X}$ (\reffig{IdAndFlip}, left).
\end{defn}
\begin{defn}[Test]
	\label{def:Test}
	It is easy to check (see \cite{BaillotMazza:LLlev}) that every \dr\ $\pi$ of \LLlev\ admits a \emph{default exponential indexing}, which is an exponential indexing (\refdef{L3Indexing}) whose range are non-negative integers and assigning $0$ to the conclusions of the net. An occurrence of a subnet of $\pi$ is said to appear at \emph{level} $n$ if all of its conclusions are assigned the integer $n$ by the default exponential indexing of $\pi$. We let the reader verify that, for every formula $A$, the \dr\ $\id_{A^\bullet}$ is in \LLlev, so the above definition of level applies in particular to these nets.
	
	Let $A$ be a formula and $k\in\Int$. The \emph{test} of index $k$ of type $A$, denoted by $\theta^A_k$, is the \dr\ of conclusions $(A^\bullet)^\bot,A^\bullet$ obtained from $\id_{A^\bullet}$ by replacing to \emph{all} occurrences of $\id_{X\ltens X}$ appearing at level $k$ the net $\gamma_{X,X}$ (\reffig{IdAndFlip}, right). We set $\theta^A=(\theta^A_k)_{k\in\Int}$.
\end{defn}
\begin{lem}
	\label{lemma:SyntTests}
	Let $A$ be a formula of linear logic. Then, $\altsem{A}_s=(A^\bullet,\theta^A)$.
\end{lem}
\begin{pf}
	A straightforward induction on $A$.\qed
\end{pf}

The interpretation of proofs is particularly easy to describe. If $\pi$ is a linear logic \dr\ of conclusion $A$, whose cut-free and $\eta$-expanded form is $\pi_0$, $\sem{\pi}_s$ is simply $\pi_0^\bullet$ with a \botl\ link juxtaposed to it. The \botl\ link is technically necessary to obtain a morphism of $\SInv(\catPN)$ whose source is $(\lone,\id_\lone)$.

We shall now prove that, for cut-free proofs, the converse of \refth{Soundness} holds in the syntactic model. We start by proving a property of the reduction of \dr s involving identity nets. We first need a couple of preliminary definitions; the first one is completely standard in rewriting theory, specialized here to \pn s following \cite{Tortora:AdditivesAndNormalizationI}.
\begin{defn}[Lift, residue]
	\label{def:Lift}
	Whenever $\pi\betared\pi'$, by simple inspection of the cut-elimination rules it is clear that any edge or link (except \cut) $x'$ of $\pi'$, comes from a unique (``the same'') edge or link $x$ of $\pi$; we say that $x$ is the \emph{lift} of $x'$, denoted by $\lift{x'}$, and that $x'$ is a \emph{residue} of $x$. When it makes sense, we define in the same way the lift and residues of a subnet.
\end{defn}
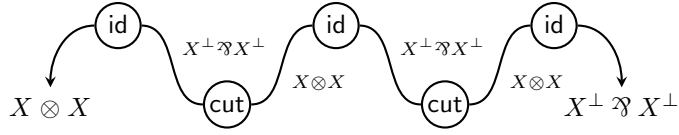
\begin{figure}[t]
	\input{Foot}
	\caption{A foot. The node $\id$ represents the net $\id_{X\ltens X}$.}
	\label{fig:Foot}
\end{figure}
\begin{defn}[Foot]
	A \emph{foot} is a net composed of a chain of three $\id_{X\ltens X}$ nets cut together as in \reffig{Foot}. The $\id_{X\ltens X}$ net whose both conclusions are premises of cuts is called the \emph{inner toe} of the foot; the other two are called \emph{outer toes}.
\end{defn}

In the following, if $\pi$ is a \ps\ of conclusions $A_1,\ldots,A_n$, none of which is a \struct, and if $\pi_1,\ldots,\pi_n$ are \ps s such that, for all $1\leq i\leq n$, $\pi_i$ has a unique occurrence of $A_i^\bot$ among its conclusions, we denote by $\mathop{\mathrm{cut}}(\pi,\pi_1,\ldots,\pi_n)$ the net obtained by juxtaposing all of the \ps s $\pi,\pi_1,\ldots,\pi_n$ and adding $n$ \cut\ links of premises $A_i,A_i^\bot$, where $A_i$ ranges over the conclusions of $\pi$.

Furthermore, we write $\pi\notaxreds\pi'$ when $\pi\betareds\pi'$ without using axiom steps (\reffig{AxStep}).
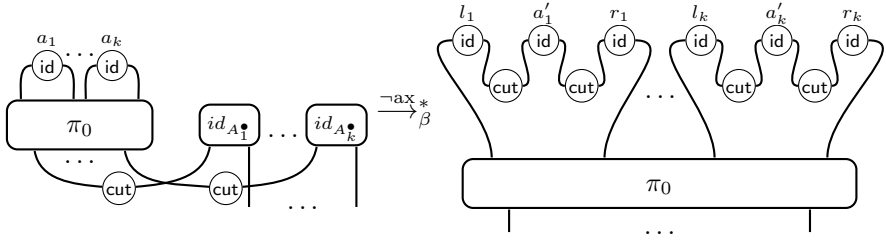
\begin{figure}[t]
%	\pict[0.75]{FeetLemma}
\input{FeetLemma}
	\caption{Graphical representation of \reflemma{Feet}. Formulas and the orientation of edges are omitted. We suppose that $\pi$ contains $k$ axioms, which become $k$ subnets of the form $\id_{X\ltens X}$ in $\pi^\bullet$, represented here by the $\id$ nodes $a_1,\ldots,a_k$; the boxes that may be present in $\pi$ are not drawn in the picture, and $\pi_0$ represents ``$\pi$ without the axioms''. For each $1\leq i\leq k$, $a_i'$ is a residue of $a_i$, and $l_i,r_i$ are residues of $\id_{X\ltens X}$ subnets of $\id_{A_i^\bullet}$.}
	\label{fig:FeetLemma}
\end{figure}
\begin{lem}
	\label{lemma:Feet}
	Let $\pi$ be a cut-free, $\eta$-expanded \ps\ of conclusions $A_1^\bullet,\ldots,A_n^\bullet$, none of which is a \struct. Then, $\mathop{\mathrm{cut}}(\pi^\bullet,\id_{A_1^\bullet},\ldots,\id_{A_n^\bullet})\notaxreds\pi'$, where $\pi'$ is obtained from $\pi^\bullet$ by replacing all occurrences of the subnet $\id_{X\ltens X}$ with feet such that, for each foot, the external toes are residues of an $\id_{X\ltens X}$ net of $\id_{A_i^\bullet}$ for some $i$, and the inner toe is the residue of the $\id_{X\ltens X}$ subnet of $\pi^\bullet$ which is replaced by that foot (see \reffig{FeetLemma}).
\end{lem}
\begin{pf}
	A straightforward induction on the size of $\pi$.\qed
\end{pf}

We now recur once more to indexings, but we modify them so that every \ps\ becomes indexable. These \emph{quasi-indexings} of course have a mere technical value: they are preserved under cut-elimination not involving axiom steps (\reflemma{QuasiInd}). Quasi-indexings were first considered by \cite{GaboardiRoversiVercelli}.
\begin{defn}[Exponential quasi-indexing]
	\label{def:QuasiInd}
	\sloppy{An \emph{exponential quasi-in\-de\-xing} is defined just as an exponential indexing (\refdef{L3Indexing}), except that it need not satisfy any constraint on axioms, \ie, if $\sigma$ is \ps, $Q$ an exponential quasi-indexing for it, and $e,e'$ the conclusions of an \axiom\ link of $\sigma$, we may have $Q(e)\neq Q(e')$.}

	Let $\sigma$ be a cut-free \ps. We may assign an index to all of its edges by proceeding as follows: we assign $0$ to all conclusions, then we ``go up'', incrementing indexes whenever we cross a \pargl, \ofcourse\ or \whynot\ link. This obviously yields an exponential quasi-indexing for $\sigma$, which is the default exponential indexing iff $\sigma$ is in \LLlev; we call it the \emph{default exponential quasi-indexing}.
\end{defn}
\begin{lem}
	\label{lemma:QuasiInd}
	Let $\pi$ be a \dr, let $Q$ be an exponential quasi-indexing for $\pi$, and let $\pi\notaxred\pi'$. Then, there exists an exponential quasi-indexing $Q'$ for $\pi'$ such that, for every edge $e$ of $\pi$, whenever $e'$ is a residue of $e$, $Q'(e')=Q(e)$.
\end{lem}
\begin{pf}
	We define $Q'$ by composing the lift function $l$, which maps edges of $\pi'$ to edges of $\pi$, with $Q$: $Q'=Q\circ l$. The claim may then be immediately verified by inspecting Figures \ref{fig:MultStep} to \ref{fig:PargStep}.\qed
\end{pf}

\begin{defn}[Swapping relation]
	\label{def:Swapping}
	We define the relation $\swapping$ on \ps s as follows: $\pi'\swapping\pi$ iff $\pi'$ is obtained from $\pi$ by replacing a non-null number of $\id_{X\ltens X}$ subnets with $\gamma_{X,X}$.
\end{defn}

We may at last prove the main result of this section:
\begin{lem}[Swapping]
	\label{lemma:Swapping}
	Let $\pi$ be a cut-free \dr\ of conclusion $A$. Suppose $\pi$ is not in \LLlev; then, there exists $k\in\Nat$ such that $\theta^A_k\circ\sem{\pi}_s\swapping\sem{\pi}_s$.
\end{lem}
\begin{pf}
	First of all, we remark that, modulo the addition of a \botl\ link, we have $\sem{\pi}_s=\pi_\eta^\bullet$, where $\pi_\eta$ is the $\eta$-expanded form of $\pi$. Observe that $\mathop{\mathrm{cut}}(\pi_\eta^\bullet,\theta^A_k)$ admits an exponential quasi-indexing $Q$ such that, if $e$ is a conclusion of an $\id_{X\ltens X}$ or $\gamma_{X,X}$ subnet of $\theta^A_k$, then $Q(e)$ is equal to the level (\refdef{Test}) of such subnet: this is obtained by joining the default exponential quasi-indexing of $\pi_\eta^\bullet$ and the default exponential indexing of $\theta^A_k$, which are compatible because they assign the same index (that is, $0$) to both premises of the \cut\ link introduced in $\mathop{\mathrm{cut}}(\pi_\eta^\bullet,\theta^A_k)$.

	Now, the fact that $\pi$ is not in \LLlev\ implies that there exists an axiom of $\pi$ whose conclusions are assigned different indexes by $Q$. The $\eta$-expansion and substitution with $\id_{X\ltens X}$ propagate the mismatch, so in $\pi_\eta^\bullet$ there is a subnet $\id_{X\ltens X}$, call it $a$, which is ``wrongly'' indexed, \ie, its conclusions $e_1,e_2$ are such that $Q(e_1)\neq Q(e_2)$.

	Let $i=Q(e_1)$ and $j=Q(e_2)$, and consider the test $\theta^A_i$ (the reader is invited to check that the same arguments would apply if we chose $\theta^A_j$); by \reflemma{Feet}, we have $\mathop{\mathrm{cut}}(\pi_\eta^\bullet,\theta^A_i)\notaxreds\pi'$ as in \reffig{FeetLemma}, modulo the fact that the outer toes of the feet present in $\pi'$ may now be $\gamma_{X,X}$ nets.

	Consider the $\id_{X\ltens X}$ subnet $a'$ of $\pi'$ which is the residue of $a$ (the ``wrongly'' indexed subnet $\id_{X\ltens X}$ of $\pi_\eta^\bullet$), and let $e_1',e_2'$ be its conclusions. By \reflemma{QuasiInd}, we have an indexing $Q'$ for $\pi'$ such that $Q'(e_1')=i$ and $Q'(e_2')=j$. Moreover, using again \reflemma{Feet}, $a'$ is the inner toe of a foot whose outer toes, which we denote by $t_1,t_2$, are residues of $\id_{X\ltens X}$ or $\gamma_{X,X}$ subnets $\lift{t_1},\lift{t_2}$ of $\theta^A_i$. We claim that $t_1$ is a $\gamma_{X,X}$ net, while $t_2$ is an $\id_{X\ltens X}$ net; this is enough to conclude, because the foot composed of $t_1,a',t_2$ reduces to a $\gamma_{X,X}$ net, and therefore the cut-free form $\pi_1$ of $\pi'$ is obtained from $\pi_\eta^\bullet$ by replacing at least one $\id_{X\ltens X}$ subnet with $\gamma_{X,X}$, which proves $\pi_1\swapping\pi_\eta^\bullet$.

	So let $d_1,d_1'$ and $d_2,d_2'$ be the conclusions of $t_1$ and $t_2$, respectively. Again by \reflemma{QuasiInd}, and by the fact that $Q$ is actually an indexing for $\theta^A_i$, we have $Q'(d_1)=Q'(d_1')$ and $Q'(d_2)=Q'(d_2')$. But even in exponential quasi-indexings the premises of \cut\ links must have the same index, so $Q'(d_1)=i$ and $Q'(d_2)=j$, which means that the levels of $\lift{t_1}$ and $\lift{t_2}$ are $i$ and $j$, respectively. Then, $t_1$ is of the form $\gamma_{X,X}$, while $t_2$ is of the form $\id_{X\ltens X}$, as claimed.\qed
\end{pf}

Since everything in the syntactic model is computed through cut-elimination, the above results give us yet another definition of \LLlev, which is of interactive nature, \ie, a net is in \LLlev\ iff it ``passes all tests''. Of course, this only works for cut-free nets.
\begin{thm}[Interactive characterization of \LLlev]
	\label{th:Interactive}
	\sloppy{Let $\pi$ be a cut-free \dr\ of conclusion $A$, whose $\eta$-expansion we denote by $\pi_\eta$.} Then, $\pi$ is in \LLlev\ iff, for all $k\in\Nat$, $\mathop{\mathrm{cut}}(\pi_\eta^\bullet,\theta^A_k)\betareds\pi_\eta^\bullet$.
\end{thm}
\begin{pf}
	We start by observing that $\sem{\pi}_s=\pi_\eta^\bullet$ (modulo the addition of a \botl\ link), so by \reflemma{SyntTests} and the definition of morphism of \obj, $\mathop{\mathrm{cut}}(\pi_\eta^\bullet,\theta^A_k)\betareds\pi_\eta^\bullet$ is equivalent to $\sem\pi_s\in\altsem{A}_s$ (recall that this notation, introduced just before \refth{Soundness}, means $\sem\pi_s\in(\SInv\catPN)[(1,\id_1),\altsem{A}_s]$). Therefore, the forward implication is a corollary of \refth{Soundness}, applied to the syntactic category \catPN. The converse is an immediate consequence of the Swapping \reflemma{Swapping}.\qed
\end{pf}

%% file: InvChain.tex
%\begin{center}
\begin{tikzpicture}[node distance = 6mm and 16mm, baseline = (current bounding box.east)]
%  \mygridto{(7,3)}
\node[lab,nopoint] (conc1)                 {$A^\perp$};
\netcut[right=of conc1,xshift=-3mm]                {cut1};
\netcut[right=of cut1,xshift=6mm]                 {cut2};
\node[lab,nopoint,right=of cut2,xshift=-2mm] (conc2)   {$A$};
\node[point,above=of conc1,yshift=1mm] (pc1) {};
\node[point,above=of conc2,yshift=1mm] (pc2) {};
\draw[thick,>=stealth,->]
     (pc1) -- (conc1);
\draw[thick,>=stealth,->]
     (pc2) -- (conc2);
\node[netpi, above=of cut1, xshift=-15mm] (sigma) {$\sigma_k$};
\node[netpi, right =of sigma, xshift=1mm] (pi) {$\pi$};
\node[netpi, right =of pi,  xshift=1mm] (tau) {$\tau_k$};
\draw[thick]
      (sigma.south) +(4mm,0) to [out=270,in=180] node [above right] {$A$}
      (cut1.west);
 \draw[thick]
      (pi.south) +(-4mm,0) to [out=270,in=0] node [above left] {$A^\perp$}
      (cut1.east);
\draw[thick]
      (pi.south) +(4mm,0) to [out=270,in=180] node [above right] {$B$}
      (cut2.west);
 \draw[thick]
      (tau.south) +(-4mm,0) to [out=270,in=0] node [above left] {$B^\perp$}
      (cut2.east);
\end{tikzpicture}
%\end{center}

%%% Local Variables: 
%%% mode: latex
%%% TeX-master: "LLtesteur"
%%% End:

%% file: IdAndFlip.tex
\begin{center}
\begin{tikzpicture}[node distance = 5mm and 8mm, baseline = (current bounding box.east)]
\netax {ax1};
\netax [above=of ax1, yshift=-3mm] {ax2};
\netpar  [below left=of ax1] {par};
\nettens [below right=of ax1] {tens};
\node[text height=1.5ex, below=of par] (conc1) {$X^\perp\parr X^\perp$};
\node[text height=1.5ex, below=of tens] (conc2) {$X\otimes X$};
\draw[thick,directed]
     (ax1) to [out=180,in=150] node [left] {$X^\perp$}
     (par);
\draw[thick,directed]
     (ax2) to [out=180,in=40] node [above left] {$X^\perp$}
     (par);
\draw[thick,directed]
     (ax1) to [out=0,in=150] node [right] {$X$}
     (tens);
\draw[thick,directed]
     (ax2) to [out=0,in=40] node [above right] {$X$}
     (tens);
\draw[->,thick,>=stealth]
     (par) edge  (conc1)
     (tens) edge (conc2);
\end{tikzpicture}\quad\quad
\begin{tikzpicture}[node distance = 5mm and 6mm, baseline = (current bounding box.east)]
\netax {ax1};
\netax [above=of ax1, yshift=-3mm] {ax2};
\netpar  [below left=of ax1] {par};
\nettens [below right=of ax1] {tens};
\node[text height=1.5ex, below=of par] (conc1) {$X^\perp\parr X^\perp$};
\node[text height=1.5ex, below=of tens] (conc2) {$X\otimes X$};
\draw[thick,directed]
     (ax1) to [out=180,in=40] node [below right] {$X^\perp$}
     (par);
\draw[thick,directed]
     (ax2) to [out=180,in=150] node [above left] {$X^\perp$}
     (par);
\draw[thick,directed]
     (ax1) to [out=0,in=150] node [right] {$X$}
     (tens);
\draw[thick,directed]
     (ax2) to [out=0,in=40] node [above right] {$X$}
     (tens);
\draw[->,thick,>=stealth]
     (par) edge  (conc1)
     (tens) edge (conc2);
\end{tikzpicture}
\end{center}

%%% Local Variables: 
%%% mode: latex
%%% TeX-master: "LLlevSem"
%%% End: 

%% file: Foot.tex
\begin{center}
\begin{tikzpicture}[node distance = 6mm and 10mm, baseline = (current bounding box.east)]
%  \mygridto{(7,3)}
\node[lab,text width=16mm,align=center] (conc1)                 {$X\otimes X$};
\netid[above right=of conc1,xshift=-12mm] {id1};
\netcut[below right=of id1] {cut1};
\netid[above right=of cut1] {id2};
\netcut[below right=of id2] {cut2};
\netid[above right=of cut2] {id3};
\node[lab,below right=of id3,text width=16mm,xshift=-12mm,align=center] (conc2)  {$X^\perp\parr X^\perp$};
\draw[thick]
      (id1) to [out=0,in=180] node [above right] {\ednd{X^\perp\parr X^\perp}}
      (cut1);
\draw[thick]
      (id2) to [out=0,in=180] node [above right] {\ednd{X^\perp\parr X^\perp}}
      (cut2);
\draw[thick,>=stealth,->]
      (id3) to [out=0,in=90] 
      (conc2);
 \draw[thick]
      (id3) to [out=180,in=0] node [below right] {\ednd{X\otimes X}}
      (cut2);
 \draw[thick]
      (id2) to [out=180,in=0] node [below right] {\ednd{X\otimes X}}
      (cut1);
 \draw[thick,>=stealth,->]
      (id1) to [out=180,in=90]
      (conc1);
\end{tikzpicture}
\end{center}

%%% Local Variables: 
%%% mode: latex
%%% TeX-master: "LLtesteur"
%%% End:

%% file: FeetLemma.tex
\begin{center}
\begin{tikzpicture}[netcircle/.style={smallnetcircle},node distance = 3mm and 2mm, baseline = (current bounding box.east)]
\node [nopoint] (piorigin) at (5.5, 3) {};
\node [netpi, below=of piorigin,text width=17mm] (pi) {$\pi_0$};
\node[nopoint,xshift=2mm] (Xconc1)  at (pi.north west)  {};
\node[nopoint,xshift=9mm] (Xconc2)  at (pi.north west) {};
\node[nopoint,xshift=-9mm] (Yconc1)  at (pi.north east)  {};
\node[nopoint,xshift=-2mm] (Yconc2)  at (pi.north east) {};
\node [nopoint, xshift=-6mm] (piconc1) at (pi.south) {};
\node [nopoint, xshift=6mm] (piconc2) at (pi.south) {};
\netid[above right=of Xconc1] {id1};
\netid[above right=of Yconc1] {id2};
\node[smallnetpi, right =of pi,xshift=4mm] (id3) {$\id_{A_1^\bullet}$};
\node[smallnetpi, right =of id3,xshift=4mm] (id4) {$\id_{A_k^\bullet}$};
\node[nopoint, xshift=1.5mm] (l3) at (id3.south west) {};
\node[nopoint, xshift=-1.5mm] (r3) at (id3.south east) {};
\node[nopoint, xshift=1.5mm] (l4) at (id4.south west) {};
\node[nopoint, xshift=-1.5mm] (r4) at (id4.south east) {};
\node[nopoint, below =of r3, yshift=-5mm] (r3c) {};
\node[nopoint, below =of r4, yshift=-5mm] (r4c) {};
\netcut[below left=of id3, xshift=-7mm, yshift=-1mm] {cut1};
\netcut[below left=of id4, xshift=-7mm, yshift=-1mm] {cut2};
\node[above =of id1, smalllab, yshift=-3mm] (labid1) {$a_1$};
\node[above =of id2, smalllab, yshift=-3mm] (labid2) {$a_k$};
\draw[thick]
      (id1) to [out=180,in=90] 
      (Xconc1);
\draw[thick]
      (id1) to [out=0,in=90] 
      (Xconc2);
\draw[thick]
      (id2) to [out=180,in=90] 
      (Yconc1);
\draw[thick]
      (id2) to [out=0,in=90] 
      (Yconc2);
\draw[thick]
      (piconc1) to [out=280, in=175]
      (cut1);
\draw[thick]
      (piconc2) to [out=280, in=175]
      (cut2);
\draw[thick]
      (l3) to [out=260, in=5]
      (cut1);
\draw[thick]
      (l4) to [out=260, in=5]
      (cut2);
\draw[thick]
      (r3) -- (r3c);
\draw[thick]
      (r4) -- (r4c);
\draw
   (r3c) edge [draw=none] node {$\ldots$} (r4c);
\draw
   (piconc1) edge [draw=none] node [below] {$\ldots$} (piconc2);
\draw
   (id3.south) edge [draw=none] node [above] {$\ldots$} (id4.south);
\draw
   (id1) edge [draw=none] node [above] {$\ldots$} (id2);
\end{tikzpicture}
$\notaxreds$\hspace{-0.5cm}
\begin{tikzpicture}[netcircle/.style={smallnetcircle},node distance = 3mm and 2mm, baseline = (current bounding box.east)]
%\mygridto{(11,5)}

%le net pi
\node [nopoint] (piorigin) at (5.5, 3) {};
\node [netpi, below=of piorigin,text width=50mm] (pi) {$\pi_0$};
\node [nopoint, xshift=-20mm] (piconc1) at (pi.south) {};
\node [nopoint, xshift=20mm] (piconc2) at (pi.south) {};
\node[nopoint,below =of piconc1] (pic1) {};
\node[nopoint,below =of piconc2] (pic2) {};
\draw[thick]
     (piconc1) -- (pic1)
     (piconc2) -- (pic2);

\node[nopoint,yshift=8mm,xshift=-29mm] (Xorigin)  at (piorigin.north)  {};
\node[nopoint,xshift=4mm] (Xconc1)  at (pi.north west)  {};
\node[nopoint,xshift=19mm] (Xconc2)  at (pi.north west) {};
\netid[above right=of Xorigin] {Xid1};
\netcut[below right=of Xid1] {Xcut1};
\netid[above right=of Xcut1] {Xid2};
\netcut[below right=of Xid2] {Xcut2};
\netid[above right=of Xcut2] {Xid3};
\node[above =of Xid1, smalllab, yshift=-3mm] (labXid1) {$l_1$};
\node[above =of Xid2, smalllab, yshift=-3mm] (labXid2) {$a'_1$};
\node[above =of Xid3, smalllab, yshift=-3mm] (labXid3) {$r_1$};
\draw[thick]
      (Xid1) to [out=0,in=180] 
      (Xcut1);
\draw[thick]
      (Xid2) to [out=0,in=180] 
      (Xcut2);
\draw[thick,>=stealth]
      (Xid3) to [out=0,in=90] 
      (Xconc2);
 \draw[thick]
      (Xid3) to [out=180,in=0] 
      (Xcut2);
 \draw[thick]
      (Xid2) to [out=180,in=0] 
      (Xcut1);
 \draw[thick,>=stealth]
      (Xid1) to [out=180,in=90]
      (Xconc1);

\node[nopoint,yshift=8mm,xshift=2mm] (Yorigin)  at (piorigin.north)  {};
\node[nopoint,xshift=-19mm] (Yconc1)  at (pi.north east)  {};
\node[nopoint,xshift=-4mm] (Yconc2)  at (pi.north east) {};
\netid[above right=of Yorigin] {Yid1};
\netcut[below right=of Yid1] {Ycut1};
\netid[above right=of Ycut1] {Yid2};
\netcut[below right=of Yid2] {Ycut2};
\netid[above right=of Ycut2] {Yid3};
\node[above =of Yid1, smalllab, yshift=-3mm] (labYid1) {$l_k$};
\node[above =of Yid2, smalllab, yshift=-3mm] (labYid2) {$a'_k$};
\node[above =of Yid3, smalllab, yshift=-3mm] (labYid3) {$r_k$};
\draw[thick]
      (Yid1) to [out=0,in=180] 
      (Ycut1);
\draw[thick]
      (Yid2) to [out=0,in=180] 
      (Ycut2);
\draw[thick,>=stealth]
      (Yid3) to [out=0,in=90] 
      (Yconc2);
 \draw[thick]
      (Yid3) to [out=180,in=0] 
      (Ycut2);
 \draw[thick]
      (Yid2) to [out=180,in=0] 
      (Ycut1);
 \draw[thick,>=stealth]
      (Yid1) to [out=180,in=90]
      (Yconc1);

\draw
   (Xcut2) edge [draw=none] node [below] {$\ldots$} (Ycut1);
\draw
   (pic1) edge [draw=none] node {$\ldots$} (pic2);

\end{tikzpicture}
\end{center}

%%% Local Variables: 
%%% mode: latex
%%% TeX-master: "LLtesteur"
%%% End:

%% file: RelComp.tex
\sloppy{Basically, the syntactic model interprets proofs as cut-free, $\eta$-expanded proofs. These latter may still be interpreted by means of a non-syntactic model, into a linear category $\catL$.} An easy but important remark is that the default interpretation of a non-syntactic model always factors through the default interpretation of the syntactic model, via a canonical interpretation of cut-free, $\eta$-expanded proofs:
\begin{displaymath}
	\xymatrix{
	\textrm{linear logic} \ar[rr]^{\textrm{\qquad synt.\ interpretation}} \ar[rrd]_{\textrm{interpretation}} & \qquad & \catPN \ar[d]^{\cX} \\
	&& \catL
	}
% 	\xymatrix{
% 	\textrm{linear logic} \ar[rr]^{\sem{\cdot}_s} \ar[rrd]_{\sem{\cdot}} & \qquad & \catPN \ar[d]^{\cX} \\
% 	&& \catL
% 	}
\end{displaymath}
Technically, the above diagram commutes in the category of $2$-graphs, recalling the fact, mentioned in the opening of \refsect{Models}, that interpretation maps may be seen as morphisms of $2$-graphs. The morphism $\cX$ is actually a functor, defined as follows.

Consider a non-trivial model of linear logic, whose linear category is $\catL$, and let $\Xi$ be an object of $\catL$ such that $\gamma_{\Xi,\Xi}\neq\id_{\Xi\ltens\Xi}$, \ie, the object used by the default assignment. We define an assignment mapping all propositional atoms to $\Xi$; then, the interpretation map of the model under this assignment yields a functor from $\catPN$ to $\catL$, which we denote by $\cX$.

When we consider \stratLL\ and the default interpretations of \refdef{DefAss} (syntactic and semantic), thanks to the $\SInv$ functor the above diagram becomes
\begin{displaymath}
	\xymatrix{
	\stratLL \ar[rr]^{\sem{\cdot}_s} \ar[rrd]_{\sem{\cdot}} & \qquad & \SInv(\catPN) \ar[d]^{\SInv(\cX)} \\
	&& \SInv(\catL)
	}
\end{displaymath}
The commutation of such a diagram is the object of the following result:
\begin{prop}[Factoring]
	\label{prop:Factoring}
	For every non-trivial model of linear logic, we have $\sem{\cdot}=\SInv(\cX)\circ\sem{\cdot}_s$ (in the category of $2$-graphs).
\end{prop}
\begin{pf}
	For clarity, we shall denote by $\odot$ the semantic tensor, \ie, the functor interpreting in $\catL$ the multiplicative conjunction $\ltens$ of linear logic.

	In the case of formulas, the proof is by induction. Let $Z$ be a propositional atom; we have $\SInv(\cX)(\sem{Z}_s)=\SInv(\cX)(X\ltens X,\xi)=(\Xi\odot\Xi,(\cX(\xi_k))_{k\in\Int})$, where $\xi_0$ is the $\gamma_{X,X}$ net and $\xi_k$ is the identity net everywhere else. But then $\cX(\xi_0)=\gamma_{\Xi,\Xi}$ and $\cX(\xi_k)$ is the identity everywhere else, so $\SInv(\cX)(\sem{Z}_s)=\sem Z$. The inductive cases are straightforward.

	For proofs, recall that $\SInv(\cX)$ acts like $\cX$ on morphisms. Then, for every \dr\ $\pi$ whose cut-free, $\eta$-expanded form is $\pi_0$, we have $\cX(\sem{\pi}_s)=\cX(\pi_0^\bullet)=\sem{\pi_0}=\sem{\pi}$.\qed
\end{pf}
The following is the semantic equivalent of \reflemma{SyntTests}, from which it follows by applying \refprop{Factoring}.
\begin{lem}
	\label{lemma:Tests}
	Consider a non-trivial model of linear logic, let $A$ be a linear logic formula, and let $\altsem A=(S,s)$. Then, for all $k\in\Int$, $s_k=\cX(\theta^A_k)$, where the nets $\theta^A_k$ are the tests introduced in \refdef{Test}.
\end{lem}
\begin{pf}
	By applying, in the order, \refdef{AltSem}, \refprop{Factoring}, \refdef{AltSem} again, \reflemma{SyntTests} and the definition of $\SInv$, we have
	\begin{align*}
		\altsem{A} &= \sem{A^+} = \SInv(\cX)(\sem{A^+}_s) = \SInv(\cX)(\altsem{A}_s) = \\
		&=\SInv(\cX)(A^\bullet,(\theta^A_k)_{k\in\Int}) = (\cX(A^\bullet),(\cX(\theta^A_k))_{k\in\Int}),
	\end{align*}
	as desired.\qed
\end{pf}

As already observed in the proof of \refth{Interactive}, the interactive characterization of \LLlev\ may be stated as follows:
\begin{center}
	$\pi$ is in \LLlev\ iff $\sem\pi_s\in\altsem A_s$,
\end{center}
where $\pi$ is a cut-free \dr\ of conclusion $A$. Thanks to \refprop{Factoring} and \reflemma{Tests}, this equivalence may be lifted to any non-syntactic model, provided the model does not ``blur out'' too many differences in proofs.
\begin{defn}[Swap-sensitivity]
	\label{def:SwapSens}
	\sloppy{A non-trivial model of linear logic is said to be \emph{swap-sensitive} if, for every cut-free, $\eta$-expanded \dr\ $\pi$, $\pi'\swapping\pi$ implies $\cX(\pi')\neq\cX(\pi)$.}
\end{defn}

Swap-sensitivity (which trivially holds in the syntactic model) is a very mild condition; it is enjoyed by all models of linear logic which we are aware of. %In \refapp{CohSwap}, we prove for instance the swap-sensitivity of
For instance, coherence spaces are swap-sensitive, which implies the swap-sensitivity of all web-based models of linear logic (relational model, finiteness spaces, etc.).

We may now give a converse of \refth{Soundness}:
\begin{thm}[Semantic characterization of \LLlev]
	\label{th:Sem}
	Let $\pi$ be a cut-free \dr\ of conclusion $A$, and fix a swap-sensitive model of linear logic. Then, $\pi$ is in \LLlev\ iff $\sem{\pi}\in\altsem{A}$.
\end{thm}
\begin{pf}
	The forward implication is \refth{Soundness}. For what concerns the converse, let $\pi$ be a cut-free linear logic \dr\ of conclusion $A$, whose $\eta$-expansion we denote by $\pi_\eta$, such that $\pi\not\in\LLlev$. By \reflemma{Swapping}, there exists $k\in\Nat$ such that $\mathop{\mathrm{cut}}(\theta^A_k,\pi_\eta^\bullet)$ reduces to a net $\pi'$ such that $\pi'\swapping\pi_\eta^\bullet$. Let $\altsem A=(S,s)$. Then, using \reflemma{Tests}, \refprop{Factoring}, and the swap-sensitivity of the model, we have
	$$s_k\circ\sem\pi=\cX(\theta^A_k)\circ\cX(\sem{\pi}_s)=\cX(\theta^A_k\circ\sem{\pi}_s)\neq\cX(\sem{\pi}_s)=\sem\pi,$$
	which proves that $\sem{\pi}\not\in\altsem{A}$.\qed
\end{pf}

%% file: CatBack.tex
\paragraph*{Monoidal functors} In the following, when declaring a category to be (symmetric) monoidal, we shall only specify the bifunctor and the unit object of the monoidal structure, leaving implicit the natural transformations accounting for associativity, left and right unit, and---if present---symmetry; these will always be denoted by $\alpha,\lambda,\rho,\gamma$, respectively. In case there is more than one monoidal structure involved, we shall use superscripts to denote to which structure these natural transformations belong; for instance, if we have two monoidal structures given by the bifunctors $\otimes$ and $\odot$, then their associativities will be denoted by $\alpha^\otimes$ and $\alpha^\odot$, respectively.

Let $(\catA,\otimes,1)$ and $(\catB,\odot,U)$ be two (symmetric) monoidal categories. A \emph{monoidal functor} from $\catA$ to $\catB$ is a triple $(F,m,u)$ where $F:\catA\rightarrow\catB$ is a functor, $m:F(-)\odot F(-)\Rightarrow F(-\otimes-)$ is a natural transformation, and $u:U\rightarrow F1$ is an arrow of $\catB$, such that the following diagrams commute in $\catB$, for all objects $A,B,C$ of $\catA$:
\begin{displaymath}
	\xymatrix{
		(FA\odot FB)\odot FC \ar[d]_{m\odot\id_{FC}} \ar[rr]^{\alpha^\odot} & & FA\odot(FB\odot FC) \ar[d]^{\id_{FA}\odot m} \\
		F(A\otimes B)\odot FC \ar[d]_{m} & & FA\odot F(B\otimes C) \ar[d]^{m} \\
		F((A\otimes B)\otimes C) \ar[rr]^{F\alpha^\otimes} & & F(A\otimes(B\otimes C))
	}
\end{displaymath}
\begin{displaymath}
	\xymatrix{
		U\odot FB \ar[d]_{u\odot\id_{FB}} \ar[r]^{\lambda^\odot} & FB \\
		F1\odot FB \ar[r]^{m} & F(1\otimes B) \ar[u]_{F\lambda^\odot}
	}
	\qquad\qquad\qquad
	\xymatrix{
		FA\odot U \ar[d]_{\id_{FA}\odot u} \ar[r]^{\rho^\odot} & FA \\
		FA\odot F1 \ar[r]^{m} & F(A\otimes 1) \ar[u]_{F\rho^\odot}
	}
\end{displaymath}
In case the natural transformation $m$ and the arrow $u$ are actually isomorphisms, we speak of a \emph{strong} monoidal functor.

If the monoidal structures on $\catA$ and $\catB$ are symmetric, one further speaks of a \emph{symmetric monoidal functor} whenever the following additional diagram commutes in $\catB$, for all objects $A,B$ of $\catA$:
\begin{displaymath}
	\xymatrix{
		FA\odot FB \ar[r]^{m} \ar[d]_{\gamma^\odot} & F(A\otimes B) \ar[d]^{F\gamma^\otimes} \\
		FB\odot FA \ar[r]^{m} & F(B\otimes A)
	}
\end{displaymath}

\paragraph*{Monoidal natural transformations} \sloppy{Let $(\catA,\otimes,1),(\catB,\odot,U)$ be (symmetric) monoidal categories,} let $(F,m,u),(G,n,v)$ be two (symmetric) monoidal functors between them, and let $\varphi:F\Rightarrow G$ be a natural transformation. We say that $\varphi$ is \emph{monoidal} if the following diagrams commute in $\catB$, for all objects $A,B$ of $\catA$:
\begin{displaymath}
	\xymatrix{
		FA\odot FB \ar[d]_{m} \ar[rr]^{\varphi_A\odot\varphi_B} & & GA\odot GB \ar[d]^{n} \\
		F(A\otimes B) \ar[rr]^{\varphi_{A\otimes B}} & & G(A\otimes B)
	}
	\qquad\qquad\qquad
	\xymatrix{
		& U \ar[dl]_{u} \ar[dr]^{v} & \\
		F1 \ar[rr]^{\varphi_1} & & G1
	}
\end{displaymath}

\paragraph*{2-categories and 2-functors} Basically, a 2-category is a category in which the homsets are themselves categories, and in which the two compositions (the one induced by the category and the one present in the homsets) interact in a certain way.

More formally, a 2-category $\catA$ is given by following data:
\begin{itemize}
	\item a collection of \emph{objects} and \emph{morphisms} between them, forming a usual category;
%	\item for each ordered pair of objects $A,B$, a collection of \emph{morphisms} which are said to be of \emph{source} $A$ and \emph{target} $B$; if $f$ is such a morphism, we write $f:A\rightarrow B$; two morphisms are said to be \emph{coinitial} (resp.\ \emph{cofinal}) if they have the same source (resp.\ target); %we say that $f$ is \emph{composable} with $g$ if the target of $f$ equals the source of $g$;
	\item for each ordered pair of morphisms $f,g:A\rightarrow B$, a collection of \emph{cells} which are said to be of \emph{horizontal source} $A$, \emph{vertical source} $f$, \emph{horizontal target} $B$, and \emph{vertical target} $g$; if $\varphi$ is such a cell, we write $\varphi:f\Rightarrow g:A\rightarrow B$; %coinitiality, cofinality, and composability are defined just as in the case of morphisms, except that for cells there are two versions of each, one horizontal, and one vertical;
% 	\item a notion of composition of morphisms which is associative and has identity elements, that is:
% 	\begin{itemize}
% 		\item for all $f:A\rightarrow B,g:B\rightarrow C$, a morphism $g\circ f:A\rightarrow C$;
% 		\item for all $f:A\rightarrow B,g:B\rightarrow C,h:C\rightarrow D$, we have
% 		\begin{eqnarray*}
% 			(h\circ g)\circ f &=& h\circ(g\circ f);
% 		\end{eqnarray*}
% 		\item for every object $A$, an \emph{identity morphism} $\id_A$ such that, for all $f:A\rightarrow B$ and $g:C\rightarrow A$, we have
% 		\begin{eqnarray*}
% 			f\circ\id_A &=& f\\
% 			\id_A\circ g &=& g;
% 		\end{eqnarray*}
% 	\end{itemize}
	\item for each morphism $f:A\rightarrow B$, an \emph{identity cell} $\iota_f:f\Rightarrow f:A\rightarrow B$;
	\item a notion of \emph{horizontal composition} of cells, denoted by $\circ$ (just like the usual composition of morphisms), giving, for all cells $\varphi:f\Rightarrow g:A\rightarrow B$ and $\psi:h\Rightarrow k:B\rightarrow C$, a cell $\psi\circ\varphi:h\circ f\Rightarrow k\circ g:A\rightarrow C$;
	\item a notion of \emph{vertical composition} of cells, denoted by $\ast$, giving, for all cells $\varphi:f\Rightarrow g:A\rightarrow B$ and $\psi:g\Rightarrow h:A\rightarrow B$, a cell $\psi\ast\varphi:f\Rightarrow h:A\rightarrow B$;
	\item both notions of composition are required to be associative;
	\item identity cells of identity morphisms are required to be neutral elements for horizontal composition, that is, for all $\varphi:f\Rightarrow g:A\rightarrow B$, we have $\varphi\circ\iota_{\id_A}=\iota_{\id_B}\circ\varphi=\varphi$;
	\item identity cells are required to be neutral elements for vertical composition, that is, for all $\varphi:f\Rightarrow g:A\rightarrow B$, we have $\varphi\ast\iota_f=\iota_g\ast\varphi=\varphi$;
	\item the two compositions must satisfy the following \emph{exchange law}, for all $\varphi:f\Rightarrow g:A\rightarrow B$, $\psi:g\Rightarrow h:A\rightarrow B$, $\theta:i\Rightarrow j:B\rightarrow C$, and $\zeta:j\Rightarrow k:B\rightarrow C$:
	$$(\zeta\circ\psi)\ast(\theta\circ\varphi)=(\zeta\ast\theta)\circ(\psi\ast\varphi).$$
\end{itemize}
In what follows, we shall often omit horizontal source and target of cells, \ie, we shall simply write $\varphi:f\Rightarrow g$ instead of $\varphi:f\Rightarrow g:A\rightarrow B$ whenever $A$ and $B$ are clear from the context. Moreover, if $f$ is a morphism, in the context of an equation concerning cells we shall abusively denote its identity cell $\iota_f$ simply by $f$.

The prototypical example of 2-category is \Cat, whose objects, morphisms and cells are categories, functors and natural transformations, respectively. One can check that symmetric monoidal categories, symmetric monoidal functors and monoidal natural transformations also form a 2-category, which is of special interest in this work, and which we denote by \SymMonCat. A further, albeit more trivial example is given by \emph{strict} monoidal categories: each such category induces a 2-category with one object, whose morphisms and cells are respectively the objects and arrows of the original category.

Just as functors are the natural notion of morphism for categories, 2-functors are the corresponding notion for 2-categories. Given two 2-categories $\catA,\catB$, a \emph{2-functor} $F$ from $\catA$ to $\catB$ is a map associating with each object $A$, morphism $f$, and cell $\varphi$ of $\catA$ an object $FA$, a morphism $Ff$, and a cell $F\varphi$ of $\catB$, in such a way that:
\begin{itemize}
	\item when restricting to objects and morphisms, $F$ is a functor in the usual sense;
	\item if $\varphi:f\Rightarrow g:A\rightarrow B$ is a cell of $\catA$, then $F\varphi:Ff\Rightarrow Fg:FA \rightarrow FB;$
	\item for every morphism $f$ of $\catA$, we have $F(\iota_f)=\iota_{Ff}$;
	\item for every horizontally-composable cells $\varphi,\psi$ of $\catA$, we have $F(\psi\circ\varphi)=F\psi\circ F\varphi$;
	\item for every vertically-composable cells $\varphi,\psi$ of $\catA$, we have $F(\psi\ast\varphi)=F\psi\ast F\varphi$.
\end{itemize}

\paragraph*{Adjunctions in 2-categories} The concept of adjunction, fundamental in category theory, can be nicely reformulated using the language of 2-categories. Let $\catA$ be a 2-category. An \emph{adjunction} in $\catA$ is a quadruple $(f,g,\eta,\varepsilon)$ where $f:\objA\rightarrow\objB$, $g:\objB\rightarrow\objA$ are morphisms ($\objA,\objB$ being objects of $\catA$) and $\eta:\id_A\Rightarrow g\circ f$, $\varepsilon:f\circ g\Rightarrow\id_B$ are cells, satisfying the so-called \emph{triangle identities} (or \emph{zig-zag identities})
\begin{eqnarray*}
	(\varepsilon\circ f)\ast(f\circ\eta) &=& f,\\
	(g\circ\varepsilon)\ast(\eta\circ g) &=& g.
\end{eqnarray*}
The morphisms $f$ and $g$ are called the left and right adjoint of the adjunction, respectively, and one writes $f\dashv g$; the endomorphisms $g\circ f$ and $f\circ g$ are called the \emph{monad} and \emph{comonad} of the adjunction; $\eta$ is called the \emph{unit} of the monad, and $\varepsilon$ the counit of the comonad.

If we apply the above formulation to \Cat, we obtain the usual notion of adjunction; moreover, the endofunctor $g\circ f$ (resp.\ $f\circ g$) is a monad in the category $A$ (resp.\ a comonad in the category $B$) in the usual sense, with $\eta$ as its unit (resp.\ $\varepsilon$ as its counit).

However, we can also apply this formulation to other 2-categories; in particular, here we are interested in \emph{symmetric monoidal adjunctions}, which are adjunctions in \SymMonCat. Spelled out, given two symmetric monoidal categories $\catA,\catB$, a symmetric monoidal adjunction is a quadruple $((F,m,u),(G,n,v),\eta,\varepsilon)$ such that $(F,m,u):\catA\rightarrow\catB$, $(G,n,v):\catB\rightarrow\catA$ are symmetric monoidal functors, $F\dashv G$ in the usual sense, but the natural transformations $\eta:\Id_\catA\Rightarrow G\circ F$, $\varepsilon:F\circ G\Rightarrow\Id_\catB$ are required to be monoidal.

An important fact that we use in our work is that 2-functors preserve adjunctions: if $\catA,\catB$ are 2-categories, $F$ is a 2-functor from $\catA$ to $\catB$, and $(f,g,\eta,\varepsilon)$ is an adjunction in $\catA$, then $(Ff,Fg,F\eta,F\varepsilon)$ is an adjunction in $\catB$, as can be immediately verified by applying $F$ to both sides of the triangle identities, and using the properties defining a 2-functor.

Another useful result, which we shall only state, is the following characterization of symmetric monoidal adjunctions among usual adjunctions (\ie, adjunctions in \SymMonCat\ among adjunctions in \Cat):
\begin{prop}
	\label{prop:SymMonAdj}
	Let $\catA,\catB$ be symmetric monoidal categories, and let $(F,m,u)$ be a symmetric monoidal functor from $\catA$ to $\catB$ such that $F$ admits a right adjoint $G$. Then, the adjunction $F\dashv G$ lifts to a symmetric monoidal adjunction $(F,m,u)\dashv(G,n,v)$ iff $(F,m,u)$ is strong.
\end{prop}